\documentclass[10pt]{article}
\usepackage[numbers]{natbib}
\usepackage{graphicx}
\usepackage{marsden_article}
\usepackage[all]{xy}

\usepackage{framed}
\usepackage{float}

\usepackage{mathrsfs}
\usepackage{epstopdf}
\usepackage{pdfsync}

\newtheorem{remark}[theorem]{Remark}

\begin{document}
\title{A Variational Principle for Extended Irreversible Thermodynamics: Heat Conducting Viscous Fluids}
\vspace{-0.2in}

\author{Fran\c{c}ois Gay-Balmaz\footnote{Division of Mathematical Sciences, Nanyang Technological University, 21 Nanyang Link, Singapore 637371; email: \url{francois.gb@ntu.edu.sg}}}


\maketitle

\vspace{-0.3in}

\begin{center}
\abstract{Extended irreversible thermodynamics is a theory that expands the classical framework of nonequilibrium thermodynamics by going beyond the local-equilibrium assumption. A notable example of this is the Maxwell-Cattaneo heat flux model, which introduces a time lag in the heat flux response to temperature gradients.
In this paper, we develop a variational formulation of the equations of extended irreversible thermodynamics by introducing an action principle for a nonequilibrium Lagrangian that treats thermodynamic fluxes as independent variables.
A key feature of this approach is that it naturally extends both Hamilton's principle of reversible continuum mechanics and the earlier variational formulation of classical irreversible thermodynamics. The variational principle is initially formulated in the material (Lagrangian) description, from which the Eulerian form is derived using material covariance (or relabeling symmetries).
The tensorial structure of the thermodynamic fluxes dictates the choice of objective rate in the Eulerian description, and plays a central role in the emergence of nonequilibrium stresses - arising from both viscous and thermal effects - that are essential to ensure thermodynamic consistency.
This framework naturally results in the Cattaneo-Christov model for heat flux. We also investigate the extension of the approach to accommodate higher-order fluxes and the general form of entropy fluxes.
The variational framework presented in this paper has promising applications in the development of structure-preserving and thermodynamically consistent numerical methods. It is particularly relevant for modeling systems where entropy production is a delicate issue that requires careful treatment to ensure consistency with the laws of thermodynamics.
\vspace{2mm}
}
\end{center}


\tableofcontents

\medskip 

\section{Introduction}

\textit{Extended irreversible thermodynamics} is a theory of nonequilibrium thermodynamics that goes beyond the local-equilibrium hypothesis. This hypothesis postulates that the local and instantaneous relations between the thermal and mechanical
properties of a physical system are the same as for a uniform system at equilibrium. The theory of nonequilibrium thermodynamics under this assumption is usually referred to as \textit{classical irreversible thermodynamics} (\citep{dGMa1969,StSc1974,Wo1975,KoPr1998}). The need of an extended theory takes its origin in the observation that the classical Fourier law of heat conduction leads to a parabolic partial differential equation for the temperature, which implies that any initial disturbance is felt instantly throughout the entire medium. This behavior, sometimes referred to as the paradox of instantaneous heat propagation, contradicts the principle of causality. While this issue can be ignored in some applications, it becomes significant for high-frequency processes and small-length scales materials. 
Similar observations apply to other processes such as diffusion or viscosity when governed by the classical Fick or Newton laws. 

To eliminate these anomalies, a damped version of Fourier's law, the Maxwell-Cattaneo law, has been proposed by introducing a heat-flux relaxation term, \citep{Ca1948,Ca1958,Ve1958}. While the resulting hyperbolic partial differential equation for the temperature solves the above paradox, it turns out that the Maxwell-Cattaneo law is not compatible with the local-equilibrium hypothesis since the classical entropy production computed under this hypothesis fails to be positive. A thermodynamically consistent theory that embodies the Maxwell-Cattaneo law can, however, be constructed that goes beyond the local-equilibrium hypothesis. The main feature of this theory, called  \textit{extended irreversible thermodynamics} \citep{JoCVLe2010}, is to promote the heat flux, and other thermodynamic fluxes, to the status of independent variables at the same level as other classical variables such as temperature. The ﬁrst approach to extended irreversible thermodynamics was made by \citep{Mu1966,Mu1967} in a classical framework and independently by \citep{Is1976} in a relativistic context.
The extension of nonequilibrium thermodynamics beyond the local-equilibrium hypothesis is an extremely lively topic, still under development, with multiple perspectives and a wide range of applications. For an overview, see \citep{Jo2020} and the references cited therein. Another prominent approach that extends nonequilibrium thermodynamics beyond the local-equilibrium assumption is \textit{rational extended thermodynamics}, as discussed in \citep{MuRu1998,RuSu2015}. This approach is based on the closure of moment equations within kinetic theory. We refer to \citep{MuWe2012}, \citep{CiJoRuVa2014}, \citep{Jo2020}, and the references therein for comparisons and discussions of various approaches to nonequilibrium theories.

The goal of the present paper is to contribute to the development of nonequilibrium thermodynamics beyond the local-equilibrium hypothesis by providing a \textit{unifying variational formulation for extended irreversible thermodynamics}. The proposed variational principle is an extension of the variational formulation for nonequilibrium thermodynamics developed in \citep{GBYo2017a,GBYo2017b,GBYo2019a} in the setting of \textit{classical irreversible thermodynamics}, i.e. under the local-equilibrium hypothesis. The latter variational formulation is itself an extension of the Hamilton principle in finite dimensional and continuum mechanics, and shares several commonalities with the Lagrange-d'Alembert principle for nonholonomic constrained systems.
Consistently with the main feature of extended irreversible thermodynamics, our variational formulation is based on a nonequilibrium Lagrangian which, in addition to the usual mechanical and thermodynamic variables, also depends on the thermodynamic fluxes of the irreversible processes involved.

Besides providing a new unifying and modeling tool in extended irreversible thermodynamics, the proposed variational formulation is also shown in this paper to make the following two contributions to the theory:
\begin{itemize}
\item[\rm (i)] It justifies the use of one particular objective rate of the heat flux, namely the Lie derivative of a vector field density, and hence directly yields the Cattaneo-Christov law of heat conduction (\citep{Ch2009}). Similar insights are also obtained for the other thermodynamic fluxes.
\item[\rm (ii)] It directly provides the nonequilibrium flux-dependent stress tensors appearing in the momentum equation besides the usual viscous stress tensors, which are needed for thermodynamic consistency in the spatial description when the objective rates for the thermodynamic fluxes are used.
\end{itemize}

The variational formulation for \textit{extended} irreversible thermodynamics that we develop shares the following features with the variational formulation for \textit{classical} irreversible thermodynamics as presented in \citep{GBYo2017a,GBYo2017b,GBYo2019a}:

\begin{itemize}
\item[\rm (i)] The variational formulation is naturally written in the material description of the continuum, while its spatial version is derived from it under the material covariance assumption;
\item[\rm (ii)] In the material description the variational formulation is a continuum version of the variational formulation for finite dimensional thermodynamic systems, and also an extension of the Hamilton principle of mechanics;
\item[\rm (iii)] The variational formulation makes crucial use of the concept of \textit{thermodynamic displacement} of an irreversible process, defined such that its rate of change equals the thermodynamic force of the process;
\item[\rm (iv)] The variational formulation is of d'Alembert type, i.e., the critical curve condition is subject to two constraints: a constraint (phenomenological constraint) on the solution curve and a constraint (variational constraint) on the variations to be considered when computing the criticality condition.
\end{itemize}

In the case of classical irreversible thermodynamics, the variational formulation can be extended to cover the treatment of open systems, see \citep{GBYo2018a}, and systems with constraints, thereby offering a relevant modeling tool in nonequilibrium thermodynamics, see \citep{GB2019,ElGB2021,GBPu2022,GBPu2024} for applications to geophysical fluids and porous media.
As shown in \citep{ElGB2020}, the variational formulation provides a systematic way to derive single (\citep{EdBe1991a,EdBe1991b}) and double (\citep{Ka1984}) generator bracket formulations in non-equilibrium thermodynamics, including the metriplectic/GENERIC double generator brackets (\citep{Mo1986,GrOt1997,OtGr1997}).

\medskip

\noindent\textbf{Variational principles related to thermodynamics.} Several variational approaches have been proposed in relation with thermodynamics. At the heart of most of these is the \textit{principle of least dissipation of energy}, as introduced in \citep{Onsager1931} and later extended in \citep{OnMa1953} and \citep{MaOn1953}, which underlies the reciprocal relations in the linear case. Another principle was formulated by \citep{Prigogine1947}, \citep{GlPr1971} as a condition on steady state processes, known as the \textit{principle of minimum entropy production}. Onsager's approach was generalized in \citep{Ziegler1968} to the case of systems with nonlinear phenomenological laws. We refer to \citep{Gyarmati1970} for reviews and developments of Onsager's variational principles, as well as for a study of the relationship between Onsager's and Prigogine's principles. In this direction, we also refer to e.g. \citep[\S6]{Lavenda1978} and \citep{Ichiyanagi1994} for overviews on variational approaches to irreversible processes. 
Unlike the variational approach we develop here, these principles do not seek to extend the Hamilton principle of classical and continuum mechanics. Rather than capturing the complete evolution of a system, they primarily focus on the entropy production equation, aiming to derive the phenomenological laws governing the irreversible processes involved. We refer to \citep{VaNy1989} and the references therein for a discussion on the construction of variational principles underlying the differential equations of physical theories.

\medskip

\noindent\textbf{Plan of the paper.} In the remaining part of the introduction we review the variational formulation for finite dimensional thermodynamic systems experiencing mechanical friction and heat conduction, as an extension of Hamilton's principle and as a preparation for its continuum version revisited in Section \ref{Sec_2} for heat conducting viscous fluids in the material description. The variational formulation of classical irreversible thermodynamics in the spatial description is then deduced under the assumption of material covariance. We also recall the general form of the energy and entropy equations for classical irreversible thermodynamics in both the material and spatial description in order to be able to later show the analogies and differences with the case of extended irreversible thermodynamics, see also Table 1. Comments are also given regarding the equivalent use of the first and second Piola-Kirchhoff stress tensors in classical irreversible thermodynamics.
In Section \ref{Sec_3} we present the variational formulation for extended irreversible thermodynamics in the material description, for continua involving the processes of viscosity and heat conduction. This formulation involves a given \textit{nonequilibrium Lagrangian density} which depends on the heat and viscous fluxes. The variational formulation is an extension of the one for classical irreversible thermodynamics reviewed in Section \ref{Sec_2}. The variational formulation in the spatial description is then derived by assuming the material covariance of the nonequilibrium Lagrangian density. The transition from the material to the spatial description directly justifies the choice of one particular objective rate for the heat flux and viscous flux. We also systematically derive from it the momentum equation for extended irreversible thermodynamics in the spatial description, which involves \textit{additional nonequilibrium stress tensors}. The general form of the energy and entropy equation in the material and spatial description are derived and compared with the corresponding expressions reviewed earlier in the classical irreversible case.

\medskip

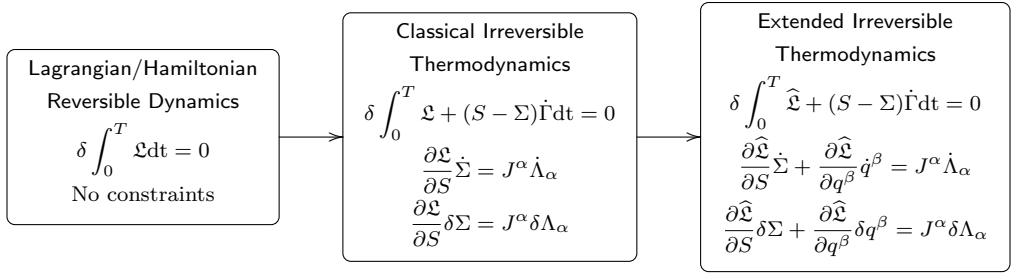
\begin{figure}[h!]
{\noindent
\footnotesize
\begin{center}
\hspace{.3cm}
\begin{xy}
\xymatrix{
*+[F-:<3pt>]{
\begin{array}{c}
\vspace{0.1cm}\textsf{Lagrangian/Hamiltonian}\\
\vspace{0.1cm}\textsf{Reversible Dynamics}\\
\vspace{0.1cm}\displaystyle\delta\int_0^T\mathfrak{L}{\rm dt}=0\\
\vspace{0.1cm}\text{No constraints}
\end{array}
}
\ar[r]
& *+[F-:<3pt>]{
\begin{array}{c}
\vspace{0.1cm}\textsf{Classical Irreversible}\\
\vspace{0.1cm}\textsf{Thermodynamics}\\
\vspace{0.1cm}\displaystyle \delta\int_0^T\mathfrak{L}+ (S-\Sigma)\dot \Gamma {\rm dt}=0\\
\vspace{0.1cm}\displaystyle \frac{\partial \mathfrak{L}}{\partial S} \dot  \Sigma =  \text{$J^\alpha \dot \Lambda_\alpha$}\\
\vspace{0.1cm}\displaystyle \frac{\partial \mathfrak{L}}{\partial S} \delta  \Sigma =  \text{$J^\alpha \delta \Lambda_\alpha$}
\end{array}
}
\ar[r]
& *+[F-:<3pt>]{
\begin{array}{c}
\vspace{0.1cm}\textsf{Extended Irreversible}\\
\vspace{0.1cm}\textsf{Thermodynamics}\\
\vspace{0.1cm}\displaystyle \delta\int_0^T\widehat{\mathfrak{L}}+ (S-\Sigma)\dot \Gamma {\rm dt}=0\\
\vspace{0.1cm}\displaystyle \frac{\partial \widehat{\mathfrak{L}}}{\partial S} \dot  \Sigma + \frac{\partial \widehat{\mathfrak{L}}}{\partial q^\beta} \dot  q^\beta =  \text{$J^\alpha \dot \Lambda_\alpha$}\\
\vspace{0.1cm}\displaystyle \frac{\partial \widehat{\mathfrak{L}}}{\partial S} \delta  \Sigma + \frac{\partial \widehat{\mathfrak{L}}}{\partial q^\beta} \delta  q^\beta =  \text{$J^\alpha \delta \Lambda_\alpha$}
\end{array}
}
&\\
}
\end{xy}
\end{center}
}
\caption{From reversible dynamics to extended irreversible thermodynamics via classical irreversible thermodynamics.}
\label{figure_1}
\end{figure}

\noindent\textbf{Illustrations.} Figure \ref{figure_1} illustrates the transition from Hamilton’s principle to the proposed variational framework for \textit{extended} irreversible thermodynamics, passing through the intermediate case of \textit{classical} irreversible thermodynamics. Here $\mathfrak{L}$ and $\widehat{\mathfrak{L}}$ denote the equilibrium and nonequilibrium Lagrangian densities, respectively. The quantities $J^\alpha$ and $X_\alpha=\dot \Lambda_\alpha$ are the thermodynamic fluxes and forces, with $\Lambda_\alpha$ being the thermodynamic displacements, associated with with certain irreversible processes $\alpha$. Additionally, $q_\beta$ denotes the thermodynamic fluxes on which the nonequilibrium Lagrangian depends. Finally $S$ and $\Sigma$ are the total and internal entropy densities. All notations will be further detailed later. It is evident that the second and third variational formulations are of the d’Alembert type. Table 1 summarizes the energy and entropy balances in each case in the material description, before any phenomenological expressions are chosen.

\pagebreak

\medskip

\begin{table}[h!]
\begin{tabular}{|c | c |} 
\hline
\textsf{$\phantom{\int^A}$Energy balance $\phantom{\int^A}$} & \textsf{Entropy balance}\\[0.25ex] 
\hline
\hline
\multicolumn{2}{|c|}{\textsf{Reversible dynamics$\phantom{\int^A}$}} \\
\hline
$\dot{\mathfrak{E}}= \operatorname{DIV} \left( \boldsymbol{P}  ^{\rm cons}\cdot \dot{ \boldsymbol{\varphi} }\right)$
&
$\phantom{\int^A \!\!\!\!}\dot S=0\phantom{\int^A \!\!\!\!}$
\\[0.25ex] 
\hline
\hline
\multicolumn{2}{|c|}{$\phantom{\int^A}$\textsf{Classical Irreversible Thermodynamics$\phantom{\int^A}$}} \\
\hline
$\dot{\mathfrak{E}}= \operatorname{DIV} \left( \boldsymbol{P}  ^{\rm tot}\cdot \dot{ \boldsymbol{\varphi} } - \mathfrak{T}  \boldsymbol{J} \right)$ &
$ \phantom{\int^A \!\!\!\!}\mathfrak{T} (\dot S + \operatorname{DIV} \boldsymbol{J}) =   \boldsymbol{S} : \boldsymbol{D}  -  \boldsymbol{J}\cdot \nabla \mathfrak{T} \geq 0\phantom{\int^A \!\!\!\!}
$
\\[0.25ex] 
\hline
\hline
\multicolumn{2}{|c|}{$\phantom{\int^A}$\textsf{Extended Irreversible Thermodynamics (classical entropy flux)$\phantom{\int^A}$}} \\
\hline
$
\dot{\widehat{\mathfrak{E}}}= \operatorname{DIV} \left( \boldsymbol{P}  ^{\rm tot}\cdot \dot{ \boldsymbol{\varphi} } - \mathfrak{T}  \boldsymbol{J} \right)
$
&
$
\phantom{\int^A \!\!\!\!}\begin{array}{l}
\phantom{\int^A \!\!\!\!}\mathfrak{T}(\dot S + \operatorname{DIV} \boldsymbol{J}  )\phantom{\int^A \!\!\!\!}\\
\phantom{\int^A \!\!\!\!}=\boldsymbol{S} : \boldsymbol{D} + \frac{\partial \widehat{\mathfrak{L}}}{\partial \boldsymbol{S} }:\dot{\boldsymbol{S}} - \boldsymbol{J}  \cdot \nabla \mathfrak{T} + \frac{\partial \widehat{\mathfrak{L}}}{\partial \boldsymbol{Q} } \cdot \dot{\boldsymbol{Q}} \geq 0\phantom{\int^A \!\!\!\!}
\end{array}\phantom{\int^A \!\!\!\!}
$
\\[0.25ex] 
\hline
\hline
\multicolumn{2}{|c|}{$\phantom{\int^A}$\textsf{Extended Irreversible Thermodynamics (general entropy flux \& higher order fluxes)$\phantom{\int^A}$}} \\
\hline
$
\hspace{0.2cm}\dot{\widehat{\mathfrak{E}}}= \operatorname{DIV} \left( \boldsymbol{P}  ^{\rm tot}\cdot \dot{ \boldsymbol{\varphi} } - \mathfrak{T}  \boldsymbol{J} \right)\hspace{0.2cm}
$
&
$
\hspace{-0.7cm}\phantom{\int^A \!\!\!\!}\begin{array}{l}
\phantom{\int^A \!\!\!\!}\mathfrak{T}(\dot S + \operatorname{DIV}( \boldsymbol{J} +\boldsymbol{J}') )\phantom{\int^A \!\!\!\!} \\
\phantom{\int^A \!\!\!\!}=\boldsymbol{S} : \boldsymbol{D} + \frac{\partial \widehat{\mathfrak{L}}}{\partial \boldsymbol{S} }:\dot{\boldsymbol{S}} - \boldsymbol{J}  \cdot \nabla \mathfrak{T} +  \frac{\partial \widehat{\mathfrak{L}}}{\partial \boldsymbol{Q} ^{(k)}} \cdot \dot{\boldsymbol{Q}}^{(k)}+ \mathfrak{T}\operatorname{DIV} \boldsymbol{J}' \geq 0\phantom{\int_A^A \!\!\!\!}
\end{array}\phantom{\int^A \!\!\!\!}\hspace{-0.7cm}
$
\\[0.25ex] 
\hline
\end{tabular}
\label{table}
\caption{Comparison of energy and entropy balances in reversible dynamics, classical, and extended thermodynamics (material frame description).}
\end{table}






\noindent\textbf{The Hamilton principle and its extension to nonequilibrium thermodynamics.} One of the most fundamental statements in classical mechanics is the principle of critical action or Hamilton's principle, according to which the motion of a mechanical system between two given positions is given by a curve that makes the integral of the Lagrangian of the system critical (see,
for instance, \citep{LaLi1969}):
\begin{equation}\label{HP_classic}
\delta \int_0^TL(q, \dot  q) {\rm d} t=0
\end{equation}
for arbitrary infinitesimal variations $ \delta q$ with $ \delta q(0)= \delta q(T)=0$. An application of \eqref{HP_classic} directly yields the Euler-Lagrange equations 
\begin{equation}\label{EL} 
\frac{d}{dt} \frac{\partial L}{\partial \dot  q ^i  } - \frac{\partial L}{\partial q ^i }=0.
\end{equation} 

\medskip 

An extension of \eqref{HP_classic} and \eqref{EL} to non-equilibrium thermodynamics was proposed in \citep{GBYo2017a}. In the simplest case of a thermo-mechanical system with mechanical variables $q, v$, one entropy variable $S$, and a friction force $F(q, \dot  q, S)$, this extension reads as follows. Find the curves $q(t)$ and $S(t)$ which are critical for the \textit{variational condition}
\begin{equation}\label{VCond}
\delta \int_0^TL(q, \dot  q, S) {\rm d} t=0,
\end{equation} 
subject to the \textit{phenomenological constraint}
\begin{equation}\label{PC} 
\frac{\partial L}{\partial S} \dot  S = F (q, \dot  q, S) \cdot \dot  q
\end{equation} 
and for variations subject to the \textit{variational constraint}
\begin{equation}\label{VC} 
\frac{\partial L}{\partial S} \delta   S = F (q, \dot  q, S) \cdot \delta   q,
\end{equation}
with $\delta q(0)=\delta q(T)=0$.

A direct application of \eqref{VCond}--\eqref{VC} yields the coupled mechanical and thermal equations
\begin{equation}\label{thermo_mech} 
\frac{d}{dt} \frac{\partial L}{\partial \dot  q ^i  } - \frac{\partial L}{\partial q ^i }=F_i \qquad\text{and}\qquad \frac{\partial L}{\partial S} \dot  S = F (q, \dot  q, S) \cdot \dot  q.
\end{equation}
In this generalized variational formulation, the temperature is defined as $T=- \frac{\partial L}{\partial S}$, which is positive for physically relevant Lagrangians. The rate of entropy production of the system is $\dot S= - \frac{1}{T}  F (q, \dot  q, S) \cdot \dot  q$ hence from the second law $F$ must be dissipative. When the Lagrangian has the standard form
\begin{equation}
L(q,v,S)= K(q,v)- U(q,S),
\end{equation}
where the kinetic energy $K$ is assumed to be independent of $S$ and $U(q, S)$ is the internal energy, then $T =-\frac{\partial L}{\partial S}=\frac{\partial U}{\partial S}$ recovers the standard definition of the temperature in thermodynamics. We refer to \citep{GBYo2017a,GBYo2019a} for comments about the structure of the variational principle \eqref{VCond}--\eqref{VC}, the relation with the Lagrange-d'Alembert principle in nonholonomic mechanics, as well as several applications.
We refer the reader to \citep[Remark 2.2]{GB2019} for a discussion of the difference between this variational formulation and the variational-like approach based on Rayleigh dissipation functions.

\medskip 

We now write the extension of \eqref{VCond}--\eqref{VC} to systems with friction and heat conduction which plays an important guiding role for the present paper, see  \citep{GBYo2017a,GBYo2019a,GBYo2023}. Assume that the thermodynamic system is made of several subsystems indexed by $A=1,...,N$, each of them characterized by a configuration variable $q_A$ and an entropy variable $S_A$. The Lagrangian is hence a function of the form $L(q_1,...,q_N, \dot  q_1,..., \dot  q_N, S_1,...,S_N)$.
The system involves the following irreversible processes: friction forces $F^{A}$ experienced by subsystem $A$ and heat exchanges between subsystems $A$ and $B$ described by fluxes $J_{AB}$ such that $J_{AB}=J_{BA}$.

To incorporate heat exchange, the new variables $\Gamma ^A$, $A=1,...,N$ are introduced, called the \textit{thermal displacements}, such that their time rate of changes coincide with the temperature of each subsystem, $\dot \Gamma ^A=T^A$, see also Remark \ref{remark_gamma}. This is an example of a   \textit{thermodynamic displacement} for an irreversible process, introduced in \citep{GBYo2017a,GBYo2019a}, defined so that its time rate of change coincides with thermodynamic force of the process. The introduction of $\Gamma^A$ is accompanied with the introduction of an entropy variable $\Sigma_A$ whose meaning and importance will be clarified later.

The variational formulation for a system with friction and heat conduction is stated as follows.
Find the curves $q_A(t)$, $S_A(t)$, $\Gamma^A(t)$, $\Sigma_A(t)$ which are critical for the \textit{variational condition}
\begin{equation}\label{VCond_conduction}
\delta \int_0^T \!\Big[ L\left(q_1,...,q_N, \dot  q_1,..., \dot  q_N,S_1,...,S_{K}\right)+  \sum_{A=1}^N\dot{\Gamma }^A( S_A- \Sigma  _A)\Big] {\rm d}t=0
\end{equation}
subject to the \textit{phenomenological constraint}
\begin{equation}\label{PC_conduction}
\frac{\partial L}{\partial S_A}\dot \Sigma_A  =  F^{A}\!\cdot\! \dot q_A   + \sum_{B=1}^NJ_{AB}\dot\Gamma^B, \quad \text{for $A=1,...,N$},
\end{equation}
and for variations subject to the \textit{variational constraint}
\begin{equation}\label{VC_conduction}
\frac{\partial L}{\partial S_A}\delta \Sigma_A  =  F^{A}\!\cdot\! \delta q_A   + \sum_{B=1}^N J_{AB}\delta\Gamma^B, \quad \text{for $A=1,...,N$},
\end{equation}
with $\delta q_A(0)=\delta q_A(T)=0$ and $ \delta \Gamma^A(0)=\delta \Gamma^A(T)=0$, $A=1,...,N$.

A direct application of \eqref{VCond_conduction}--\eqref{VC_conduction} yields the coupled mechanical and thermal equations
\begin{equation}\label{thermo_mech_nonsimple} 
\frac{d}{dt}\frac{\partial L}{\partial \dot q_A}- \frac{\partial L}{\partial q_A}= F^{A}\quad\text{and}\quad \frac{\partial L}{\partial S_A}\dot S_A=  F^{A} \!\cdot\!\dot q_A - \sum_{B=1}^P J_{AB}\left(\frac{\partial L}{\partial S_B} - \frac{\partial L}{\partial S_A}\right),
\end{equation}
for all $A=1,...,N$. In particular, the variations $ \delta S_A$ and $ \delta \Gamma ^A$ give
\begin{equation}\label{two_relations} 
\delta S_A: \;\; \frac{\partial L}{\partial S_A} = - \dot  \Gamma ^A, \qquad \delta  \Gamma ^A :\;\; \dot  S_A= \dot  \Sigma _A + \sum_B J_{AB}, \quad A=1,...,N.
\end{equation} 

Recalling that $- \frac{\partial L}{\partial S_A} =T^A$ is the temperature of the $A$-th subsystem, its entropy balance is found from \eqref{thermo_mech_nonsimple} as
\begin{equation}
\dot  S_A= - \frac{1}{T^A} F^A \cdot \dot  q_A + \sum_B J_{AB} \frac{T^A-T^B}{T^A} , \quad A=1,...,N.
\end{equation}
The second law does not imply $ \dot  S_A \geq 0$ since the $A$-th subsystem is not adiabatically closed. To apply the second law, one has to identify the rate of internal (as opposed to total) entropy production. In our variational approach, the rate of internal entropy production is identified with $ \dot  \Sigma _A$ and from the second condition \eqref{two_relations} one directly obtains the rate of internal entropy production as
\begin{equation}
\dot  \Sigma _A = - \frac{1}{T^A} F^A \cdot \dot  q_A -  \sum_B J_{AB} \frac{T^B}{T^A},
\end{equation}
which, from the second law must satisfy $ \dot  \Sigma _A\geq 0$. In terms of the Prigogine equation $ {\rm d} S_A= {\rm d} _iS_A+ {\rm d} _eS_A$ of the $A$-th subsystem, we have $ {\rm d} S_A= \dot  S_A {\rm d} t$ and $ {\rm d} _iS_A= \dot  \Sigma _A {\rm d} t$.
We refer to \citep{GBYo2023} for more details, including the phenomenological relations for $F^{A}$ and $J_{AB}$.

\section{Geometric variational formulation of classical irreversible thermodynamics}\label{Sec_2}

In this section, we revisit the variational formulation of fluid dynamics in the \textit{reversible case} and its extension to \textit{classical irreversible thermodynamics}, as presented by \citep{GBYo2017b}, with an emphasis on covariance assumptions. Covariance plays a crucial role in linking the material and Eulerian descriptions of nonequilibrium thermodynamic fluxes and forces. This discussion also serves to introduce several concepts and notations that will be used in the context of extended irreversible thermodynamics. Based on the framework outlined here, we will later highlight the analogies and differences between the classical and extended cases, particularly regarding the Lagrangian variational framework and the balance of energy and entropy.  A systematic comparison of these cases is essential for guiding the development of extended irreversible thermodynamics, see also Table 1.

For both the reversible and irreversible cases, we begin by presenting the variational formulation in the material description, since it is in this description that it takes its simpler form, namely, it is a direct extension of the variational formulation \eqref{HP_classic} for finite-dimensional \textit{mechanical} systems and \eqref{VCond_conduction}--\eqref{VC_conduction} for finite dimensional \textit{thermodynamic} systems. While our focus in this paper is on fluids, the approach naturally extends to more general continua, see Remark \ref{rmk_continua}.

To simplify the exposition we work with domains in $ \mathbb{R} ^3  $ and Euclidean metrics. The intrinsic differential geometric formulation of continuum mechanics in terms of Riemannian metrics is however crucial to discuss the notions of covariance and their implications on variational formulations, see Remark \ref{rmk_FI}.

\subsection{Review of the reversible case}

We consider here the variational formulation underlying fluid motion in the reversible case, both in the material and spatial descriptions.

\medskip

\noindent\textbf{Material description.} In the material description the configuration of a continuum, fluid or solid, at time $t \in [0,T]$ is described by a time dependent map $\boldsymbol{\varphi}(t) : \mathcal{B} \rightarrow \mathbb{R} ^3  $ assumed in this paper to be a smooth embedding. This map gives the current location $ \mathbf{x} = \boldsymbol{\varphi} (t, \mathbf{X} ) \in \mathbb{R} ^3$ of each point of the continuum, with $ \mathbf{X} \in  \mathcal{B} $ the labels of such points. From the configuration map, the Lagrangian velocity $ \boldsymbol{V}(t, \mathbf{X} )$ and the deformation gradient $ \boldsymbol{F}(t, \mathbf{X} )$ are defined as 
\begin{equation}
\boldsymbol{V}(t, \mathbf{X} )= \dot { \boldsymbol{\varphi} }(t, \mathbf{X} )= \frac{d}{dt} \boldsymbol{\varphi} (t, \mathbf{X} ), \qquad \boldsymbol{F}(t, \mathbf{X} )= \nabla \boldsymbol{\varphi} (t, \mathbf{X} ). 
\end{equation}

The material description also involves the mass density $ \varrho ( t,\mathbf{X} )$ and entropy density $S( t, \mathbf{X} )$ of the continuum. These fields are constant in time in the reversible case in the material description, i.e. $ \varrho (t, \mathbf{X} )= \varrho ( \mathbf{X} )$ and $S(t, \mathbf{X} )= S( \mathbf{X} )$. We can thus write
\begin{equation}\label{mat_cons}
\frac{d}{dt} \varrho =0, \qquad \frac{d}{dt} S=0,
\end{equation}
which are the statements of mass and entropy conservation in the material frame.

For the case of an Euler fluid, the Lagrangian density in the material description takes the form
\begin{equation}\label{Mat_Lag_Rev} 
\mathfrak{L}( \boldsymbol{\varphi}  ,\dot{\boldsymbol{\varphi}} , \nabla{ \boldsymbol{\varphi} } , \varrho , S)= \frac{1}{2} \varrho | \dot{\boldsymbol{\varphi}} | ^2 - \mathscr{E}(\nabla{ \boldsymbol{\varphi} }, \varrho , S)\varrho.
\end{equation}
The two terms in \eqref{Mat_Lag_Rev} represent the kinetic energy density and internal energy density, with $\mathscr{E}$ written here as a function of the deformation gradient $ \boldsymbol{F}= \nabla \boldsymbol{\varphi} $, mass density $ \varrho $, and entropy density $S$.

\medskip

\noindent\textbf{Variational formulation in the material description.} The equations of evolution follow from Hamilton's principle
\begin{equation}\label{HP} 
\left. \frac{d}{d\varepsilon}\right|_{\varepsilon=0}  \int_0^ T\!\! \int_ \mathcal{B} \mathfrak{L}( \boldsymbol{\varphi} _ \varepsilon  ,\dot{\boldsymbol{\varphi}}_ \varepsilon  , \nabla{ \boldsymbol{\varphi} }_ \varepsilon  , \varrho , S) {\rm d} ^3 \mathbf{X}\, {\rm d} t=0
\end{equation} 
with respect to variations $ \boldsymbol{\varphi} _ \varepsilon (t, \mathbf{X} )$ of the configuration map that are held fixed at $t=0,T$. This principle is the continuum counterpart of Hamilton's principle \eqref{HP_classic} in classical mechanics. This becomes most evident by defining the Lagrangian function
\begin{equation}\label{Lag_function} 
L( \boldsymbol{\varphi} , \dot{ \boldsymbol{\varphi} })= \int_ \mathcal{B} \mathfrak{L}( \boldsymbol{\varphi},\dot{\boldsymbol{\varphi}}, \nabla{ \boldsymbol{\varphi} } , \varrho , S) {\rm d} ^3 \mathbf{X},
\end{equation} 
associated to $\mathfrak{L}$, which is the analogue to $L(q, \dot  q)$ in \eqref{HP_classic}. Definition \eqref{Lag_function} also highlights that $ \boldsymbol{\varphi} $ is the only dynamic field in the material description, while $ \varrho $, $S$ are treated as fixed fields in Hamilton’s principle and are therefore not subject to variation.
 
The variational principle \eqref{Mat_Lag_Rev} produces the Euler-Lagrange equations and boundary conditions
\begin{equation}\label{EL_BC_mat} 
\frac{d}{dt}\frac{ \partial   \mathfrak{L}  }{ \partial   \dot{ \boldsymbol{\varphi} } }   - \frac{ \partial   \mathfrak{L} }{\partial \boldsymbol{\varphi}  }= -\operatorname{DIV} \frac{\partial \mathfrak{L}}{\partial \nabla  \boldsymbol{\varphi}  } \quad\text{on}\quad  \mathcal{B},  \qquad \frac{\partial \mathfrak{L}}{\partial  \boldsymbol{\varphi}_{,A}^a  } \boldsymbol{N}_A \delta \boldsymbol{\varphi} ^a=0  \quad\text{on}\quad  \partial \mathcal{B},
\end{equation} 
with $ \boldsymbol{N}$ the outward pointing unit normal vector field to $\mathcal{B} $ and $\operatorname{DIV}$ the divergence operator in the material frame.
The actual content of the second condition depends on the set of allowed variations at the boundary $ \partial \mathcal{B} $, see Remark \ref{rmk_BC}. For the fluid Lagrangian \eqref{Mat_Lag_Rev} we get
\begin{equation}
\varrho\, \ddot { \boldsymbol{\varphi} } = \operatorname{DIV}\big( \boldsymbol{P}  ^{\rm cons} \big) \quad\text{on}\quad  \mathcal{B},  \qquad  (\boldsymbol{P}  ^{\rm cons})^A_a\boldsymbol{N}_A \delta \boldsymbol{\varphi} ^a=0  \quad\text{on}\quad  \partial \mathcal{B},
\end{equation}
where we introduced the \textit{first Piola-Kirchhoff stress tensor}
\begin{equation}\label{PK1_fluid} 
\boldsymbol{P}^{\rm cons} =- \frac{\partial \mathfrak{L}}{\partial \boldsymbol{F} }=\varrho \frac{\partial \mathscr{E}}{ \partial   \boldsymbol{F}}, \qquad (\boldsymbol{P}^{\rm cons})^A_a =- \frac{\partial \mathfrak{L}}{\partial \boldsymbol{F}^a_A }=\varrho \frac{\partial \mathscr{E}}{ \partial   \boldsymbol{F}^a_A}.
\end{equation}
The notation "cons" refers to a conservative stress tensor to distinguish it from dissipative (irreversible) stress tensors that will be introduced later. For fluids, $ \boldsymbol{P}^{\rm cons}$ is related to the pressure, see \eqref{P_cons_p} below. 

\begin{remark}[Material frame indifference]\label{rmk_FI}{\rm The Lagrangian of a continuum must satisfy the axiom of \textit{material frame indifference}, whose most general statement is given by covariance under diffeomorphisms of the ambient space $\mathbb{R} ^3$, \citep{MaHu1983}. For this notion of material frame indifference, the diffeomorphisms act on the configuration maps $ \boldsymbol{\varphi} $ by composition on the left, and also on the Riemannian metric of the ambient space, given here by the Euclidean metric $ \delta _{ab}$ on $ \mathbb{R} ^3  $. This form of material frame indifference and its implications are best described by formulating continuum mechanics in an intrinsic differential geometric way, in terms of an arbitrary Riemannian metric $g$ on $ \mathbb{R} ^3  $. We refer to \citep{MaHu1983} for such a differential geometric treatment of frame indifference and material covariance (see below) and to \citep{GBMaRa2012} for the relations between these notions of covariance and the material, convective, and spatial variational formulations for reversible continuum mechanics. 
To simplify the exposition, in the present paper we have chosen to work with domains in $ \mathbb{R} ^3  $ and Euclidean metrics. The differential geometric treatment of the approach developed here for extended irreversible thermodynamics will be carried out elsewhere.}
\end{remark} 

One of the consequence of material frame indifference is the following symmetry of the first Piola-Kirchhoff stress tensor
\begin{equation}\label{prop_P}
(\boldsymbol{P}^{\rm cons}) ^{aA} \boldsymbol{F} ^b_A=(\boldsymbol{P}^{\rm cons})^{bA} \boldsymbol{F} ^a_A,
\end{equation} 
where $(\boldsymbol{P}^{\rm cons}) ^{aA}= \delta^{ab} (\boldsymbol{P}^{\rm cons}) ^A_b$.
This is equivalent to the symmetry of the \textit{second Piola-Kirchhoff stress tensor}, which is another frequently used measure of stress, defined by
\begin{equation}\label{def_PK2_cons} 
(\boldsymbol{S}^{\rm cons})^{AB}=  (\boldsymbol{P}^{\rm cons}) ^{aA} (\boldsymbol{F}^{-1} )^B_a.
\end{equation}

\begin{remark}[Boundary conditions]\label{rmk_BC}{\rm If we consider the case of a continuum moving in a fixed domain $ \mathcal{D} = \boldsymbol{\varphi} (t, \mathcal{B} )$ for all $t$ (tangential boundary conditions), then the variations $ \delta \boldsymbol{\varphi} |_{ \partial \mathcal{B} }$ in \eqref{EL_BC_mat} are arbitrary vector fields tangent to the boundary. This is what is frequently assumed in the reversible case. In presence of viscosity, one usually further assumes that the boundary is held pointwise fixed (no-slip boundary conditions), in which case one has $ \delta \boldsymbol{\varphi} |_{ \partial \mathcal{B} }=0$ and hence the second condition in \eqref{EL_BC_mat} vanishes.}
\end{remark} 

\noindent\textbf{Spatial description.} Besides material frame indifference, isotropic continua also satisfy \textit{material covariance}, i.e. covariance with respect to diffeomorphisms of $ \mathcal{B} $. In terms of the material Lagrangian density $\mathfrak{L}$, material covariance reads
\begin{equation}\label{mat_cov} 
\mathfrak{L}\big( \boldsymbol{\varphi} \circ \boldsymbol{\psi}  ,\dot{\boldsymbol{\varphi}}\circ \boldsymbol{\psi}  , \nabla( \boldsymbol{\varphi} \circ \boldsymbol{\psi} ) , (\varrho \circ \boldsymbol{\psi} )J_{ \boldsymbol{\psi} } , (S \circ \boldsymbol{\psi} )J_{ \boldsymbol{\psi} }\big)=\big(\mathfrak{L}\big( \boldsymbol{\varphi}  ,\dot{\boldsymbol{\varphi}} , \nabla{ \boldsymbol{\varphi} } , \varrho , S\big) \circ \boldsymbol{\psi} \big)J_{ \boldsymbol{\psi} },
\end{equation} 
for all diffeomorphisms $ \boldsymbol{\psi} : \mathcal{B} \rightarrow \mathcal{B}$. We note that such a diffeomorphism $ \boldsymbol{\psi} $ acts on the configuration map $ \boldsymbol{\varphi} $ by composition on the right, i.e., by relabeling, and on the densities $ \varrho $, $S$, $\mathfrak{L}$ by composition on the right followed by multiplication by the Jacobian $J_ { \boldsymbol{\psi} }= \operatorname{det}( \nabla \boldsymbol{\psi} )$.

From the material covariance property \eqref{mat_cov} we can associate to $\mathfrak{L}$ a Lagrangian density $\ell$ in the spatial description defined by
\begin{equation}\label{def_ell} 
\mathfrak{L}\big( \boldsymbol{\varphi}  ,\dot{\boldsymbol{\varphi}} , \nabla{ \boldsymbol{\varphi} } , \varrho , S\big) =\big(\ell( \boldsymbol{u}  , \rho  , s) \circ \boldsymbol{\varphi} \big)J_{ \boldsymbol{\varphi} }
\end{equation}
with $\boldsymbol{u}(t, \mathbf{x} )$ the velocity $ \rho (t, \mathbf{x} ) $ the mass density and $s(t, \mathbf{x} )$ the entropy density in the spatial description, given by
\begin{equation}\label{mat_to_spat}
\boldsymbol{u} = \dot  {\boldsymbol{\varphi}} \circ \boldsymbol{\varphi} ^{-1} , \quad \rho  = (\varrho \circ \boldsymbol{\varphi} ^{-1} ) J_{ \boldsymbol{\varphi} ^{-1} } , \quad s  = (S \circ \boldsymbol{\varphi} ^{-1} ) J_{ \boldsymbol{\varphi}^{-1}  }.
\end{equation}
These relations are more explicitly written as
\begin{equation}\label{mat_to_spat_explicit} 
\begin{aligned}
\boldsymbol{u}( t,\boldsymbol{\varphi} (t, \mathbf{X} ))&= \dot { \boldsymbol{\varphi} }(t, \mathbf{X} )\\
\rho  (t, \boldsymbol{\varphi} (t, \mathbf{X} ))J_{ \boldsymbol{\varphi} }(t,X)&= \varrho ( \mathbf{X} )\\
s  (t, \boldsymbol{\varphi} (t, \mathbf{X} ))J_{ \boldsymbol{\varphi} }(t,X)&= S ( \mathbf{X} )
\end{aligned} 
\end{equation} 
and represent the standard relations between the material and spatial quantities.

For the Lagrangian \eqref{Mat_Lag_Rev}, since the first term is clearly material covariant, the property \eqref{mat_cov} is equivalent to the following condition on the internal energy function
\begin{equation}
\mathscr{E}(\nabla( \boldsymbol{\varphi}  \circ \boldsymbol{\psi }) , (\varrho \circ \boldsymbol{\psi} )J_{ \boldsymbol{\psi} }, (S \circ \boldsymbol{\psi} )J_{ \boldsymbol{\psi} })= \mathscr{E}(\nabla{ \boldsymbol{\varphi} }, \varrho , S)\circ \boldsymbol{\psi} 
\end{equation}
for all diffeomorphisms $ \boldsymbol{\psi} : \mathcal{B} \rightarrow \mathcal{B} $. Thus, we can write it as
\begin{equation}
\mathscr{E}(\nabla{ \boldsymbol{\varphi} }, \varrho , S)\varrho  =\big( \varepsilon ( \rho  , s) \circ \boldsymbol{\varphi} \big)J_{ \boldsymbol{\varphi} }
\end{equation}
in terms of an internal energy density expression $ \varepsilon ( \rho  , s)$. In particular, material covariance for fluids recovers the fact that the internal energy function $\mathscr{E}(\nabla{ \boldsymbol{\varphi} }, \varrho , S)$ in the material description depends on the deformation gradients only via its Jacobian $J_ {\boldsymbol{\varphi}}= \operatorname{det}( \nabla \boldsymbol{\varphi} )$. 

In conclusion the spatial Lagrangian associated to \eqref{Mat_Lag_Rev} via \eqref{def_ell} has the standard form
\begin{equation}\label{ell} 
\ell( \boldsymbol{u}, \rho  , s)= \frac{1}{2} \rho  | \boldsymbol{u}| ^2 - \varepsilon ( \rho  , s).
\end{equation}

\noindent\textbf{Variational formulation in the spatial description.} In terms of the spatial variables, Hamilton's principle \eqref{HP} reads
\begin{equation}\label{red_HP} 
\delta \int_0^ T\!\! \int_ \mathcal{B}\ell( \boldsymbol{u}, \rho  , s) {\rm d} ^3 \mathbf{x}    \, {\rm d} t=0 
\end{equation} 
for variations $ \delta \boldsymbol{u}$, $ \delta \rho  $, and $ \delta s$ induced by the arbitrary variations $ \delta \boldsymbol{\varphi} = \left. \frac{d}{d\varepsilon}\right|_{\varepsilon=0} \boldsymbol{\varphi} _ \varepsilon $ of the configuration map $ \boldsymbol{\varphi} $ in \eqref{HP}. From \eqref{mat_to_spat}, or \eqref{mat_to_spat_explicit}, one obtains the constrained spatial variations
\begin{equation}\label{red_var} 
\delta \boldsymbol{u}= \partial _t \boldsymbol{\zeta} + \boldsymbol{u} \cdot \nabla \boldsymbol{\zeta}- \boldsymbol{\zeta } \cdot \nabla \boldsymbol{u}  , \qquad \delta   \rho  = - \operatorname{div}( \rho  \boldsymbol{\zeta} ), \qquad \delta s= - \operatorname{div}( s \boldsymbol{\zeta} )
\end{equation} 
with
\begin{equation}\label{zeta} 
\boldsymbol{\zeta} = \delta \boldsymbol{\varphi} \circ \boldsymbol{\varphi} ^{-1} \quad\text{i.e.}\quad \boldsymbol{\zeta} (t, \boldsymbol{\varphi} (t, \mathbf{X} ))= \delta \boldsymbol{\varphi} (t, \mathbf{X} ).
\end{equation}
A direct application of the variational principle \eqref{red_HP}--\eqref{red_var}  yields the fluid momentum equation in spatial description
\begin{equation}\label{spat_EL}
\partial _t \frac{\partial \ell}{\partial \boldsymbol{u} }  + \pounds _ {\boldsymbol{u}} \frac{\partial \ell}{\partial \boldsymbol{u} }= \rho  \nabla \frac{\partial \ell}{\partial \rho   }+ s \nabla \frac{\partial \ell}{\partial s}
\end{equation} 
while from \eqref{mat_to_spat_explicit} follows the mass and entropy conservation laws in the spatial description
\begin{equation}
\partial _t \rho  + \operatorname{div}( \rho  \boldsymbol{u})=0, \qquad \partial _t s + \operatorname{div}( s \boldsymbol{u})=0.    
\end{equation}
In \eqref{spat_EL} the second term is the Lie derivative of the fluid momentum density $ \frac{\partial \ell}{\partial \boldsymbol{u}}$ in the direction $ \boldsymbol{u}$, explicitly given by
\begin{equation}\label{Lie_derivative} 
\pounds _ {\boldsymbol{u}} \frac{\partial \ell}{\partial \boldsymbol{u} }= \boldsymbol{u} \cdot \nabla \frac{\partial \ell}{\partial \boldsymbol{u} } + \nabla \boldsymbol{u}^\mathsf{T} \frac{\partial \ell}{\partial \boldsymbol{u} } + \frac{\partial \ell}{\partial \boldsymbol{u} } \operatorname{div}( \boldsymbol{u}).
\end{equation}
Although the form of the constrained variations \eqref{red_var} induced by the free variations $\delta \boldsymbol{\varphi} $ follow from a direct computation using the relations \eqref{mat_to_spat_explicit}, additional insight is gained by interpreting the transition from Hamilton principle \eqref{Lag_function} to the principle \eqref{red_HP}--\eqref{red_var} as an application of the process of Euler-Poincar\'e reduction for systems on Lie groups, see \citep{HoMaRa1998}. This perspective originates from the geometric interpretation of the Euler equations on diffeomorphism groups, as introduced by \citep{Ar1966}.

For the fluid Lagrangian \eqref{ell}, the momentum equation \eqref{spat_EL} recovers the compressible Euler equation
\begin{equation}
\rho  ( \partial _t  \boldsymbol{u}+ \boldsymbol{u} \cdot \nabla \boldsymbol{u})=- \nabla p   
\end{equation}
with $p= \rho  \frac{\partial \varepsilon }{\partial \rho  }+s  \frac{\partial \varepsilon }{\partial s  }- \varepsilon $ the pressure. Going back to the material description, for fluids the first Piola-Kirchhoff stress tensor is found as
\begin{equation}\label{P_cons_p} 
(\boldsymbol{P}^{\rm cons})^A_a= - (p \circ \boldsymbol{\varphi } )J_ { \boldsymbol{\varphi}}  (\boldsymbol{F} ^{-1} )^A_a .
\end{equation}

\begin{remark}[General continua]\label{rmk_continua}{\rm For a unified treatment of fluids and elasticity, one needs to introduce on $ \mathcal{B} $ a Riemannian metric $ \mathbf{G}( \mathbf{X} )= \mathbf{G} _{AB} {\rm d} X^A  {\rm d} X^B $, besides the densities $ \varrho ( \mathbf{X} )$ and $S( \mathbf{X} )$. Material covariance \eqref{mat_cov} naturally extends to this additional dependence and yields the introduction of the Cauchy deformation tensor $c(t, \mathbf{x} )$ in the spatial description. This generalization can be easily carried out throughout the present paper, see \citep{GBMaRa2012} for the reversible case.}
\end{remark}

\subsection{Classical irreversible thermodynamics}

We review here how the Hamilton principle \eqref{HP} and its spatial version \eqref{red_HP}--\eqref{red_var} are extended to classical irreversible thermodynamics by following \citep{GBYo2017b}. We focus on the case of a heat conducting viscous fluid and assume that the fluid motion is in a fixed domain, with no-slip boundary conditions:
\begin{equation}\label{No_slip} 
\frac{d}{dt} \boldsymbol{\varphi}(t, \mathbf{X} ) =0, \quad \text{for all} \quad  \mathbf{X} \in  \partial  \mathcal{B}.
\end{equation} 

\medskip

\noindent\textbf{Variational formulation in the material description.} Besides the Lagrangian, in the irreversible case one needs to introduce the thermodynamic fluxes associated with the irreversible processes considered. In the case of viscosity and heat conduction the corresponding fluxes in the material description are given by the viscous Piola-Kirchhoff tensor $ \boldsymbol{P}$ and the material entropy flux $ \boldsymbol{J}$. In a similar way to the finite dimensional case recalled in \eqref{VCond_conduction}--\eqref{VC_conduction}, the variational formulation needs the introduction of the \textit{thermal displacement} $ \Gamma (t, \mathbf{X} )$, see Remark \ref{remark_gamma}, whose time rate of change is identified with the temperature in the material frame, i.e., $ \frac{d}{dt} \Gamma (t, \mathbf{X} )= \mathfrak{T}(t, \mathbf{X} )$, as well as the additional entropy density variable $ \Sigma (t, \mathbf{X} )$ whose time rate of change is identified with the \textit{rate of internal entropy production of the continua} $ \frac{d}{dt} \Sigma (t, \mathbf{X} )= \mathfrak{I}(t, \mathbf{X} )\geq 0$.

The variational formulation reads
\begin{equation}\label{VCond_continuum} 
\delta \int_0^T \!\!\int_ \mathcal{B} \left[\mathfrak{L}( \boldsymbol{\varphi}  , \dot{\boldsymbol{\varphi}} ,  \nabla  \boldsymbol{\varphi}, \varrho ,S) + (S- \Sigma ) \dot \Gamma \right] {\rm d}^3  \mathbf{X} \, {\rm d} t=0
\end{equation} 
subject to the \textit{phenomenological constraint}
\begin{equation}\label{PC_continuum}
\frac{\partial \mathfrak{L}}{\partial S}\dot \Sigma = - \boldsymbol{P} : \nabla \dot{\boldsymbol{\varphi}} + \boldsymbol{J}  \cdot  \nabla  \dot \Gamma  
\end{equation}
and for variations subject to the \textit{variational constraint}
\begin{equation}\label{VC_continuum}
\frac{\partial \mathfrak{L}}{\partial S} \delta  \Sigma = - \boldsymbol{P} : \nabla \delta{\boldsymbol{\varphi} } + \boldsymbol{J}  \cdot\nabla  \delta  \Gamma  .
\end{equation}
The analogy with the finite dimensional case \eqref{VCond_conduction}--\eqref{VC_conduction} is clear. This principle consistently recovers the Hamilton principle \eqref{HP} when the thermodynamic fluxes vanishes.
A direct application of \eqref{VCond_continuum}--\eqref{VC_continuum} yields the evolution equations
\begin{equation}\label{NSF_material} 
\left\{
\begin{array}{l}
\vspace{0.2cm}\displaystyle\frac{d}{dt}\frac{ \partial   \mathfrak{L}  }{ \partial   \dot{\boldsymbol{\varphi}} }   - \frac{ \partial   \mathfrak{L} }{\partial \boldsymbol{\varphi}  }= \operatorname{DIV}\Big( - \frac{\partial \mathfrak{L}}{\partial \nabla{\boldsymbol{\varphi} } } +  \boldsymbol{P}  \Big) \\
\displaystyle
-\frac{\partial \mathfrak{L} }{ \partial    S}(\dot S + \operatorname{DIV} \boldsymbol{J}  ) = \boldsymbol{P} : \nabla\dot{\boldsymbol{\varphi}}  + \boldsymbol{J}  \cdot \nabla \frac{\partial  \mathfrak{L} }{\partial    S},
\end{array}
\right.
\end{equation}
for a viscous heat conducting continuum with Lagrangian $\mathfrak{L}$, together with the conditions
\begin{equation}\label{Gamma_Sigma} 
\dot \Gamma = - \frac{\partial   \mathfrak{L} }{ \partial   S} \qquad\text{and}\qquad \dot \Sigma = \dot S + \operatorname{DIV} \boldsymbol{J},
\end{equation} 
found from the variations $ \delta S$, $ \delta \Gamma$, which attribute to the fields $\Gamma$ and $ \Sigma $ the desired physical meaning (see \eqref{two_relations} for the finite dimensional case).
The variations $ \delta \Gamma$ on the boundary enforce the insulated boundary condition
\begin{equation}\label{adiab_cl} 
\boldsymbol{J}  \cdot \boldsymbol{N}  =0 \quad\text{on}\quad \partial \mathcal{B} 
\end{equation} 
meaning that the continuum is adiabatically closed, see Remark \ref{BC_NSF} for other types of boundary conditions.

System \eqref{NSF_material} provides the standard equations for continuum thermodynamics in material description written in terms of a general Lagrangian density. These equations must be supplemented with the phenomenological expressions for the thermodynamic fluxes, here $ \boldsymbol{P}$ and $ \boldsymbol{J}$, which are more often given in the spatial description.  The material momentum equation in \eqref{NSF_material} involves the total first Piola-Kirchhoff stress tensor
\begin{equation}\label{P_tot} 
\boldsymbol{P} ^{\rm tot}= - \frac{\partial \mathfrak{L} }{\partial \nabla \boldsymbol{\varphi}  }  + \boldsymbol{P} = \boldsymbol{P} ^{\rm cons}+\boldsymbol{P} .
\end{equation} 

For the Lagrangian density \eqref{Mat_Lag_Rev}, system \eqref{NSF_material} takes the familiar form 
\begin{equation}
\varrho\, \ddot { \boldsymbol{\varphi} } = \operatorname{DIV}\big(  \boldsymbol{P}  ^{\rm cons}+ \boldsymbol{P}  \big), \qquad  \mathfrak{T}(\dot S + \operatorname{DIV} \boldsymbol{J}  ) = \boldsymbol{P} : \nabla\dot{\boldsymbol{\varphi}}  - \boldsymbol{J}  \cdot \nabla \mathfrak{T},
\end{equation} 
with $ \boldsymbol{P} ^{\rm cons}= \varrho \,\partial \mathfrak{E}/ \partial \boldsymbol{F}$ and $ \mathfrak{T}= \varrho \,\partial \mathfrak{E}/ \partial S$.

\begin{remark}[Boundary conditions on the temperature]\label{BC_NSF}{\rm The variational formulation \eqref{VCond_continuum}-- \eqref{VC_continuum} yields the insulated boundary conditions \eqref{adiab_cl}. This is consistent with the fact that the class of variational formulation for thermodynamics recalled here only covers the case of adiabatically closed systems. Other important classes of boundary condition for viscous heat conducting fluids are given by prescribing the temperature $\mathfrak{T} |_{ \partial \mathcal{B} }= \mathfrak{T} _0$ or the heat flux $\mathfrak{T} \boldsymbol{J}  \cdot \boldsymbol{N} |_{ \partial \mathcal{B} }=Q_0$ on the boundary, in which case the system is no more adiabatically closed. 
These boundary conditions can be obtained by using an extension of the variational formulation recalled here to open thermodynamic systems. We refer to \citep{GBYo2018a} for the variational formulation of finite dimensional open systems with heat and mass transfer, and to  \citep{GaGB2024} for heat conducting viscous fluids with application to Rayleigh-B\'enard convection for Dirichlet and Neumann thermal boundary conditions.}
\end{remark}

\begin{remark}[Recovering variational formulations in the dissipationless reversible case]\label{remark_reversible}{\rm There are several ways to specialize the variational formulation \eqref{VCond_continuum}--\eqref{VC_continuum} to the reversible case.\\
(A) First, as expected, setting the thermodynamic fluxes to zero, i.e., $ \boldsymbol{P}=0$, $ \boldsymbol{J}=0$, one recovers the standard Euler-Lagrange equations for $ \boldsymbol{\varphi}$, together with the entropy conservation law $\dot S=0$. While this variational approach may seem artificial in the absence of dissipative processes - since it still involves a variational constraint - it nevertheless demonstrates consistency with the reversible case.\\
(B) Second, in addition to setting $ \boldsymbol{P}=0$ and $ \boldsymbol{J}=0$, one may also ignore the variable $\Sigma$, and thus the associated constraints. One then simply obtains the critical point condition:
\begin{equation}\label{VCond_continuum_reversible} 
\delta \int_0^T \!\!\int_ \mathcal{B} \left[\mathfrak{L}( \boldsymbol{\varphi}  , \dot{\boldsymbol{\varphi}} ,  \nabla  \boldsymbol{\varphi}, \varrho ,S) + S \dot \Gamma \right] {\rm d}^3  \mathbf{X} \, {\rm d} t=0.
\end{equation} 
In this case, entropy conservation $\dot S=0$ follows from the Euler-Lagrange equation associated with $\Gamma$.  This setting is particularly relevant for open fluid models; see \citep{ElGBWu2024}.\\
(C) Finally, in the reversible case, one may directly assume that $S$ is constant in time, as is done for $\varrho$, and omit the variables $\Gamma$ and $\Sigma$ altogether. The variational formulation \eqref{VCond_continuum}--\eqref{VC_continuum} then reduces to the classical Hamilton principle of continuum mechanics, as given in \eqref{HP}.}
\end{remark}

\begin{remark}[Thermal displacement]\label{remark_gamma}{\rm The concept of thermal displacement $\Gamma$ plays a crucial role in the variational formulation for both classical and extended irreversible thermodynamics. However, it should be noted that $\Gamma$ is not a state variable in the heat conduction models considered in this paper, as the final equations involve only the temperature $T=\dot \Gamma$. Although $\Gamma$ does not appear explicitly in the final governing equations, its inclusion is essential for deriving them from a variational principle. The notion of thermal displacement was originally introduced in \citep{vH1884}; see also  \citep{vL1921,Ta1949,He1955} for early discussions of this concept. This notion has since been extended to other irreversible processes beyond heat transfer, as considered in \citep{GBYo2017a,GBYo2019a}, where the generalized variable is referred to as a \textit{thermodynamic displacement}, $\Lambda_\alpha$, such that its time derivative $\dot{\Lambda}_\alpha= X_\alpha$ corresponds to the thermodynamic force (or affinity) $X_\alpha$ associated with the irreversible process $\alpha$.
}
\end{remark}

\medskip

\noindent\textbf{Energy and entropy balance in the material frame.} For future comparison with the extended case later we formulate these balance laws for the general evolution equations \eqref{NSF_material} associated to an arbitrary Lagrangian density $\mathfrak{L}( \boldsymbol{\varphi}  , \dot{\boldsymbol{\varphi}} ,  \nabla  \boldsymbol{\varphi}, \varrho ,S)$. The corresponding total energy density $\mathfrak{E}( \boldsymbol{\varphi}  , \dot{\boldsymbol{\varphi}} ,  \nabla  \boldsymbol{\varphi}, \varrho ,S)$ of the continuum is defined by
\begin{equation}
\mathfrak{E} = \frac{\partial \mathfrak{L} }{\partial \dot{ \boldsymbol{\varphi} } } \cdot \dot{ \boldsymbol{\varphi} } - \mathfrak{L} 
\end{equation}
from which we directly get, by using \eqref{NSF_material}, the energy balance
\begin{equation}\label{energy_balance_CIT} 
\dot{\mathfrak{E}}= \operatorname{DIV} \left( \boldsymbol{P}  ^{\rm tot}\cdot \dot{ \boldsymbol{\varphi} } - \mathfrak{T}  \boldsymbol{J} \right) ,
\end{equation} 
with $ \boldsymbol{P}  ^{\rm tot}$ the total first Piola-Kirchhoff stress tensor defined in \eqref{P_tot} and $ \mathfrak{T}=-\frac{\partial \mathfrak{L} }{\partial S}$ the material temperature.
Using the boundary condition \eqref{No_slip} this yields the integrated form
\begin{equation}\label{integrated_E} 
\frac{d}{dt} \int_ \mathcal{B} \mathfrak{E}\,{\rm d}^3 \mathbf{X}  = - \int _{ \partial \mathcal{B} } \mathfrak{T} \boldsymbol{J} \cdot \boldsymbol{N}  {\rm d}A.
\end{equation}
The second law imposes the inequality
\begin{equation}\label{2nd_law} 
\mathfrak{T} (\dot S + \operatorname{DIV} \boldsymbol{J}) =   \boldsymbol{P}: \nabla \dot{ \boldsymbol{\varphi} } -  \boldsymbol{J}\cdot \nabla \mathfrak{T} \geq 0
\end{equation}
on the entropy balance equation, found from the second equation in \eqref{NSF_material}, which gives the integrated form
\begin{equation}\label{integrated_S} 
\frac{d}{dt} \int_ \mathcal{B} S\, {\rm d} ^3  \mathbf{X}  = \int_ \mathcal{B} \frac{1}{\mathfrak{T} } \left(  \boldsymbol{P}: \nabla \dot{ \boldsymbol{\varphi} } -  \boldsymbol{J} \cdot \nabla \mathfrak{T}\right){\rm d} ^3  \mathbf{X}  - \int_{ \partial \mathcal{B} }  \boldsymbol{J} \cdot  \boldsymbol{N} {\rm d}A .
\end{equation} 
In \eqref{integrated_E} and \eqref{integrated_S} the boundary integrals vanish in the adiabatically closed case \eqref{adiab_cl}. Note that the inequality \eqref{2nd_law} can be equivalently written $\dot \Sigma \geq 0$, from \eqref{Gamma_Sigma}, thereby attributing to $ \dot\Sigma$ (which doesn't coincide with $\dot  S$) the meaning of rate of internal entropy production, with
\begin{equation}\label{Sigma_expresssion_CIT} 
\dot  \Sigma =\frac{1}{\mathfrak{T} } \left(  \boldsymbol{P}: \nabla \dot{ \boldsymbol{\varphi} } -  \boldsymbol{J} \cdot \nabla \mathfrak{T}\right).
\end{equation} 

\medskip

\noindent\textbf{Formulation with the second Piola-Kirchhoff stress tensor.} The variational formulation \eqref{VCond_continuum}--\eqref{VC_continuum} can be equivalently expressed in terms of the second Piola-Kirchhoff stress tensor associated to $ \boldsymbol{P}$ defined, as in \eqref{def_PK2_cons}, by 
\begin{equation}\label{def_PK2} 
\boldsymbol{S}^{AB}=  \boldsymbol{P} ^{aA} (\boldsymbol{F}^{-1} )^B_a .
\end{equation} 
It is assumed that $ \boldsymbol{P} $ satisfies \eqref{prop_P} and hence $ \boldsymbol{S}^{AB}= \boldsymbol{S}^{BA}$. One notes the equalities 
\begin{equation}\label{identity_power_P_S} 
\boldsymbol{P} : \nabla \dot{\boldsymbol{\varphi}}= \boldsymbol{S} : \boldsymbol{D}(  \dot{\boldsymbol{\varphi}}) \quad\text{and}\quad \boldsymbol{P} : \nabla \delta \boldsymbol{\varphi}= \boldsymbol{S} : \boldsymbol{D}(\delta \boldsymbol{\varphi})
\end{equation} 
with $ \boldsymbol{D}  (\dot{\boldsymbol{\varphi}} ) $ the \textit{material rate of deformation tensor} and $\boldsymbol{D}(\delta \boldsymbol{\varphi})$ its variational correspondent, defined by
\begin{equation}\label{def_D} 
\begin{aligned} 
\boldsymbol{D}  (\dot{\boldsymbol{\varphi}})_{AB}&= \frac{1}{2} \delta _{bc}\big( \boldsymbol{F} ^b_B \dot{\boldsymbol{\varphi}}  ^c_{,A}+ \boldsymbol{F} ^b_A \dot{\boldsymbol{\varphi}}^c _{,B} \big) \\
\boldsymbol{D}  ( \delta \boldsymbol{\varphi})_{AB}&= \frac{1}{2} \delta _{bc}\big( \boldsymbol{F} ^b_B \delta \boldsymbol{\varphi} ^c_{,A}+ \boldsymbol{F} ^b_A \delta \boldsymbol{\varphi}^c _{,B} \big).
\end{aligned}
\end{equation} 
Using \eqref{def_D} in the phenomenological and variational constraints \eqref{PC_continuum} and \eqref{VC_continuum} allows us to rewrite the variational principle \eqref{VCond_continuum}--\eqref{VC_continuum} uniquely in terms of $ \boldsymbol{S}$ without reference to $ \boldsymbol{P}$.
It is this version that is best suited to treat extended irreversible thermodynamics, see below.

\medskip

\noindent\textbf{Variational formulation in the spatial description.} The variational formulation in the spatial description is directly obtained by converting \eqref{VCond_continuum}--\eqref{VC_continuum} in spatial variables. This process uses the relation 
 \eqref{def_ell} between the Lagrangian densities in material and spatial descriptions (which assumes \eqref{mat_cov}), as well as the relations
 \begin{equation}\label{fluxes_conversion}
\begin{aligned} 
\boldsymbol{j}^a& =( \boldsymbol{F}^a_A \boldsymbol{J}^A) \circ \boldsymbol{\varphi } ^{-1}  \, J_{ \boldsymbol{\varphi}^{-1} } \\
\boldsymbol{\sigma}^{ab}&= ( \boldsymbol{F}^a_A \boldsymbol{P}^{bA}) \circ \boldsymbol{\varphi }^{-1}   \, J_{ \boldsymbol{\varphi}^{-1} } =( \boldsymbol{F}^a_A \boldsymbol{S}^{AB} \boldsymbol{F}^b_B) \circ \boldsymbol{\varphi } ^{-1}  \, J_{ \boldsymbol{\varphi}^{-1} } \\
\end{aligned} 
\end{equation} 
between the thermodynamic fluxes $ \boldsymbol{J}$, $ \boldsymbol{S}$ in the material descriptions and the corresponding fluxes $ \boldsymbol{j}$, $ \boldsymbol{\sigma }$ in the spatial description.

In addition to the material-to-spatial relations \eqref{mat_to_spat} and \eqref{fluxes_conversion}, we also consider the spatial versions $ \boldsymbol{\gamma} (t, \mathbf{x} )$, $ \varsigma ( t, \mathbf{x} )$ of $ \Gamma (t, \mathbf{X} )$, $ \Sigma (t, \mathbf{X} )$ given by
\begin{equation}\label{def_gamma_sigma} 
\gamma = \Gamma \circ \boldsymbol{\varphi} ^{-1} , \qquad \varsigma = (\Sigma \circ \boldsymbol{\varphi} ^{-1} )J_{ \boldsymbol{\varphi} ^{-1} }.
\end{equation}
For convenience, we also introduce the Eulerian time derivative and Eulerian variations of a scalar $f(t, \mathbf{x} )$ and density $g(t, \mathbf{x} )$ field as
\begin{equation}
\begin{aligned}
D_tf&= \partial _tf + \boldsymbol{u} \cdot \nabla f & & D_ \delta f= \delta f+ \boldsymbol{\zeta} \cdot \nabla f \\
\bar D_tg&= \partial _t g+ \operatorname{div}( f \boldsymbol{u} ) & & \bar D_ \delta g= \delta g+ \operatorname{div}(g \boldsymbol{\zeta}),
\end{aligned}
\end{equation} 
where $ \boldsymbol{\zeta} (t, \mathbf{x} )$ is defined in \eqref{zeta}.

This results in the variational formulation
\begin{equation}\label{VCond_continuum_spat} 
\delta \int_0^T \!\!\int_ \mathcal{D} \left[\ell( \boldsymbol{u}  ,  \rho  , s) + (s- \varsigma  ) D_t \gamma  \right] {\rm d}^3  \mathbf{x} \, {\rm d} t=0
\end{equation} 
subject to the \textit{phenomenological constraint}
\begin{equation}\label{PC_continuum_spat}
\frac{\partial \ell}{\partial s}\bar D_t \varsigma  = - \boldsymbol{ \sigma } : \nabla \boldsymbol{u }  + \boldsymbol{j}  \cdot  \nabla  D_t \gamma
\end{equation}
and for variations subject to the \textit{variational constraint}
\begin{equation}\label{VC_continuum_spat}
\frac{\partial \ell}{\partial s}\bar D_\delta  \varsigma  = - \boldsymbol{ \sigma } : \nabla \boldsymbol{\zeta }  + \boldsymbol{j}  \cdot  \nabla  D_\delta  \gamma.
\end{equation}
as well as the usual Eulerian variations
\begin{equation}\label{red_var_2} 
\delta \boldsymbol{u}= \partial _t \boldsymbol{\zeta} + \boldsymbol{u} \cdot \nabla \boldsymbol{\zeta}- \boldsymbol{\zeta } \cdot \nabla \boldsymbol{u}  , \qquad \delta   \rho  = - \operatorname{div}( \rho  \boldsymbol{\zeta} ).
\end{equation}

Note that besides \eqref{mat_to_spat}, \eqref{fluxes_conversion}, and \eqref{def_gamma_sigma}, the material-to-spatial conversion from \eqref{VCond_continuum}--\eqref{VC_continuum} to \eqref{VCond_continuum_spat}--\eqref{VC_continuum_spat} also uses formulas like
\begin{equation}
\big( \boldsymbol{J} \cdot \nabla \dot  \Gamma  \big) \circ \boldsymbol{\varphi} ^{-1} J_{ \boldsymbol{\varphi} ^{-1} }=\boldsymbol{j}  \cdot  \nabla  D_t \gamma \quad\text{and}\quad \big( \boldsymbol{J} \cdot \nabla \delta \Gamma  \big) \circ \boldsymbol{\varphi} ^{-1} J_{ \boldsymbol{\varphi} ^{-1} }=\boldsymbol{j}  \cdot  \nabla  D_\delta  \gamma.
\end{equation}

This principle \eqref{VCond_continuum_spat}--\eqref{red_var_2} consistently recovers the reduced Hamilton principle \eqref{red_HP}--\eqref{red_var}  when the thermodynamic fluxes vanish.
A direct application of \eqref{VCond_continuum_spat}--\eqref{VC_continuum_spat} yields the evolution equations
\begin{equation}\label{NSF_spat} 
\left\{
\begin{array}{l}
\vspace{0.2cm}\displaystyle \partial _t \frac{\partial \ell}{\partial \boldsymbol{u} }  + \pounds _ {\boldsymbol{u}} \frac{\partial \ell}{\partial \boldsymbol{u} }= \rho  \nabla \frac{\partial \ell}{\partial \rho   }+ s \nabla \frac{\partial \ell}{\partial s} + \operatorname{div} \boldsymbol{\sigma} \\
\displaystyle
-\frac{\partial \ell }{ \partial  s}(\bar D_t s + \operatorname{div} \boldsymbol{j}  ) = \boldsymbol{ \sigma } : \nabla \boldsymbol{u}   + \boldsymbol{j}  \cdot \nabla \frac{\partial  \ell }{\partial   s},
\end{array}
\right.
\end{equation}
for a viscous heat conducting continuum with Lagrangian $\ell$, together with the conditions
\begin{equation}\label{Gamma_Sigma_spat} 
D_t \gamma  = - \frac{\partial   \ell }{ \partial   s} \qquad\text{and}\qquad \bar D_t \varsigma  = \bar D_t s + \operatorname{div} \boldsymbol{j} ,
\end{equation} 
which follow from the variations $ \delta s$, $ \delta \gamma$. These conditions assign the expected physical meanings to the fields $\gamma$ and $ \varsigma$, namely, $ \gamma (t, \mathbf{x} )$ represents the \textit{temperature displacement in spatial description}, while the Eulerian time rate of change $\bar D_t \varsigma (t, \mathbf{x} )= \mathfrak{i}(t, \mathbf{x} )\geq 0$ corresponds to the \textit{internal entropy production in the spatial description}. The variations $ \delta \gamma$ on the boundary enforce the insulated boundary condition
\begin{equation}\label{adiab_cl_Euler} 
\boldsymbol{j}  \cdot \boldsymbol{n}  =0 \quad\text{on}\quad \partial \mathcal{D} .
\end{equation} 
We refer to \citep{GBYo2017b} and \citep{GB2019} for the detailed computations leading from \eqref{VCond_continuum_spat}--\eqref{VC_continuum_spat} to \eqref{NSF_spat}--\eqref{adiab_cl_Euler}. See also \citep{GaGB2024} for other thermal boundary conditions.

For the Lagrangian \eqref{ell} one gets the viscous heat conducting fluid equations
\begin{equation}
\rho  ( \partial _t \boldsymbol{u} + \boldsymbol{u} \cdot \nabla \boldsymbol{u})= - \nabla p + \operatorname{div} \boldsymbol{\sigma} , \qquad T (\bar D_t s + \operatorname{div} \boldsymbol{j}  ) = \boldsymbol{ \sigma } : \nabla \boldsymbol{u}   - \boldsymbol{j}  \cdot \nabla T,
\end{equation} 
with $p= \rho  \frac{\partial \varepsilon }{\partial \rho  }+s  \frac{\partial \varepsilon }{\partial s  }- \varepsilon $ and $T= \frac{\partial \varepsilon }{\partial s}$.
These equations are closed using the usual phenomenological expressions for the thermodynamic fluxes $ \boldsymbol{\sigma}$ and $ \boldsymbol{j}$  in classical irreversible thermodynamics (Stokes', Newton's, and Fourier's laws), see, e.g.,  \citep{dGMa1969,Wo1975,KoPr1998}.

\medskip

\begin{remark}[Recovering variational formulations in the dissipationless reversible case]{\rm In parallel with Remark \ref{remark_reversible}, we observe that the Eulerian variational formulation \eqref{VCond_continuum_spat}--\eqref{red_var_2} specializes to the reversible case in three natural ways, corresponding to the Eulerian (or symmetry-reduced) analogs of the items (A)-(B)-(C) discussed in Remark \ref{remark_reversible}.\\
(A) First, by setting $\boldsymbol{\sigma}=0$ and $\boldsymbol{j}=0$ in \eqref{VCond_continuum_spat}--\eqref{red_var_2}, one recovers the reversible fluid equations, while still involving both the phenomenological and variational constraints.\\
(B) Second, one may further eliminate the variable $\varsigma$, thereby removing the constraints. This leads to the critical point condition:
\begin{equation}\label{VCond_continuum_spat_rev} 
\delta \int_0^T \!\!\int_ \mathcal{D} \left[\ell( \boldsymbol{u}  ,  \rho  , s) + sD_t \gamma  \right] {\rm d}^3  \mathbf{x} \, {\rm d} t=0,
\end{equation} 
as used, e.g., in \citep{ElGBWu2024} for open fluid models. In this case, the entropy conservation $\bar D_ts=0$ arises as an Euler-Lagrange equation.\\
(C) Finally, assuming that $S$ is constant in time in the material frame directly implies $\bar D_ts=0$ in the Eulerian frame, together with the constrained variations $\delta s= - \operatorname{div}(s\boldsymbol{\zeta})$. This yields the Eulerian version of the Hamilton principle, also known as the Euler-Poincar\'e principle, in which the entropy density $s$ is treated analogously to the mass density $\rho$.}
\end{remark}

\noindent\textbf{Energy and entropy balance in the spatial frame.} We derive here these balance laws for an arbitrary Lagrangian density $\ell( \boldsymbol{u}, \rho  , s)$ for future comparison with the extended case.
In this general setting the total energy density $e( \boldsymbol{u}, \rho  , s)$ is defined in the spatial description as
\begin{equation}
e= \frac{\partial \ell}{\partial \boldsymbol{u} } \cdot \boldsymbol{u} -\ell.
\end{equation}
From this expression and the general equations \eqref{NSF_spat}, the energy balance can be elegantly written in terms of $\ell$ as
\begin{equation}\label{energy_cons_spat_CIT} 
\bar D_t e = \operatorname{div}\left( \left( \Big(\rho  \frac{\partial \ell}{\partial \rho  } +  s  \frac{\partial \ell}{\partial s  } - \ell\Big) \delta + \boldsymbol{\sigma} \right)  \cdot \boldsymbol{u}+ \boldsymbol{j} \frac{\partial \ell}{\partial s} \right) 
\end{equation} 
yielding the integrated form
\begin{equation}\label{integrated_energy_spat_CIT} 
\frac{d}{dt} \int_ \mathcal{D} e \,{\rm d} \mathbf{x} = - \int_{ \partial \mathcal{D} } T\boldsymbol{j}\cdot \boldsymbol{n}\, {\rm d} A  ,
\end{equation} 
with $T= - \frac{\partial \ell}{\partial s}$ the temperature in the spatial frame, and we used $ \boldsymbol{u}|_{ \partial \mathcal{D} }=0$. The entropy balance can be written as
\begin{equation}\label{2nd_law_spat} 
\bar D_t s + \operatorname{div} \boldsymbol{j}  =  \frac{1}{T} \left( \boldsymbol{ \sigma } : \nabla \boldsymbol{u}   - \boldsymbol{j}  \cdot \nabla T \right) \geq 0,
\end{equation} 
giving the integrated form
\begin{equation}
\frac{d}{dt} \int_ \mathcal{D} s\, {\rm d} \mathbf{x} = \int_ \mathcal{D}  \frac{1}{T} \left( \boldsymbol{ \sigma } : \nabla \boldsymbol{u}   - \boldsymbol{j}  \cdot \nabla T \right)  {\rm d} \mathbf{x} - \int_{ \partial \mathcal{D} } \boldsymbol{j} \cdot \boldsymbol{n} \,{\rm d} A .
\end{equation}
Note that the inequality \eqref{2nd_law_spat} can be equivalently written $\bar D_t \varsigma \geq 0$, from \eqref{Gamma_Sigma_spat}, thereby attributing to $\bar D_t \varsigma $ (which doesn't coincide with $\bar D_t s$) the meaning of spatial rate of internal entropy production, with
\begin{equation}
\bar D_t \varsigma  =\frac{1}{T} \left( \boldsymbol{ \sigma } : \nabla \boldsymbol{u}   - \boldsymbol{j}  \cdot \nabla T \right) .
\end{equation}

\section{Extended irreversible thermodynamics of viscous heat conducting fluid}\label{Sec_3}

The basic idea of extended irreversible thermodynamics is to abandon the local thermodynamic equilibrium hypothesis by including the thermodynamic fluxes as state variables. 
From a variational Lagrangian point of view, it is therefore natural to extend the previous considered Lagrangian density to include the dependence on the fluxes. This is done by preserving the two covariances of the Lagrangian densities, namely, the covariance associated to material frame indifference, which is as an axiom of continuum mechanics, and the covariance with respect to material diffeomorphisms, which is an assumption on the continuum implying isotropy.

\subsection{Variational formulation in the material description}

To cover the case of extended irreversible thermodynamics, we generalize the class of Lagrangian densities from $\mathfrak{L}( \boldsymbol{\varphi} , \dot{ \boldsymbol{\varphi} }, \nabla \boldsymbol{\varphi} , \varrho , S)$ to \textit{nonequilibrium Lagrangian densities}
\begin{equation}
\widehat{\mathfrak{L}}( \boldsymbol{\varphi} , \dot{ \boldsymbol{\varphi} }, \nabla \boldsymbol{\varphi} , \varrho , S, \boldsymbol{S}, \boldsymbol{Q})
\end{equation}
which explicitly depend on the thermodynamic fluxes $ \boldsymbol{S}$ and $ \boldsymbol{Q}$, associated with viscosity and heat transfer, respectively. Note that we do not impose the relation $ \boldsymbol{Q}= \mathfrak{T} \boldsymbol{J}$ at this stage, as this becomes a phenomenological assumption, and more general relations may emerge. For further discussion, see \citep{JoCVLe2010}, as well as Remark \ref{choice_flux} for comments on the choice of fluxes.

The proposed variational formulation of extended irreversible thermodynamics reads
\begin{equation}\label{VCond_continuum_ext} 
\delta \int_0^T \!\!\int_ \mathcal{B} \left[\widehat{\mathfrak{L}}( \boldsymbol{\varphi}  , \dot{\boldsymbol{\varphi}} ,  \nabla  \boldsymbol{\varphi}, \varrho ,S, \boldsymbol{S}, \boldsymbol{Q}) + (S- \Sigma ) \dot \Gamma \right] {\rm d}^3  \mathbf{X} \, {\rm d} t=0
\end{equation} 
subject to the \textit{phenomenological constraint}
\begin{equation}\label{PC_continuum_ext}
\frac{\partial \widehat{\mathfrak{L}}}{\partial S}\dot \Sigma+\frac{\partial \widehat{\mathfrak{L}}}{\partial \boldsymbol{S} }:\dot{\boldsymbol{S}} +\frac{\partial \widehat{\mathfrak{L}}}{\partial \boldsymbol{Q} } \cdot \dot{ \boldsymbol{Q}}  = - \boldsymbol{S} : \boldsymbol{D} (\dot{\boldsymbol{\varphi}} )+ \boldsymbol{J}  \cdot  \nabla  \dot \Gamma  
\end{equation}
and for variations subject to the \textit{variational constraint}
\begin{equation}\label{VC_continuum_ext}
\frac{\partial \widehat{\mathfrak{L}}}{\partial S} \delta  \Sigma +\frac{\partial \widehat{\mathfrak{L}}}{\partial \boldsymbol{S} }:\delta \boldsymbol{S} +\frac{\partial \widehat{\mathfrak{L}}}{\partial \boldsymbol{Q} }\cdot  \delta  \boldsymbol{Q}= - \boldsymbol{S} : \boldsymbol{D} (\delta{\boldsymbol{\varphi} } )+ \boldsymbol{J}  \cdot\nabla  \delta  \Gamma  .
\end{equation}
This principle consistently recovers \eqref{VCond_continuum}--\eqref{VC_continuum} when the Lagrangian does not depend on the fluxes $ \boldsymbol{S}$ and $ \boldsymbol{Q}$, i.e., for classical irreversible thermodynamics, see \eqref{identity_power_P_S}.

We have opted not to present the detailed computation leading to the critical conditions, as it closely follows that for \eqref{VCond_continuum}--\eqref{VC_continuum}, which has already been thoroughly discussed in earlier works. More details will be provided later for its spatial version, which is more involved. An application of \eqref{VCond_continuum_ext}--\eqref{VC_continuum_ext} yields the evolution equations
\begin{equation}\label{NSF_material_ext} 
\left\{
\begin{array}{l}
\vspace{0.2cm}\displaystyle\frac{d}{dt}\frac{ \partial \widehat{\mathfrak{L}} }{ \partial   \dot{\boldsymbol{\varphi}} }   - \frac{ \partial \widehat{\mathfrak{L}}}{\partial \boldsymbol{\varphi}  }= \operatorname{DIV}\Big( - \frac{\partial\widehat{\mathfrak{L}}}{\partial \nabla{\boldsymbol{\varphi} } } +  \boldsymbol{P}  \Big) \\
\displaystyle
-\frac{\partial \widehat{\mathfrak{L}}}{ \partial    S}(\dot S + \operatorname{DIV} \boldsymbol{J}  )- \frac{\partial \widehat{\mathfrak{L}}}{\partial \boldsymbol{S} }:\dot{\boldsymbol{S}} -\frac{\partial \widehat{\mathfrak{L}}}{\partial \boldsymbol{Q} } \cdot \dot{ \boldsymbol{Q}}  =  \boldsymbol{S} : \boldsymbol{D} (\dot{\boldsymbol{\varphi}} ) + \boldsymbol{J}  \cdot \nabla \frac{\partial  \widehat{\mathfrak{L}} }{\partial    S},
\end{array}
\right.
\end{equation}
for the extended irreversible thermodynamics of a viscous heat conducting continuum with Lagrangian $\widehat{\mathfrak{L}}$, together with the conditions
\begin{equation}\label{Gamma_Sigma_ext} 
\dot \Gamma = - \frac{\partial  \widehat{\mathfrak{L}} }{ \partial   S} \qquad\text{and}\qquad \dot \Sigma = \dot S + \operatorname{DIV} \boldsymbol{J},
\end{equation} 
found from the variations $ \delta S$, $ \delta \Gamma$, which attribute to the fields $\Gamma$ and $ \Sigma $ the desired physical meaning.
As before, arbitrary variations $ \delta \Gamma |_{ \partial \mathcal{B} }$ enforce the boundary condition
\begin{equation}\label{adiab_cl_ext} 
\boldsymbol{J}  \cdot \boldsymbol{N}  =0,
\end{equation} 
in which case the continuum is adiabatically closed.

\medskip

\noindent\textbf{Standard Lagrangian densities.} The natural extension of the typical Lagrangian density \eqref{Mat_Lag_Rev} to the nonequilibrium setting takes the form
\begin{equation}\label{Mat_Lag_Rev_ext} 
\widehat{\mathfrak{L}}( \boldsymbol{\varphi}  ,\dot{\boldsymbol{\varphi}} , \nabla{ \boldsymbol{\varphi} } , \varrho , S, \boldsymbol{S}, \boldsymbol{Q})= \frac{1}{2} \varrho | \dot{\boldsymbol{\varphi}} | ^2 - \widehat{\mathscr{E}}(\nabla{ \boldsymbol{\varphi} }, \varrho , S, \boldsymbol{S}, \boldsymbol{Q})\varrho,
\end{equation}
from which one defines the \textit{nonequilibrium first Piola-Kirchhoff stress tensor} (conservative part) $ \boldsymbol{P} ^{\rm cons}$ and the \textit{nonequilibrium temperature} $\mathfrak{T}$ as
\begin{equation}
\boldsymbol{P} ^{\rm cons}= \varrho \frac{\partial \widehat{\mathscr{E}}}{\partial  \boldsymbol{F} }\quad\text{and}\quad \mathfrak{T} = \varrho \frac{\partial \widehat{\mathscr{E}}}{\partial S}.
\end{equation}

For the Lagrangian density \eqref{Mat_Lag_Rev_ext}, system \eqref{NSF_material_ext} takes the form 
\begin{equation}
\varrho\, \ddot { \boldsymbol{\varphi} } = \operatorname{DIV}\big(  \boldsymbol{P}  ^{\rm cons}+ \boldsymbol{P}  \big), \qquad  \mathfrak{T}(\dot S + \operatorname{DIV} \boldsymbol{J}  ) =\boldsymbol{S} : \boldsymbol{D}(  \dot{\boldsymbol{\varphi}})+ \frac{\partial \widehat{\mathfrak{L}}}{\partial \boldsymbol{S} }:\dot{\boldsymbol{S}} - \boldsymbol{J}  \cdot \nabla \mathfrak{T} + \frac{\partial \widehat{\mathfrak{L}}}{\partial \boldsymbol{Q} } \cdot \dot{\boldsymbol{Q}}.
\end{equation}

\medskip

\noindent\textbf{Material frame indifference.} Similarly to the Lagrangian density $\mathfrak{L}( \boldsymbol{\varphi} , \dot{ \boldsymbol{\varphi} }, \nabla \boldsymbol{\varphi} , \varrho , S)$ used in the reversible case and the classical irreversible case in Section \ref{Sec_2}, the nonequilibrium Lagrangian density $\widehat{\mathfrak{L}}( \boldsymbol{\varphi} , \dot{ \boldsymbol{\varphi} }, \nabla \boldsymbol{\varphi} , \varrho , S, \boldsymbol{S}, \boldsymbol{Q}  )$ is also assumed to satisfy the axiom of \textit{material indifference}. As we recalled in Remark \ref{rmk_FI}, this axiom is given by covariance of the Lagrangian density with respect to the diffeomorphisms of the ambient space $ \mathbb{R} ^3  $, which act by composition on the left on $ \boldsymbol{\varphi} $ and also on the ambient metric. We note that such diffeomorphisms do not act on the material fields $ \varrho (t, \mathbf{X} )$, $S(t, \mathbf{X} )$, and on the thermodynamic fluxes $ \boldsymbol{S}(t, \mathbf{X} )$ and $ \boldsymbol{Q}(t, \mathbf{X} )$. Thus, when written in the material description, material frame indifference for $\mathfrak{L}$ and $\widehat{\mathfrak{L}}$ do not substantially differ. This is no more the case when this axiom is written in the spatial description.

\subsection{Energy and entropy balance in the material description}

The total energy density in extended irreversible thermodynamics takes the general form $\widehat{\mathfrak{E}}( \boldsymbol{\varphi}  , \dot{\boldsymbol{\varphi}} ,  \nabla  \boldsymbol{\varphi}, \varrho ,S, \boldsymbol{S}, \boldsymbol{Q})$ and is defined from the Lagrangian density as before by
\begin{equation}
\widehat{\mathfrak{E}}= \frac{\partial \widehat{\mathfrak{L}}}{\partial \dot{ \boldsymbol{\varphi} } } \cdot \dot{ \boldsymbol{\varphi} } - \widehat{\mathfrak{L}}.
\end{equation}
Remarkably, despite the additional flux dependencies, the energy balance in the material description takes the same form as in classical irreversible thermodynamics, namely
\begin{equation}\label{energy_balance_EIT} 
\dot{\widehat{\mathfrak{E}}}= \operatorname{DIV} \left( \boldsymbol{P}  ^{\rm tot}\cdot \dot{ \boldsymbol{\varphi} } - \mathfrak{T}  \boldsymbol{J} \right),
\end{equation}
with the notable difference that $ \boldsymbol{P}= - \frac{\partial \widehat{\mathfrak{L}}}{\partial \nabla \boldsymbol{\varphi} } $ and $\mathfrak{T}= - \frac{\partial  \widehat{\mathfrak{L}}}{\partial S} $ are the nonequilibrium version of $ \boldsymbol{P}$ and $ \mathfrak{T}$ appearing earlier. 

Taking the time derivative of $\widehat{\mathfrak{E}}$ and using the first equation in \eqref{NSF_material_ext}, we get
\begin{equation}
\frac{d}{dt} \widehat{\mathfrak{E}}  = \operatorname{DIV} \Big( \boldsymbol{P} \cdot \dot  { \boldsymbol{\varphi} }- \frac{\partial \widehat{\mathfrak{L}}}{\partial \nabla \boldsymbol{\varphi} }  \cdot \dot  { \boldsymbol{\varphi} }\Big) - \boldsymbol{P} : \nabla \dot  { \boldsymbol{\varphi} }-  \frac{\partial \widehat{\mathfrak{L}}}{\partial S}\dot S -  \frac{\partial \widehat{\mathfrak{L}}}{\partial \boldsymbol{S} }\dot{\boldsymbol{S}} -\frac{\partial \widehat{\mathfrak{L}}}{\partial \boldsymbol{Q} }\dot{ \boldsymbol{Q}} .
\end{equation}
Then \eqref{energy_balance_EIT} follows by using the second equation in \eqref{NSF_material_ext} together with the first equality \eqref{identity_power_P_S}. The integrated form follows as earlier, giving
\begin{equation}\label{integrated_E_EIT} 
\frac{d}{dt} \int_ \mathcal{B} \widehat{\mathfrak{E}}\,{\rm d}^3 \mathbf{X}  = - \int _{ \partial \mathcal{B} }  \mathfrak{T}\boldsymbol{J}  \cdot \boldsymbol{N}  {\rm d}A.
\end{equation}
We note that $\mathfrak{T}  \boldsymbol{J}$ appears as in \eqref{energy_balance_CIT} and \eqref{integrated_E}, but is now defined using the nonequilibrium temperature.
The second law imposes the inequality
\begin{equation}\label{2nd_law_EIT} 
\mathfrak{T}(\dot S + \operatorname{DIV} \boldsymbol{J}  ) =\boldsymbol{S} : \boldsymbol{D}(  \dot{\boldsymbol{\varphi}})+ \frac{\partial \widehat{\mathfrak{L}}}{\partial \boldsymbol{S} }:\dot{\boldsymbol{S}} - \boldsymbol{J}  \cdot \nabla \mathfrak{T} + \frac{\partial \widehat{\mathfrak{L}}}{\partial \boldsymbol{Q} } \cdot \dot{\boldsymbol{Q}} \geq 0
\end{equation}
on the entropy balance equation, found from the second equation in \eqref{NSF_material_ext}, which gives the integrated form
\begin{equation}\label{integrated_S_EIT} 
\frac{d}{dt} \int_ \mathcal{B} S\, {\rm d} ^3  \mathbf{X}  = \int_ \mathcal{B} \frac{1}{\mathfrak{T} } \Big( \boldsymbol{S} : \boldsymbol{D}(  \dot{\boldsymbol{\varphi}})+ \frac{\partial \widehat{\mathfrak{L}}}{\partial \boldsymbol{S} }:\dot{\boldsymbol{S}} - \boldsymbol{J}  \cdot \nabla \mathfrak{T} + \frac{\partial \widehat{\mathfrak{L}}}{\partial \boldsymbol{Q} } \cdot \dot{\boldsymbol{Q}} \Big){\rm d} ^3  \mathbf{X}  - \int_{ \partial \mathcal{B} }  \boldsymbol{J} \cdot  \boldsymbol{N} {\rm d}A .
\end{equation} 
As previously, we note that the inequality \eqref{2nd_law_EIT} can be equivalently written $\dot \Sigma \geq 0$, from \eqref{Gamma_Sigma_ext}, thereby attributing to $ \dot\Sigma$ the meaning of rate of internal entropy production in extended irreversible thermodynamics, with
\begin{equation}\label{Sigma_expresssion_EIT} 
\dot  \Sigma =\frac{1}{\mathfrak{T} } \Big( \boldsymbol{S} : \boldsymbol{D}(  \dot{\boldsymbol{\varphi}})+ \frac{\partial \widehat{\mathfrak{L}}}{\partial \boldsymbol{S} }:\dot{\boldsymbol{S}} - \boldsymbol{J}  \cdot \nabla \mathfrak{T} + \frac{\partial \widehat{\mathfrak{L}}}{\partial \boldsymbol{Q} } \cdot \dot{\boldsymbol{Q}} \Big).
\end{equation}
This expression has to be compared with its analog \eqref{Sigma_expresssion_CIT} in the classical irreversible case.

\subsection{Variational formulation in the spatial description}\label{subsec_3_3}

\noindent\textbf{Material covariance.} In order to obtain the spatial description, we shall extend the notion of material covariance given in \eqref{mat_cov} to the class of nonequilibrium Lagrangians. We say that $\widehat{\mathfrak{L}}\big( \boldsymbol{\varphi}  ,\dot{\boldsymbol{\varphi}} , \nabla{ \boldsymbol{\varphi} } , \varrho , S,\boldsymbol{S}, \boldsymbol{Q}  \big)$ is material covariant if it satisfies
\begin{equation}\label{mat_cov_ext} 
\begin{aligned} 
&\widehat{\mathfrak{L}}\big( \boldsymbol{\varphi} \circ \boldsymbol{\psi}  ,\dot{\boldsymbol{\varphi}}\circ \boldsymbol{\psi}  , \nabla( \boldsymbol{\varphi} \circ \boldsymbol{\psi} ) , (\varrho \circ \boldsymbol{\psi} )J_{ \boldsymbol{\psi} } , (S \circ \boldsymbol{\psi} )J_{ \boldsymbol{\psi} }, (\boldsymbol{ \psi } ^* \boldsymbol{S} )J_{ \boldsymbol{\psi} },( \boldsymbol{ \psi } ^* \boldsymbol{Q} )J_{ \boldsymbol{\psi} }  \big)\\
&=\big(\widehat{\mathfrak{L}}\big( \boldsymbol{\varphi}  ,\dot{\boldsymbol{\varphi}} , \nabla{ \boldsymbol{\varphi} } , \varrho , S, \boldsymbol{S}, \boldsymbol{Q} \big) \circ \boldsymbol{\psi} \big)J_{ \boldsymbol{\psi} },
\end{aligned} 
\end{equation} 
for all diffeomorphisms $ \boldsymbol{\psi} : \mathcal{B} \rightarrow \mathcal{B}$. Here $\boldsymbol{ \psi } ^* \boldsymbol{S}$, resp., $\boldsymbol{ \psi } ^* \boldsymbol{Q}$ denote the pull-back of a two contravariant symmetric tensor field, resp., of a vector field, given in local coordinates as follows
\begin{equation}
(\boldsymbol{ \psi } ^* \boldsymbol{S})^{AB}= (\boldsymbol{S}^{CD} (\boldsymbol{ \psi }  ^{-1} )_{,C}^A(\boldsymbol{ \psi }  ^{-1} )_{,D}^B) \circ \boldsymbol{ \psi } 
\quad\text{and}\quad (\boldsymbol{ \psi } ^* \boldsymbol{Q})^A= (\boldsymbol{Q}^B (\boldsymbol{ \psi }  ^{-1} )_{,B}^A) \circ \boldsymbol{ \psi }.
\end{equation}
The occurrence of the transformation $ \boldsymbol{S} \mapsto (\boldsymbol{ \psi } ^* \boldsymbol{S} )J_{ \boldsymbol{\psi} }$ follows from the nature of $ \boldsymbol{S}$ as a $2$-contravariant tensor field density. Similarly, the transformation $ \boldsymbol{Q} \mapsto (\boldsymbol{ \psi } ^* \boldsymbol{Q} )J_{ \boldsymbol{\psi} }$ follows from the nature of $ \boldsymbol{Q}$ as a vector field density.

These transformations will determine two subsequent important steps
\begin{itemize}
\item[\rm (i)] The definition of the spatial versions of these two fluxes;
\item[\rm (ii)] The time rate of change of the resulting spatial fluxes.
\end{itemize}

\medskip

\noindent\textbf{Nonequilbrium Lagrangian in the spatial description.}
From the material covariance property \eqref{mat_cov} we can associate to $\widehat{\mathfrak{L}}$ a nonequilibrium Lagrangian density $\widehat{\ell}$ in the spatial description defined by
\begin{equation}\label{def_ell_ext} 
\widehat{\mathfrak{L}}\big( \boldsymbol{\varphi}  ,\dot{\boldsymbol{\varphi}} , \nabla{ \boldsymbol{\varphi} } , \varrho , S, \boldsymbol{S}, \boldsymbol{Q} \big) =\big(\widehat{\ell}\,( \boldsymbol{u}  , \rho  , s, \boldsymbol{\sigma} , \boldsymbol{q} ) \circ \boldsymbol{\varphi} \big)J_{ \boldsymbol{\varphi} }
\end{equation}
with $\boldsymbol{u}(t, \mathbf{x} )$, $ \rho (t, \mathbf{x} ) $, and $s(t, \mathbf{x} )$ the velocity, mass density, and entropy density in the spatial description, see \eqref{mat_to_spat} and \eqref{mat_to_spat_explicit}, and with $ \boldsymbol{\sigma}(t, \mathbf{x} )$ and $ \boldsymbol{q}(t, \mathbf{x} )$ the spatial versions of $ \boldsymbol{S}(t ,\mathbf{X} )$ and $ \boldsymbol{Q}(t, \mathbf{X} )$ given by
\begin{equation}\label{mat_to_spat_SJ} 
\boldsymbol{\sigma} = (\boldsymbol{\varphi} _* \boldsymbol{S})J_{ \boldsymbol{\varphi} ^{-1} }  , \quad \boldsymbol{q}= (\boldsymbol{\varphi} _* \boldsymbol{Q}) J_{ \boldsymbol{\varphi}^{-1}  }.
\end{equation}
In coordinates, these relations are written as
\begin{equation}\label{mat_to_spat_SJ_explicit} 
\begin{aligned}
\boldsymbol{\sigma }^{ab}( t,\boldsymbol{\varphi} (t, \mathbf{X} ))J_{ \boldsymbol{\varphi} }(t,X)&=  \boldsymbol{\varphi }^a_{,A}(t, \mathbf{X} ) \boldsymbol{\varphi }^b_{,B}(t, \mathbf{X} ) \boldsymbol{S}^{AB}(t, \mathbf{X} )\\
\boldsymbol{q}^a(t, \boldsymbol{\varphi} (t, \mathbf{X} ))J_{ \boldsymbol{\varphi} }(t,X)&=  \boldsymbol{\varphi }^a_{,A}( t,\mathbf{X} ) \boldsymbol{Q}^A( t,\mathbf{X} ).
\end{aligned} 
\end{equation} 
It is easily checked that, under the material covariance of $\widehat{\mathfrak{L}}$, there is a unique spatial Lagrangian density $\widehat{\ell}$ satisfying \eqref{mat_to_spat_SJ}.  This is conceptually the same step as the one used to pass from the material to the spatial Lagrangian densities in the reversible case in \eqref{def_ell}. The main conclusion here are the resulting relations \eqref{mat_to_spat_SJ} between the material and spatial thermodynamic fluxes. In particular, \eqref{mat_to_spat_SJ} represents the natural covariant transformation of a material contravariant symmetric tensor field density, resp., a material vector field density,  into their spatial versions. These are the natural tensorial generalizations of the material-to-spatial transformation of mass and entropy densities $ \varrho \mapsto \rho  =(\varrho \circ \boldsymbol{\varphi} ^{-1}) J_{ \boldsymbol{\varphi} ^{-1} }$, $S \mapsto s=(S \circ \boldsymbol{\varphi} ^{-1}) J_{ \boldsymbol{\varphi} ^{-1} }$ appearing earlier.

\medskip

\noindent\textbf{Standard Lagrangian densities.}
For the Lagrangian \eqref{Mat_Lag_Rev_ext}, since the first term is clearly material covariant, the property \eqref{mat_cov_ext} is equivalent to the following condition on the material nonequilibrium  internal energy function
\begin{equation}
\widehat{\mathscr{E}}\big(\nabla( \boldsymbol{\varphi}  \circ \boldsymbol{\psi }) , (\varrho \circ \boldsymbol{\psi} )J_{ \boldsymbol{\psi} }, (S \circ \boldsymbol{\psi} )J_{ \boldsymbol{\psi} },(\boldsymbol{ \psi } ^* \boldsymbol{S} )J_{ \boldsymbol{\psi} },( \boldsymbol{ \psi } ^* \boldsymbol{Q} )J_{ \boldsymbol{\psi} } \big)= \widehat{\mathscr{E}}\big(\nabla{ \boldsymbol{\varphi} }, \varrho , S, \boldsymbol{S}, \boldsymbol{Q}\big)\circ \boldsymbol{\psi} 
\end{equation}
for all diffeomorphisms $ \boldsymbol{\psi} : \mathcal{B} \rightarrow \mathcal{B} $. Thus, we can write it as
\begin{equation}
\widehat{\mathscr{E}}\big(\nabla{ \boldsymbol{\varphi} }, \varrho , S, \boldsymbol{S}, \boldsymbol{Q}\big)\varrho  =\big( \widehat{\varepsilon}\,( \rho  , s, \boldsymbol{\sigma} , \boldsymbol{q}) \circ \boldsymbol{\varphi} \big)J_{ \boldsymbol{\varphi} }
\end{equation}
in terms of a nonequilibrium internal energy density expression $ \widehat{\varepsilon}\,( \rho  , s, \boldsymbol{\sigma} , \boldsymbol{q})$.

In conclusion the spatial nonequilibrium Lagrangian density associated to \eqref{Mat_Lag_Rev_ext} via \eqref{def_ell_ext} has the standard form
\begin{equation}\label{ell_ext} 
 \widehat{\ell}( \boldsymbol{u}, \rho  , s, \boldsymbol{\sigma}, \boldsymbol{q} )= \frac{1}{2} \rho  | \boldsymbol{u}| ^2 - \widehat{\varepsilon}\,( \rho  , s, \boldsymbol{\sigma}, \boldsymbol{q}).
\end{equation}

\medskip

\noindent\textbf{Derivation of the variational formulation.} We derive here the spatial version of the variational formulation \eqref{VCond_continuum_ext}--\eqref{VC_continuum_ext}. This is achieved by converting each of the three conditions, as done in {\bf (i)},{\bf (ii)},{\bf (iii)} below.

\medskip

\noindent{\bf (i)} The variational condition \eqref{VCond_continuum_ext} can be easily converted into its spatial version as done earlier, by using the relations \eqref{def_ell_ext}, \eqref{mat_to_spat}, \eqref{def_gamma_sigma}, \eqref{mat_to_spat_SJ}.

\medskip

\noindent{\bf (ii)} To get the spatial version of the phenomenological constraint \eqref{PC_continuum_ext}, we note that when $ \boldsymbol{S}(t, \mathbf{X} )$ and $\boldsymbol{Q}(t, \mathbf{X} )$ are related to $ \boldsymbol{\sigma} (t, \mathbf{x} )$ and $ \boldsymbol{q}(t, \mathbf{x} )$ via the formulas \eqref{mat_to_spat_SJ} for some time dependent diffeomorphism $ \boldsymbol{\varphi} (t, \mathbf{X} )$, we can show the following identities
\begin{align}
\frac{\partial \widehat{\mathfrak{L}}}{\partial \boldsymbol{S} }:\dot{\boldsymbol{S}} &= \bigg( \Big(\frac{\partial \widehat{\ell}}{\partial \boldsymbol{\sigma} } :\mathfrak{D}_t \boldsymbol{\sigma}  \Big) \circ \boldsymbol{\varphi} \bigg) J_{\boldsymbol{\varphi}} \label{F1}\\
\frac{\partial \widehat{\mathfrak{L}}}{\partial \boldsymbol{Q} } \cdot \dot{ \boldsymbol{Q}}&= \bigg( \Big(\frac{\partial \widehat{\ell}}{\partial \boldsymbol{q} }\cdot  \mathfrak{D}_t \boldsymbol{q}  \Big) \circ \boldsymbol{\varphi} \bigg)  J_{\boldsymbol{\varphi}}, \label{F2}
\end{align}  
where $\mathfrak{D}_t$ denotes the time rate of change of Eulerian contravariant tensor field densities (or Truesdell rate), given by
\begin{align}
\mathfrak{D}_t \boldsymbol{\sigma}&=\partial _t \boldsymbol{\sigma }+ \pounds _ {\boldsymbol{u}} \boldsymbol{\sigma }  + \boldsymbol{\sigma} \operatorname{div} \boldsymbol{u}\\
\mathfrak{D}_t \boldsymbol{q}&= \partial _t \boldsymbol{q}+ \pounds _{ \boldsymbol{u}} \boldsymbol{q}   + \boldsymbol{q} \operatorname{div} \boldsymbol{u}
\end{align}
with $(\pounds _{ \boldsymbol{u}} \boldsymbol{\sigma })^{ab} = \boldsymbol{u}^c \partial _c \boldsymbol{\sigma}^{ab}- \boldsymbol{ \sigma }^{cb} \partial _c \boldsymbol{u}^a- \boldsymbol{ \sigma }^{ac} \partial _c \boldsymbol{u}^b$ and $(\pounds _{ \boldsymbol{u}} \boldsymbol{q})^a = \boldsymbol{u}^c \partial _c \boldsymbol{q}^a- \boldsymbol{q}^c \partial _c \boldsymbol{u}^a$ the Lie derivative of a contravariant tensor field and a vector field. To prove \eqref{F1}--\eqref{F2} we first take the partial derivative of the equality \eqref{def_ell_ext} with respect to $ \boldsymbol{Q}$ in some direction $ \boldsymbol{V}$, and get  
\begin{equation}
\frac{\partial \widehat{\mathfrak{L}}}{\partial \boldsymbol{Q}} \cdot  \boldsymbol{V}  = \Big( \frac{\partial \widehat{\ell}}{\partial \boldsymbol{q}} \cdot \boldsymbol{\varphi} _* \boldsymbol{V}J_{ \boldsymbol{\varphi} ^{-1} } \Big) \circ \boldsymbol{\varphi} J_{ \boldsymbol{\varphi} } = \Big( \frac{\partial \widehat{\ell}}{\partial \boldsymbol{q}} \cdot \boldsymbol{\varphi} _*  \boldsymbol{V}   \Big) \circ \boldsymbol{\varphi}=  \boldsymbol{\varphi} ^*  \frac{\partial \widehat{\ell}}{\partial \boldsymbol{q}} \cdot  \boldsymbol{V}.
\end{equation}
In the second equality we simplified the Jacobian factors while the last equality can be checked in coordinates by using expressions like \eqref{mat_to_spat_SJ_explicit}. Similar steps give
\begin{equation}
\frac{\partial \widehat{\mathfrak{L}}}{\partial \boldsymbol{S}} : \boldsymbol{W} = \Big( \frac{\partial \widehat{\ell}}{\partial \boldsymbol{\sigma }} : \boldsymbol{\varphi} _*  \boldsymbol{W} J_{ \boldsymbol{\varphi} ^{-1} } \Big) \circ \boldsymbol{\varphi} J_{ \boldsymbol{\varphi} } = \Big( \frac{\partial \widehat{\ell}}{\partial \boldsymbol{\sigma}} : \boldsymbol{\varphi} _*  \boldsymbol{W} \Big) \circ \boldsymbol{\varphi}=  \boldsymbol{\varphi} ^*  \frac{\partial \widehat{\ell}}{\partial \boldsymbol{\sigma}} :  \boldsymbol{W}.
\end{equation}
We thus get
\begin{equation}\label{result_1} 
\frac{\partial \widehat{\mathfrak{L}}}{\partial \boldsymbol{Q}}=  \boldsymbol{\varphi} ^*  \frac{\partial \widehat{\ell}}{\partial \boldsymbol{q}}  \quad\text{and}\quad \frac{\partial \widehat{\mathfrak{L}}}{\partial \boldsymbol{S}}=  \boldsymbol{\varphi} ^*  \frac{\partial \widehat{\ell}}{\partial \boldsymbol{\sigma }} .
\end{equation}
Using again \eqref{mat_to_spat_SJ}, we compute
\begin{equation}\label{result_2}
\frac{d}{dt} \boldsymbol{Q}= \frac{d}{dt}  ( \boldsymbol{\varphi} ^* \boldsymbol{q}) J_{ \boldsymbol{\varphi} } = ( \boldsymbol{\varphi} ^* \mathfrak{D}_t\boldsymbol{q}) J_{ \boldsymbol{\varphi} } \quad\text{and}\quad \frac{d}{dt} \boldsymbol{S}= \frac{d}{dt}  ( \boldsymbol{\varphi} ^* \boldsymbol{ \sigma }) J_{ \boldsymbol{\varphi} } = ( \boldsymbol{\varphi} ^* \mathfrak{D}_t\boldsymbol{\sigma }) J_{ \boldsymbol{\varphi} }
\end{equation} 
which follows either from a local computation or from the properties of the Lie derivative, \citep{MaHu1983}.
Using \eqref{result_1} and \eqref{result_2}  we finally obtain
\begin{equation}
\frac{\partial \widehat{\mathfrak{L}}}{\partial \boldsymbol{S} }:\dot{\boldsymbol{S}} =  \boldsymbol{\varphi} ^*  \frac{\partial \widehat{\ell}}{\partial \boldsymbol{\sigma }} \cdot ( \boldsymbol{\varphi} ^* \mathfrak{D}_t\boldsymbol{\sigma }) J_{ \boldsymbol{\varphi} }=\bigg( \Big(\frac{\partial \widehat{\ell}}{\partial \boldsymbol{\sigma} } :\mathfrak{D}_t \boldsymbol{\sigma}  \Big) \circ \boldsymbol{\varphi} \bigg) J_{\boldsymbol{\varphi}} 
\end{equation}
and
\begin{equation}
\frac{\partial \widehat{\mathfrak{L}}}{\partial \boldsymbol{Q} } \cdot \dot{\boldsymbol{Q}} =  \boldsymbol{\varphi} ^*  \frac{\partial \widehat{\ell}}{\partial \boldsymbol{q}} \cdot ( \boldsymbol{\varphi} ^* \mathfrak{D}_t\boldsymbol{q}) J_{ \boldsymbol{\varphi} }=\bigg( \Big(\frac{\partial \widehat{\ell}}{\partial \boldsymbol{q} } \cdot \mathfrak{D}_t \boldsymbol{q}  \Big) \circ \boldsymbol{\varphi} \bigg) J_{\boldsymbol{\varphi}} 
\end{equation}
as desired.

\medskip

\noindent{\bf (iii)} Finally, to get the spatial version of the variational constraint \eqref{VC_continuum_ext}, we use similar steps as before. For variations $ \boldsymbol{S}_ \varepsilon (t, \mathbf{X} )$ and $\boldsymbol{Q}_ \varepsilon (t, \mathbf{X} )$ related to $ \boldsymbol{\sigma} _ \varepsilon (t, \mathbf{x} )$ and $ \boldsymbol{q}_ \varepsilon (t, \mathbf{x} )$ via the formulas \eqref{mat_to_spat_SJ} by diffeomorphisms $ \boldsymbol{\varphi} _ \varepsilon (t, \mathbf{x} )$, we get
\begin{align}
\frac{\partial \widehat{\mathfrak{L}}}{\partial \boldsymbol{S} }: \delta \boldsymbol{S} &= \bigg( \Big(\frac{\partial \widehat{\ell}}{\partial \boldsymbol{\sigma} } :\mathfrak{D}_ \delta  \boldsymbol{\sigma}  \Big) \circ \boldsymbol{\varphi} \bigg) J_{\boldsymbol{\varphi}} \label{F3}\\
\frac{\partial \widehat{\mathfrak{L}}}{\partial \boldsymbol{Q} } \cdot \delta  \boldsymbol{Q}&= \bigg( \Big(\frac{\partial \widehat{\ell}}{\partial \boldsymbol{q} }\cdot  \mathfrak{D}_\delta  \boldsymbol{q}  \Big) \circ \boldsymbol{\varphi} \bigg)  J_{\boldsymbol{\varphi}}, \label{F4}
\end{align}  
where $\mathfrak{D}_ \delta $ denotes the variational rate of change of Eulerian contravariant tensor field densities
\begin{align}
\mathfrak{D}_\delta  \boldsymbol{\sigma}&=\delta \boldsymbol{\sigma }+ \pounds _ {\boldsymbol{\zeta}} \boldsymbol{\sigma }  + \boldsymbol{\sigma} \operatorname{div} \boldsymbol{\zeta}\\
\mathfrak{D}_\delta  \boldsymbol{q}&= \delta  \boldsymbol{q}+ \pounds _{ \boldsymbol{\zeta}} \boldsymbol{q}   + \boldsymbol{q} \operatorname{div} \boldsymbol{\zeta},
\end{align}
for $ \boldsymbol{\zeta} = \delta \boldsymbol{\varphi} \circ \boldsymbol{\varphi} ^{-1} $, see \eqref{zeta}.

\medskip

\noindent\textbf{The variational formulation.} By using \eqref{F1},  \eqref{F2},  \eqref{F3},  \eqref{F4}, we can convert the variational formulation \eqref{VCond_continuum_ext}--\eqref{VC_continuum_ext} into the spatial description and get the following.

Find the critical fields of
\begin{equation}\label{VCond_continuum_spat_ext} 
\delta \int_0^T \!\!\int_ \mathcal{D} \left[\widehat{\ell}( \boldsymbol{u}  ,  \rho  , s, \boldsymbol{\sigma}, \boldsymbol{q} ) + (s- \varsigma  ) D_t \gamma  \right] {\rm d}^3  \mathbf{x} \, {\rm d} t=0
\end{equation} 
subject to the \textit{phenomenological constraint}
\begin{equation}\label{PC_continuum_spat_ext}
\frac{\partial \widehat{\ell}}{\partial s}\bar D_t \varsigma+\frac{\partial \widehat{\ell}}{\partial \boldsymbol{\sigma} }:\mathfrak{D}_t \boldsymbol{\sigma} +\frac{\partial \widehat{\ell}}{\partial \boldsymbol{q} } \cdot \mathfrak{D}_t \boldsymbol{q}   = - \boldsymbol{ \sigma } : \nabla \boldsymbol{u }  + \boldsymbol{j}  \cdot  \nabla  D_t \gamma
\end{equation}
and for variations subject to the \textit{variational constraint}
\begin{equation}\label{VC_continuum_spat_ext}
\frac{\partial \widehat{\ell}}{\partial s}\bar D_ \delta  \varsigma+\frac{\partial \widehat{\ell}}{\partial \boldsymbol{\sigma} }:\mathfrak{D}_\delta \boldsymbol{\sigma} +\frac{\partial \widehat{\ell}}{\partial \boldsymbol{q} } \cdot \mathfrak{D}_\delta \boldsymbol{q}  = - \boldsymbol{ \sigma } : \nabla \boldsymbol{\zeta }  + \boldsymbol{j}  \cdot  \nabla  D_\delta  \gamma,
\end{equation}
as well as the usual Eulerian variations
\begin{equation}\label{red_var_2_ext} 
\delta \boldsymbol{u}= \partial _t \boldsymbol{\zeta} + \boldsymbol{u} \cdot \nabla \boldsymbol{\zeta}- \boldsymbol{\zeta } \cdot \nabla \boldsymbol{u}  , \qquad \delta   \rho  = - \operatorname{div}( \rho  \boldsymbol{\zeta} ).
\end{equation}

\medskip

\noindent\textbf{Derivation of the critical conditions.} By taking the variations in \eqref{VCond_continuum_spat_ext}, we get
\begin{equation}\label{comput1}
\begin{aligned}
&\int_0^T\int_ \mathcal{D} \Big[\frac{\partial \widehat{\ell}}{\partial \boldsymbol{u} } \cdot \delta \boldsymbol{u}  + \frac{\partial \widehat{\ell}}{\partial \rho  }   \delta \rho  + \frac{\partial \widehat{\ell}}{\partial s }   \delta s + \frac{\partial \widehat{\ell}}{\partial \boldsymbol{\sigma } } : \delta \boldsymbol{\sigma }  + \frac{\partial \widehat{\ell}}{\partial \boldsymbol{q} } \cdot \delta \boldsymbol{q} + \delta s  D_t \gamma  - \delta  \varsigma   D_t \gamma \\
& \hspace{5cm} - \bar D_t s \delta \gamma  - \bar D_t \varsigma  \delta \gamma + (s- \varsigma ) \nabla \gamma \cdot \delta \boldsymbol{u}\Big] {\rm d}^3   \mathbf{x} \,{\rm d} t=0,
\end{aligned}
\end{equation}
for variations subject to \eqref{VC_continuum_spat_ext} and \eqref{red_var_2}.
Since the variations $ \delta s$ are arbitrary, we get
\begin{equation}\label{interm1} 
\frac{\partial \widehat{\ell}}{\partial s} =- D_t \gamma .
\end{equation}
This equality can be used in the seventh term in \eqref{comput1}, which can then be replaced by using the variational constraint \eqref{VC_continuum_spat_ext}. Having made this replacement, we then isolate $ \boldsymbol{\zeta} $ and $ \delta \gamma $ by using the divergence theorem for the terms $ \boldsymbol{\sigma} : \nabla \boldsymbol{\zeta} $ and $ \boldsymbol{j} \cdot \nabla D_ \delta \gamma $, together with $ \boldsymbol{\zeta} |_{ \partial \mathcal{D} }=0$. By collecting the terms proportional to $ \delta \gamma $, we get the conditions
\begin{equation}\label{interm2} 
\bar D_t \varsigma = \bar D_t s + \operatorname{div} \boldsymbol{j} \quad\text{and}\quad (\boldsymbol{j} \cdot  \boldsymbol{n}) \delta \gamma =0 \quad\text{on}\quad \partial \mathcal{D}.
\end{equation} 
So, from \eqref{red_var_2} we are left with the condition
\begin{equation}\label{comput2}
\begin{aligned}
&\int_0^T\int_ \mathcal{D} \Big[\Big( \frac{\partial \widehat{\ell}}{\partial \boldsymbol{u} } +(s- \varsigma ) \nabla \gamma \Big) \cdot \delta \boldsymbol{u}  + \frac{\partial \widehat{\ell}}{\partial \rho  }   \delta \rho  + \Big( \nabla \frac{\partial \widehat{\ell}}{\partial s} \varsigma + \operatorname{div} \boldsymbol{\sigma} - \operatorname{div} \boldsymbol{j} \nabla \gamma \Big) \cdot \boldsymbol{\zeta}  \\
& \hspace{2cm}  + \frac{\partial \widehat{\ell}}{\partial \boldsymbol{\sigma } } : \delta \boldsymbol{\sigma }  + \frac{\partial \widehat{\ell}}{\partial \boldsymbol{q} } \cdot \delta \boldsymbol{q}  - \frac{\partial \widehat{\ell}}{\partial \boldsymbol{\sigma} }:\mathfrak{D}_\delta \boldsymbol{\sigma} -\frac{\partial \widehat{\ell}}{\partial \boldsymbol{q} } \cdot \mathfrak{D}_\delta \boldsymbol{q}  \Big] {\rm d}^3   \mathbf{x} \,{\rm d} t=0,
\end{aligned}
\end{equation}  
for variations subject to \eqref{red_var_2}. It remains to isolate $ \boldsymbol{\zeta} $. To treat the first line in \eqref{comput2} we use \eqref{red_var_2} then we integrate by part in time and use the divergence theorem on $\mathcal{D}$, to finally get   
\begin{equation}\label{result1}
\int_0^T\int_ \mathcal{D}\Big[ -\partial _t  \frac{\partial \widehat{\ell}}{\partial \boldsymbol{u} } + \pounds _{ \boldsymbol{u}} \frac{\partial \widehat{\ell}}{\partial \boldsymbol{u} } +  \rho  \nabla \frac{\partial \widehat{\ell}}{\partial \rho   } +  s  \nabla \frac{\partial \widehat{\ell}}{\partial s   } + \operatorname{div} \boldsymbol{\sigma} \Big] \cdot \boldsymbol{\zeta} \,{\rm d} ^3 \mathbf{x} \, {\rm d} t .
\end{equation} 
To treat the second line in \eqref{comput2} we note the following formulas
\begin{equation}\label{formula1} 
\begin{aligned} 
&\int_ \mathcal{D}\Big[\frac{\partial \widehat{\ell}}{\partial \boldsymbol{q} } \cdot \mathfrak{D}_\delta \boldsymbol{q} - \frac{\partial \widehat{\ell}}{\partial \boldsymbol{q} } \cdot \delta \boldsymbol{q}\Big] {\rm d} ^3 \mathbf{x} = \int_ \mathcal{D}\Big[\frac{\partial \widehat{\ell}}{\partial \boldsymbol{q} } \cdot (\pounds _ {\boldsymbol{\zeta} }\boldsymbol{q}+ \boldsymbol{q}\operatorname{div} \boldsymbol{\zeta} ) \Big] {\rm d} ^3 \mathbf{x}\\
&=\int_ \mathcal{D} \Big[ \frac{\partial \widehat{\ell}}{\partial \boldsymbol{q} } \cdot \nabla \boldsymbol{q}- \operatorname{div}( \boldsymbol{\tau }_{ \boldsymbol{q}} )  \Big] \cdot \boldsymbol{\zeta} \,{\rm d} ^3 \mathbf{x} + \int_{ \partial \mathcal{D} } \boldsymbol{\tau}_ {\boldsymbol{q}} \cdot \boldsymbol{\zeta} \cdot \boldsymbol{n} \,{\rm d} A
\end{aligned} 
\end{equation} 
and
\begin{equation}\label{formula2} 
\begin{aligned} 
&\int_ \mathcal{D}\Big[\frac{\partial \widehat{\ell}}{\partial \boldsymbol{\sigma} } : \mathfrak{D}_\delta \boldsymbol{\sigma} - \frac{\partial \widehat{\ell}}{\partial \boldsymbol{\sigma} } : \delta \boldsymbol{\sigma}\Big] {\rm d} ^3 \mathbf{x} = \int_ \mathcal{D}\Big[\frac{\partial \widehat{\ell}}{\partial \boldsymbol{\sigma} } \cdot (\pounds _ {\boldsymbol{\zeta} }\boldsymbol{\sigma}+ \boldsymbol{\sigma}\operatorname{div} \boldsymbol{\zeta} ) \Big] {\rm d} ^3 \mathbf{x}\\
&=\int_ \mathcal{D} \Big[ \frac{\partial \widehat{\ell}}{\partial \boldsymbol{\sigma} } : \nabla \boldsymbol{\sigma}- \operatorname{div}( \boldsymbol{\tau }_{ \boldsymbol{\sigma}} )  \Big] \cdot \boldsymbol{\zeta} \,{\rm d} ^3 \mathbf{x} + \int_{ \partial \mathcal{D} } \boldsymbol{\tau}_ {\boldsymbol{\sigma}} \cdot \boldsymbol{\zeta} \cdot \boldsymbol{n} \,{\rm d} A,
\end{aligned} 
\end{equation}
where we defined the \textit{nonequilibrium stresses}
\begin{equation}\label{NES}
\begin{aligned} 
\boldsymbol{\tau}_{\boldsymbol{q}}:&= \Big(\frac{\partial \widehat{\ell}}{\partial \boldsymbol{q} }\cdot \boldsymbol{q}\Big) \delta -  \frac{\partial \widehat{\ell}}{\partial \boldsymbol{q} } \otimes \boldsymbol{q} , & \qquad (\boldsymbol{\tau}_{\boldsymbol{q}})^a_b&= \frac{\partial \widehat{\ell}}{\partial \boldsymbol{q}^c }\boldsymbol{q}^c \delta ^a_b - \frac{\partial \widehat{\ell}}{\partial \boldsymbol{q}^b}\cdot \boldsymbol{q}^a,\\
\boldsymbol{\tau}_{\boldsymbol{\sigma}}:&= \Big(\frac{\partial \widehat{\ell}}{\partial \boldsymbol{\sigma} }: \boldsymbol{\sigma}\Big) \delta -  2\frac{\partial \widehat{\ell}}{\partial \boldsymbol{\sigma } } \cdot  \boldsymbol{\sigma}, &\qquad (\boldsymbol{\tau}_{\boldsymbol{\sigma}})^a_b&=\Big(\frac{\partial \widehat{\ell}}{\partial \boldsymbol{\sigma} ^{cd}} \boldsymbol{\sigma}^{cd}\Big) \delta _b^a- 2 \frac{\partial \widehat{\ell}}{\partial \boldsymbol{\sigma }^{bd} }  \boldsymbol{\sigma}^{ad}.
\end{aligned}
\end{equation}
Hence, using $ \boldsymbol{\zeta} |_{ \partial \mathcal{D} }=0$, the second line in \eqref{comput2} reduces to
\begin{equation}\label{result2}
\int_0^T\int_ \mathcal{D}\Big[ - \frac{\partial \widehat{\ell}}{\partial \boldsymbol{\sigma} } : \nabla \boldsymbol{\sigma} - \frac{\partial \widehat{\ell}}{\partial \boldsymbol{q} } \cdot  \nabla \boldsymbol{q} + \operatorname{div} (\boldsymbol{\tau}_{\boldsymbol{\sigma }}+ \boldsymbol{\tau}_{\boldsymbol{q}})  \Big] \cdot \boldsymbol{\zeta} \,{\rm d} ^3 \mathbf{x} \, {\rm d} t .
\end{equation} 
So, finally, by using \eqref{result1}, \eqref{result2}, \eqref{PC_continuum_spat_ext}, \eqref{interm1}, and \eqref{interm2} we get
\begin{equation}\label{NSF_spat_ext} 
\left\{
\begin{array}{l}
\vspace{0.2cm}\displaystyle \partial _t \frac{\partial \widehat{\ell}}{\partial \boldsymbol{u} }  + \pounds _ {\boldsymbol{u}} \frac{\partial \widehat{\ell}}{\partial \boldsymbol{u} }= \rho  \nabla \frac{\partial \widehat{\ell}}{\partial \rho   }+ s \nabla \frac{\partial \widehat{\ell}}{\partial s} - \frac{\partial \widehat{\ell}}{\partial \boldsymbol{\sigma} } : \nabla \boldsymbol{\sigma} - \frac{\partial \widehat{\ell}}{\partial \boldsymbol{q} } \cdot  \nabla \boldsymbol{q} + \operatorname{div} (\boldsymbol{\sigma} +\boldsymbol{\tau}_{\boldsymbol{\sigma }}+ \boldsymbol{\tau}_{\boldsymbol{q}})\\
\vspace{0.2cm}\displaystyle
-\frac{\partial  \widehat{\ell} }{ \partial  s}(\bar D_t s + \operatorname{div} \boldsymbol{j}  ) = \boldsymbol{ \sigma } : \nabla \boldsymbol{u}   + \boldsymbol{j}  \cdot \nabla \frac{\partial   \widehat{\ell} }{\partial   s}+\frac{\partial \widehat{\ell}}{\partial \boldsymbol{\sigma} }:\mathfrak{D}_t \boldsymbol{\sigma} +\frac{\partial \widehat{\ell}}{\partial \boldsymbol{q} } \cdot \mathfrak{D}_t \boldsymbol{q}\\
\displaystyle\bar D_t \rho  =0.
\end{array}
\right.
\end{equation}

A reformulation of the fluid momentum equation that is useful for energy balance is given by
\begin{equation}
D_t \frac{\partial \widehat{\ell}}{\partial \boldsymbol{u} } + \frac{\partial \widehat{\ell}}{\partial \boldsymbol{u} }  \operatorname{div} \boldsymbol{u}=  \operatorname{div} \bigg[\Big(   \frac{\partial \widehat{\ell}}{\partial \rho  } \rho +   \frac{\partial \widehat{\ell}}{\partial s  } s+  \frac{\partial \widehat{\ell}}{\partial \boldsymbol{q}} \cdot  \boldsymbol{q} +    \frac{\partial \widehat{\ell}}{\partial \boldsymbol{\sigma }   }: \boldsymbol{\sigma } -\ell\Big)  \delta - \frac{\partial \widehat{\ell}}{\partial \boldsymbol{q} } \otimes \boldsymbol{q} - 2\frac{\partial \widehat{\ell}}{\partial \boldsymbol{\sigma } } \cdot  \boldsymbol{\sigma} + \boldsymbol{\sigma } \bigg] ,
\end{equation} 
which directly follows by using the expression \eqref{Lie_derivative} of the Lie derivative of the fluid momentum.

\medskip

In the following proposition, we summarize the variational formulations of extended irreversible thermodynamics developed above.

\begin{proposition}[Variational principles for extended irreversible thermodynamics]
Consider a viscous heat conducting fluid with nonequilibrium Lagrangian density
\[
\widehat{\mathfrak{L}}( \boldsymbol{\varphi}  ,\dot{\boldsymbol{\varphi}} , \nabla{ \boldsymbol{\varphi} } , \varrho , S, \boldsymbol{S}, \boldsymbol{Q})
\]
expressed in terms of the configuration map $\boldsymbol{\varphi}$ and its first derivatives, in terms of the mass and entropy densities $\varrho$ and $S$, and in terms of the thermodynamic fluxes $\boldsymbol{S}$ and $\boldsymbol{Q}$. 

\begin{itemize}
\item \textsf{Material description}: The variational principle \eqref{VCond_continuum_ext}--\eqref{VC_continuum_ext} yields the equation of motion
\begin{equation}
\left\{
\begin{array}{l}
\vspace{0.2cm}\displaystyle\frac{d}{dt}\frac{ \partial \widehat{\mathfrak{L}} }{ \partial   \dot{\boldsymbol{\varphi}} }   - \frac{ \partial \widehat{\mathfrak{L}}}{\partial \boldsymbol{\varphi}  }= \operatorname{DIV}\Big( - \frac{\partial\widehat{\mathfrak{L}}}{\partial \nabla{\boldsymbol{\varphi} } } +  \boldsymbol{P}  \Big) \\
\displaystyle
-\frac{\partial \widehat{\mathfrak{L}}}{ \partial    S}(\dot S + \operatorname{DIV} \boldsymbol{J}  )- \frac{\partial \widehat{\mathfrak{L}}}{\partial \boldsymbol{S} }:\dot{\boldsymbol{S}} -\frac{\partial \widehat{\mathfrak{L}}}{\partial \boldsymbol{Q} } \cdot \dot{ \boldsymbol{Q}}  =  \boldsymbol{S} : \boldsymbol{D} (\dot{\boldsymbol{\varphi}} ) + \boldsymbol{J}  \cdot \nabla \frac{\partial  \widehat{\mathfrak{L}} }{\partial    S},
\end{array}
\right.
\end{equation}
together with the insulated boundary condition $\boldsymbol{J}  \cdot \boldsymbol{N}  =0$. Furthermore, the variational principle also determines the following relationships for the variables $\Gamma$ and $\Sigma$ in \eqref{VCond_continuum_ext}--\eqref{VC_continuum_ext}:
\begin{equation}
\dot \Gamma = - \frac{\partial  \widehat{\mathfrak{L}} }{ \partial   S} \qquad\text{and}\qquad \dot \Sigma = \dot S + \operatorname{DIV} \boldsymbol{J}.
\end{equation}
\item \textsf{Spatial description}: When the nonequilibrium Lagrangian density $\widehat{\mathfrak{L}}$ satisfies the material covariance assumption, it has an associated nonequilibrium Eulerian Lagrangian density $\widehat{\ell}\,( \boldsymbol{u}, \rho  , s, \boldsymbol{\sigma }, \boldsymbol{q})$ and \eqref{VCond_continuum_ext}--\eqref{VC_continuum_ext} yields the Eulerian variational formulation \eqref{VCond_continuum_spat_ext}--\eqref{VC_continuum_spat_ext}. This variational principle yields the equations of motion 
\begin{equation}
\left\{
\begin{array}{l}
\vspace{0.2cm}\displaystyle \partial _t \frac{\partial \widehat{\ell}}{\partial \boldsymbol{u} }  + \pounds _ {\boldsymbol{u}} \frac{\partial \widehat{\ell}}{\partial \boldsymbol{u} }= \rho  \nabla \frac{\partial \widehat{\ell}}{\partial \rho   }+ s \nabla \frac{\partial \widehat{\ell}}{\partial s} - \frac{\partial \widehat{\ell}}{\partial \boldsymbol{\sigma} } : \nabla \boldsymbol{\sigma} - \frac{\partial \widehat{\ell}}{\partial \boldsymbol{q} } \cdot  \nabla \boldsymbol{q} + \operatorname{div} (\boldsymbol{\sigma} +\boldsymbol{\tau}_{\boldsymbol{\sigma }}+ \boldsymbol{\tau}_{\boldsymbol{q}})\\
\vspace{0.2cm}\displaystyle
-\frac{\partial  \widehat{\ell} }{ \partial  s}(\bar D_t s + \operatorname{div} \boldsymbol{j}  ) = \boldsymbol{ \sigma } : \nabla \boldsymbol{u}   + \boldsymbol{j}  \cdot \nabla \frac{\partial   \widehat{\ell} }{\partial   s}+\frac{\partial \widehat{\ell}}{\partial \boldsymbol{\sigma} }:\mathfrak{D}_t \boldsymbol{\sigma} +\frac{\partial \widehat{\ell}}{\partial \boldsymbol{q} } \cdot \mathfrak{D}_t \boldsymbol{q}\\
\displaystyle\bar D_t \rho  =0,
\end{array}
\right.
\end{equation}
which incorporate the nonequilibrium stresses \eqref{NES},
together with the insulated boundary condition $\boldsymbol{j}  \cdot \boldsymbol{n}  =0$. Furthermore, the variational principle also determines the following relationships for the variables $\gamma$ and $\varsigma$ in \eqref{VCond_continuum_spat_ext}--\eqref{VC_continuum_ext}:
\begin{equation}
D_t \gamma=- \frac{\partial \widehat{\ell}}{\partial s} \qquad\text{and}\qquad \bar D_t \varsigma = \bar D_t s + \operatorname{div} \boldsymbol{j}.
\end{equation}
\end{itemize}
\end{proposition}

\medskip

\noindent\textbf{Equation for standard Lagrangians.} For the Lagrangian density of the standard form \eqref{ell_ext}, one gets from \eqref{NSF_spat_ext} the system
\begin{equation}\label{NSF_spat_ext_standard} 
\left\{
\begin{array}{l}
\vspace{0.2cm}\displaystyle \rho  ( \partial _t \boldsymbol{u}+ \boldsymbol{u} \cdot \nabla \boldsymbol{u} )  = - \nabla p + \operatorname{div}( \boldsymbol{\tau }_{ \boldsymbol{q}}+\boldsymbol{\tau }_{ \boldsymbol{ \sigma }}+ \boldsymbol{\sigma} )   \\
\vspace{0.2cm}\displaystyle
T(\bar D_t s + \operatorname{div} \boldsymbol{j}  ) = \boldsymbol{ \sigma } : \nabla \boldsymbol{u}   - \boldsymbol{j}  \cdot \nabla T-\frac{\partial \widehat{\varepsilon }}{\partial \boldsymbol{\sigma} }:\mathfrak{D}_t \boldsymbol{\sigma} -\frac{\partial \widehat{\varepsilon }}{\partial \boldsymbol{q} } \cdot \mathfrak{D}_t \boldsymbol{q}\\
\displaystyle\bar D_t \rho  =0
\end{array}
\right.
\end{equation}  
with \textit{nonequilibrium stresses}
\begin{equation}\label{expression_stresses}
\boldsymbol{\tau}_{\boldsymbol{q}}:= - \Big(\frac{\partial \widehat{\varepsilon }}{\partial \boldsymbol{q} }\cdot \boldsymbol{q}\Big) \delta +  \frac{\partial \widehat{\varepsilon }}{\partial \boldsymbol{q} } \otimes \boldsymbol{q} \quad\text{and}\quad \boldsymbol{\tau}_{\boldsymbol{\sigma}}:= -\Big(\frac{\partial \widehat{\varepsilon }}{\partial \boldsymbol{\sigma} }: \boldsymbol{\sigma}\Big) \delta +  2\frac{\partial \widehat{\varepsilon }}{\partial \boldsymbol{\sigma } } \cdot  \boldsymbol{\sigma}
\end{equation}
and with nonequilibrium temperature and pressure given as $T= \frac{\partial\widehat{ \varepsilon}}{\partial s}$ and $p= \rho\frac{\partial \widehat{ \varepsilon}}{\partial\rho} + s\frac{\partial \widehat{ \varepsilon}}{\partial s} - \widehat{ \varepsilon}$.

\begin{remark}[Energy representation]{\rm Our variational formulation is best expressed by using the nonequilibrium internal energy function, with the nonequilibrium entropy as an independent variable, see \eqref{ell_ext}. Following the approach in \citep{GBYo2019b} it is possible to convert the variational formulation in other representations, such as the nonequilibrium free energy with the nonequilibrium temperature as an independent variable, see also \citep{ElGB2021} for other representations.
In extended irreversible thermodynamics, the nonequilibrium relations are typically first expressed in the entropy representation, where internal energy is treated as an independent variable, \citep{JoCVLe2010}, from which the energy representation used in our formulation above can be deduced. See for example \citep{LeJoCV1980} and \citep{LeJoGr2016} for the use of the energy representation in extended irreversible thermodynamics.
}
\end{remark}

\begin{remark}[Choice of fluxes]\label{choice_flux}{\rm Extended irreversible thermodynamics offers several possible choices for thermodynamic fluxes. The selection of the heat flux $\boldsymbol{q}$ in this work is not exclusive, although it is the most commonly used. Alternative choices, such as the entropy flux $\boldsymbol{j}$ or other conjugate quantities, are also viable, see also Section \ref{sec_relevant_ext}. The question of selecting independent variables in extended irreversible thermodynamics has not yet received a definitive answer, and current discussions provide only partial responses. For example, \citep{LeJoGr2016} discuss the use of the entropy flux in this context.}
\end{remark}

\begin{remark}[Expression of the nonequilibrium stresses]\label{stresses}{\rm {\bf(1)} A key insight of the variational setting is the derivation of the nonequilibrium stresses $\boldsymbol{\tau}_{\boldsymbol{q}}$ and $\boldsymbol{\tau}_{\boldsymbol{\sigma}}$, which enter the momentum equation and are essential for ensuring energetic and thermodynamic consistency (see also Remark \ref{remark_energy}).
{\bf(2)} From a mathematical perspective, their expressions in \eqref{expression_stresses} in terms of $\widehat{\epsilon}$ is reminiscent of the form of conservative stresses that arise in the Eulerian description of continuous media involving tensorial transport equations. A well-known example is the Eulerian formulation of nonlinear elasticity, expressed in terms of the Finger (or left Cauchy-Green) tensor $\boldsymbol{b}^{ij}$, which satisfies the transport equation $\partial_t\boldsymbol{b}+ \pounds_{\boldsymbol{u}}\boldsymbol{b}=0$. 
When the system is described by an internal+elastic energy density $\varepsilon(\rho, s, \boldsymbol{b})$, the corresponding momentum equation takes the form:
\begin{equation}
\rho(\partial_t\boldsymbol{u} + \boldsymbol{u} \cdot \nabla \boldsymbol{u}) = - \nabla p+ \operatorname{div}(\boldsymbol{\tau}_{\boldsymbol{b}}), \qquad \boldsymbol{\tau}_{\boldsymbol{b}}= 2 \frac{\partial\varepsilon}{\partial \boldsymbol{b} }\cdot \boldsymbol{b}
\end{equation}
with the pressure given by $p=   \rho\frac{\partial\varepsilon}{\partial\rho} + s\frac{\partial\varepsilon}{\partial s}-\varepsilon$. One notices the similarity between the mathematical form of $\boldsymbol{\tau}_{\boldsymbol{b}}$ and $\boldsymbol{\tau}_{\boldsymbol{q}}$ in terms of their respective energy density, with the slight difference between them due to the fact that $\boldsymbol{\sigma}$ is a tensorial density, whereas $\boldsymbol{b}$ is an ordinary tensor (if $\rho\boldsymbol{b}$ would be considered instead, then the same mathematical form would be achieved).  A comparable observation holds for $\boldsymbol{\tau}_{\boldsymbol{q}}$ in the case of a conservative continuum involving the tensorial transport of a vector field density. Despite these formal analogies, the physical nature of the stresses is fundamentally different. In the case of nonlinear elasticity, 
$\boldsymbol{\tau}_{\boldsymbol{b}}$ (at least in the equations above) arises from a reversible, conservative process, with 
$\boldsymbol{b}$  evolving according to a continuity-like equation. In contrast, the stresses
$\boldsymbol{\tau}_{\boldsymbol{q}}$ and $\boldsymbol{\tau}_{\boldsymbol{\sigma}}$  in \eqref{expression_stresses} describe extended irreversible processes, and the underlying fields do not satisfy such transport equations. This underscores the distinct physical origins of these stresses, despite their superficial mathematical resemblance. 
{\bf(3)} It is possible to reorganise the terms in the momentum balance in \eqref{NSF_spat_ext} in a slightly different way as
\begin{equation}\label{alternative_momentum}
\partial _t \frac{\partial \widehat{\ell}}{\partial \boldsymbol{u} }  + \pounds _ {\boldsymbol{u}} \frac{\partial \widehat{\ell}}{\partial \boldsymbol{u} }= \rho  \nabla \frac{\partial \widehat{\ell}}{\partial \rho   }+ s \nabla \frac{\partial \widehat{\ell}}{\partial s} +  \boldsymbol{\sigma}:\nabla \frac{\partial \widehat{\ell}}{\partial \boldsymbol{\sigma} }  +  \boldsymbol{q}\cdot \nabla\frac{\partial \widehat{\ell}}{\partial \boldsymbol{q} }  + \operatorname{div} (\boldsymbol{\sigma} +\boldsymbol{\tau}_{\boldsymbol{\sigma }}'+ \boldsymbol{\tau}_{\boldsymbol{q}}')
\end{equation}
with modified stresses
\begin{equation}
\boldsymbol{\tau}_{\boldsymbol{q}}'=  - \frac{\partial \widehat{\ell }}{\partial \boldsymbol{q} } \otimes \boldsymbol{q} \quad\text{and}\quad \boldsymbol{\tau}_{\boldsymbol{\sigma}}'= - 2\frac{\partial \widehat{\ell }}{\partial \boldsymbol{\sigma } } \cdot  \boldsymbol{\sigma}
\end{equation}
to be compared with \eqref{NES}. In equation \eqref{NSF_spat_ext_standard} this corresponds to rewriting the momentum equation as $\rho  ( \partial _t \boldsymbol{u}+ \boldsymbol{u} \cdot \nabla \boldsymbol{u} )  = - \nabla \widehat{p} + \operatorname{div}( \boldsymbol{\tau }_{ \boldsymbol{q}}'+\boldsymbol{\tau }_{ \boldsymbol{ \sigma }}'+ \boldsymbol{\sigma} )  $ with respect to the modified pressure $\widehat{p}= \rho\frac{\partial\varepsilon}{\partial\rho} + s\frac{\partial\varepsilon}{\partial s} + \boldsymbol{q}\cdot \frac{\partial\varepsilon}{\partial \boldsymbol{q}} +\boldsymbol{\sigma}\cdot \frac{\partial\varepsilon}{\partial \boldsymbol{\sigma}} -\varepsilon$ instead of $p=   \rho\frac{\partial\varepsilon}{\partial\rho} + s\frac{\partial\varepsilon}{\partial s}-\varepsilon$.
{\bf(4)} Finally, it is worth noting that in the material description such nonequilibrium stresses do not explicitly appear as additional terms involving $\partial \mathfrak{L}/\partial \boldsymbol{Q}$ and $\partial \mathfrak{L}/\partial \boldsymbol{S}$, see \eqref{NSF_material_ext}. Their effect is already encoded in $\partial \mathfrak{L}/\partial \nabla \boldsymbol{\varphi}$. The same comment does this time apply to the stress $\boldsymbol{\tau}_{\boldsymbol{b}}$.}
\end{remark}

\begin{remark}[Evolution equations for the fluxes]{\rm In line with the philosophy of extended irreversible thermodynamics, the evolution equations for the fluxes $\boldsymbol{q}$ and $\boldsymbol{\sigma}$ are of phenomenological origin. They are constructed to fulfill the second law of thermodynamics within the extended framework. Accordingly, these evolution equations will be addressed later in \S\ref{subsec_phenom} in the context of phenomenological modeling.}
\end{remark}

\subsection{Energy and entropy balance in the spatial description}

While in the material description the energy balance takes the same form in the classical and extended irreversible thermodynamic settings, see \eqref{energy_balance_EIT} and \eqref{energy_balance_CIT}, this is no more true in the spatial description. The total nonequilibrium energy density $\widehat{e}( \boldsymbol{u}, \rho  , s,  \boldsymbol{\sigma} , \boldsymbol{q})$ in the spatial description is defined from the nonequilibrium Lagrangian density by the usual formula
\begin{equation}
\widehat{e}= \frac{\partial \widehat{\ell}}{\partial \boldsymbol{u} } \cdot \boldsymbol{u} -\widehat{\ell}.
\end{equation}
It is then a direct computation to check that along the solution of \eqref{NSF_spat_ext}, the following nonequilibrium energy balance holds
\begin{align}\label{energy_balance_EIT_spat} 
\bar D_t \widehat{e}&= \operatorname{div}\bigg[  \bigg(\Big(   \frac{\partial \widehat{\ell}}{\partial \rho  } \rho +   \frac{\partial \widehat{\ell}}{\partial s  } s+  \frac{\partial \widehat{\ell}}{\partial \boldsymbol{q}} \cdot  \boldsymbol{q} +    \frac{\partial \widehat{\ell}}{\partial \boldsymbol{\sigma }   }: \boldsymbol{\sigma } -\ell\Big)  \delta - \frac{\partial \widehat{\ell}}{\partial \boldsymbol{q}} \otimes \boldsymbol{q} - 2\frac{\partial \widehat{\ell}}{\partial \boldsymbol{\sigma } } \cdot  \boldsymbol{\sigma} + \boldsymbol{\sigma } \bigg) \cdot \boldsymbol{u} + \boldsymbol{j} \frac{\partial \widehat{\ell}}{\partial s} \, \bigg]\nonumber\\
&=\operatorname{div}\bigg[  \bigg(\Big(   \frac{\partial \widehat{\ell}}{\partial \rho  } \rho +   \frac{\partial \widehat{\ell}}{\partial s  } s -\ell\Big)  \delta +\boldsymbol{\tau}_{\boldsymbol{q}}  + \boldsymbol{\tau}_{\boldsymbol{\sigma}} +\boldsymbol{\sigma} \bigg)\cdot \boldsymbol{u}+ \boldsymbol{j} \frac{\partial \widehat{\ell}}{\partial s} \, \bigg].
\end{align}
While we do not carry it in details, the computation reveals how the additional nonequilibrium stresses appearing on the fluid momentum equation, appropriately combine with the tensorial rate of changes $\mathfrak{D}_t$ of the thermodynamic fluxes $ \boldsymbol{\sigma }$ and $ \boldsymbol{q}$, in order to yield a local conservation law.
It is interesting to compare this formula with its classical analog \eqref{energy_cons_spat_CIT}.
The integrated form is identical to \eqref{integrated_energy_spat_CIT}, see also Table 1.

The entropy balance can be written as
\begin{equation}\label{2nd_law_spat_EIT} 
\bar D_t s + \operatorname{div} \boldsymbol{j}  =  \frac{1}{T} \bigg( \boldsymbol{ \sigma } : \nabla \boldsymbol{u}   - \boldsymbol{j}  \cdot \nabla T + \frac{\partial \widehat{\ell}}{\partial \boldsymbol{\sigma} }:\mathfrak{D}_t \boldsymbol{\sigma} +\frac{\partial \widehat{\ell}}{\partial \boldsymbol{q}} \cdot \mathfrak{D}_t \boldsymbol{q}\bigg) \geq 0,
\end{equation} 
giving the integrated form
\begin{equation}
\frac{d}{dt} \int_ \mathcal{D} s\, {\rm d} \mathbf{x} = \int_ \mathcal{D}  \frac{1}{T} \bigg( \boldsymbol{ \sigma } : \nabla \boldsymbol{u}   - \boldsymbol{j}  \cdot \nabla T + \frac{\partial \widehat{\ell}}{\partial \boldsymbol{\sigma} }:\mathfrak{D}_t \boldsymbol{\sigma} +\frac{\partial \widehat{\ell}}{\partial \boldsymbol{q}} \cdot \mathfrak{D}_t \boldsymbol{q}\bigg)  {\rm d} \mathbf{x} - \int_{ \partial \mathcal{D} } \boldsymbol{j} \cdot \boldsymbol{n} \,{\rm d} A .
\end{equation} 
Note that the inequality \eqref{2nd_law_spat_EIT} can be equivalently written $\bar D_t \varsigma \geq 0$, from \eqref{Gamma_Sigma_spat}, thereby attributing to $\bar D_t \varsigma $ (which doesn't coincide with $\bar D_t s$) the meaning of spatial rate of internal entropy production, with
\begin{equation}
\bar D_t \varsigma  =\frac{1}{T} \bigg( \boldsymbol{ \sigma } : \nabla \boldsymbol{u}   - \boldsymbol{j}  \cdot \nabla T + \frac{\partial \widehat{\ell}}{\partial \boldsymbol{\sigma} }:\mathfrak{D}_t \boldsymbol{\sigma} +\frac{\partial \widehat{\ell}}{\partial \boldsymbol{q} } \cdot \mathfrak{D}_t \boldsymbol{q}\bigg)  .
\end{equation} 

\medskip

\noindent\textbf{Balance laws for standard Lagrangians.} The energy and entropy balance laws for the system described by equation \eqref{NSF_spat_ext_standard} take the following form
\begin{equation}
\bar D_t \widehat{e}=\operatorname{div} \big( (-p \delta +\boldsymbol{\tau}_{\boldsymbol{q}}  + \boldsymbol{\tau}_{\boldsymbol{\sigma}} +\boldsymbol{\sigma} )\cdot \boldsymbol{u} - \boldsymbol{j}T \big)
\end{equation} 
\begin{equation}
\bar D_t s + \operatorname{div} \boldsymbol{j}  =  \frac{1}{T} \Big( \boldsymbol{ \sigma } : \nabla \boldsymbol{u}   - \boldsymbol{j}  \cdot \nabla T - \frac{\partial \widehat{\varepsilon}}{\partial \boldsymbol{\sigma} }:\mathfrak{D}_t \boldsymbol{\sigma} -\frac{\partial \widehat{\varepsilon}}{\partial \boldsymbol{q} } \cdot \mathfrak{D}_t \boldsymbol{q}\Big) \geq 0
\end{equation} 
with total energy density given by the standard form
\begin{equation}\label{standard_energy_ext}
\widehat{e}= \frac{1}{2}\rho |\boldsymbol{u}|^2 + \widehat{\varepsilon}(\rho, s, \boldsymbol{\sigma}, \boldsymbol{q}).
\end{equation}

\begin{remark}[More on energy balance]\label{remark_energy}{\rm Rather than applying the most general setting in \eqref{energy_balance_EIT_spat}, it is insightful to analyse the energy balance directly from the special case of system \eqref{NSF_spat_ext_standard}. We focus on the flux $\boldsymbol{q}$ as the treatment of $\boldsymbol{\sigma}$ is similar. Computing the time derivative of the total energy density in \eqref{standard_energy_ext}, we get
\begin{equation}
D_t \widehat{e}= \frac{1}{2}D_t\rho |\boldsymbol{u}|^2 + \rho \boldsymbol{u} \cdot D_t \boldsymbol{u} + \frac{\partial\varepsilon}{\partial\rho} D_t\rho + \frac{\partial\varepsilon}{\partial s} D_ts+ \frac{\partial\varepsilon}{\partial \boldsymbol{q}} \cdot D_t\boldsymbol{q} 
\end{equation}
Using the momentum, mass, and entropy equations, we get
\begin{equation}
D_t \widehat{e}= -\widehat{e} \operatorname{div}\boldsymbol{u} + \operatorname{div}(-p\boldsymbol{u}  - T\boldsymbol{j})+  \boldsymbol{u} \cdot \operatorname{div}(\boldsymbol{\tau}_{\boldsymbol{q}} ) - \frac{\partial \varepsilon }{\partial \boldsymbol{q}}\cdot (\mathfrak{D}_t \boldsymbol{q} - D_t \boldsymbol{q}).
\end{equation}
This illustrates how the nonequilibrium stress  $\boldsymbol{\tau}_{\boldsymbol{q}} $ is crucial in compensating for the discrepancy between $\mathfrak{D}_t \boldsymbol{q}$ and  $D_t \boldsymbol{q}$, ensuring that the last two terms combine to yield the divergence of $\boldsymbol{\tau}_{\boldsymbol{q}}\cdot \boldsymbol{u}$ when the first formula in  \eqref{expression_stresses} is applied. In particular, if $D_t\boldsymbol{q}$ were used in the entropy equation instead of $\mathfrak{D}_t \boldsymbol{q}$, the nonequilibrium stress would not be required. However, such system of equations would lack both a material formulation and the desired covariance properties.
}
\end{remark}

\subsection{Phenomenology: Maxwell and Cattaneo-Christov laws}\label{subsec_phenom}

We briefly show here how thermodynamic consistency is achieved in extended irreversible thermodynamics, see \citep{JoCVLe2010}, leading to the evolution equations of Maxwell type for the viscous and heat fluxes. A key feature of our framework is that it directly provides the appropriate objective rate of change to be used in these flux equations. This section will help the reader make a concrete connection between the systems discussed so far and the standard expressions encountered in extended irreversible thermodynamics.

We focus on the standard Lagrangian given in \eqref{ell_ext}, but the approach can handle general Lagrangians. It is generally assumed that the nonequilibrium internal energy density satisfies
\begin{equation}\label{assumptions_ne}
\frac{\partial \widehat{\varepsilon}}{\partial \boldsymbol{q}}= \alpha \boldsymbol{q}, \qquad \frac{\partial \widehat{\varepsilon}}{\partial \boldsymbol{\sigma}}= \beta \boldsymbol{\sigma}
\end{equation}
for some state functions $\alpha$ and $\beta$. We note that a Riemannian metric (here the Euclidean one) is needed to write \eqref{assumptions_ne}. For now we consider that the entropy flux $\boldsymbol{j}$ is given by $\boldsymbol{j}= \boldsymbol{q}/T$, see \S\ref{sec_relevant_ext} for generalizations. In this case, thermodynamic consistency \eqref{2nd_law_spat_EIT} requires
\begin{equation}
\boldsymbol{ \sigma } : \big(\operatorname{Def} \boldsymbol{u}   -\beta \mathfrak{D}_t \boldsymbol{\sigma}\big) +\boldsymbol{q}\cdot \big(-  \frac{1}{T}\nabla T - \alpha  \mathfrak{D}_t \boldsymbol{q}\big)\geq 0,
\end{equation}
which is most naturally achieved by assuming the following flux-force relations
\begin{equation}\label{flux_force}
\boldsymbol{ \sigma } = \mathbb{L}\big(\operatorname{Def} \boldsymbol{u}   -\beta \mathfrak{D}_t \boldsymbol{\sigma}\big),\qquad \boldsymbol{q}= \mathbb{K} \big(-  \frac{1}{T}\nabla T - \alpha  \mathfrak{D}_t \boldsymbol{q} \big),
\end{equation}
for some positive tensor $\mathbb{L}$ and $\mathbb{K}$. 

\medskip

\noindent\textbf{Viscosity.} Comparison with the Newton-Stokes law yields from the first relation in \eqref{flux_force} and particular choice of $\mathbb{L}$ an extended \textit{compressible Maxwell model with Truesdell rate}
\begin{equation}\label{Maxwell_Truesdell}
\left\{
\begin{array}{l}
\displaystyle\vspace{0.2cm}\tau_2 (\mathfrak{D}_t\boldsymbol{\sigma})^{(0)}  + \boldsymbol{\sigma}^{(0)} = 2\eta (\operatorname{Def} \boldsymbol{u})^{(0)}\\
\displaystyle\tau_0 \frac{1}{3}\operatorname{Tr}\mathfrak{D}_t\boldsymbol{\sigma} + \frac{1}{3}  \operatorname{Tr}\boldsymbol{\sigma} = \zeta \operatorname{div}\boldsymbol{u},
\end{array}
\right.
\end{equation}
where $\eta$ and $\zeta$ denote the shear and bulk viscosity coefficients, respectively, and the relaxation coefficients are given by $\tau_2= 2 \eta \beta$, $\tau_0= 3\zeta \beta$. Here the superscript ${(0)}$ denotes the traceless part of the tensors. It is important to note that the time rate $\mathfrak{D}_t$ does not preserve the isotropic and traceless decomposition of symmetric tensor fields. As a result, system \eqref{Maxwell_Truesdell} exhibits coupling between the volumetric and deviatoric stresses. In more details, introducing the volumetric pressure component $p^v= \frac{1}{3} \operatorname{Tr}\boldsymbol{\sigma} $, equation \eqref{Maxwell_Truesdell} can be rewritten as
\begin{equation}\label{decomposed}
\left\{
\begin{array}{l}
\displaystyle\vspace{0.2cm}\tau_2 \left( \mathfrak{D}_t\boldsymbol{\sigma}^{(0)}  + \frac{2}{3} \boldsymbol{\sigma}^{(0)}: (\operatorname{Def} \boldsymbol{u})^{(0)}\delta - 2 p^v(\operatorname{Def} \boldsymbol{u})^{(0)} \right)+ \boldsymbol{\sigma}^{(0)} = 2\eta (\operatorname{Def} \boldsymbol{u})^{(0)}\\
\displaystyle\tau_0\left(  D_t p_v  + \frac{1}{3} p^v \operatorname{div} \boldsymbol{u} - \frac{2}{3}\boldsymbol{\sigma}^{(0)}: (\operatorname{Def} \boldsymbol{u})^{(0)} \right) + p_v = \zeta \operatorname{div}\boldsymbol{u}.
\end{array}
\right.
\end{equation}
Note that a particular case of \eqref{decomposed} arises when $\zeta= \frac{2\eta}{3}$, leading $\tau_2=\tau_1$, in which case the system simplifies to:
\begin{equation}\label{UCM}
\tau_2 \mathfrak{D}_t\boldsymbol{\sigma} + \boldsymbol{\sigma}= 2\eta \operatorname{Def} \boldsymbol{u}.
\end{equation}
This is the upper convected Maxwell model for tensorial densities, which involves the Truesdell rate.
Note that in our framework, the type of objective rate to be used naturally arises from the tensorial nature of the stress flux. This flux, in turn, emerges from the geometric setting underlying the Lagrangian-to-Eulerian reduction approach described in \S\ref{subsec_3_3}, see especially formula \eqref{F1}. This choice is a consequence of selecting the second Piola-Kirchhoff stress tensor $\boldsymbol{S}$  as the preferred measure of stress in the material description.

The Maxwell model with Truesdell rate in \eqref{UCM} is commonly used in compressible viscoelastic fluids, see, e.g.,   \citep{BeEd1994}, \citep{BoPh2012}. However, its energetic and thermodynamic role in those contexts differs from its role here, in the framework of extended irreversible thermodynamics.
Stress transport governed by the Truesdell rate also appears in the hydrodynamic theory of an electron gas, \citep{ToPa1999}.

\medskip

\noindent\textbf{Heat conduction.} Comparison of the second relation in \eqref{flux_force} with Fourier's law yields $\mathbb{K}= \kappa T\delta $, with $\kappa$ the thermal conductivity, so that this relation reduces to the \textit{Cattaneo-Christov heat flux model}:
\begin{equation}
\tau_1\mathfrak{D}_t\boldsymbol{q} + \boldsymbol{q}= - \kappa\nabla T,
\end{equation}
with the relaxation coefficient $\tau_1= \kappa T\alpha$. In this model, see \citep{Ch2009}, the time derivative in the
Maxwell-Cattaneo heat flux model $\tau_1\partial_t\boldsymbol{q} +  \boldsymbol{q}= - \kappa\nabla T$ is replaced by the Lie derivative of a vector field density, motivated by the requirement that this derivative be objective. In our setting, this model naturally emerges from the tensorial nature of the heat flux and from the Lagrangian-to-Eulerian reduction approach described in \S\ref{subsec_3_3}, see especially formula \eqref{F2}.

Previous couplings of compressible fluid dynamics with the Cattaneo-Christov heat flux model has been investigated in \citep{St2010}, \citep{AnMaPl2020}, \citep{An2022}, \citep{CBKaXu2024}, among others. In these works, the system comprises the standard balance equations for energy and momentum, with Fourier's law for heat conduction replaced by the Cattaneo-Christov model \eqref{flux_force}. However, this system differs from the one considered here, as it does not enforce thermodynamic consistency in the sense of extended irreversible thermodynamics. In particular, it deviates from our system \eqref{NSF_spat_ext_standard} both in the fluid momentum equation and the entropy evolution equation.

Extensions of classical heat transfer equations also arise from frameworks other than extended irreversible thermodynamics, such as the Green-Naghdi theories of nonclassical heat conduction based on the concept of thermal displacement \citep{GrNa1991}.
More recently, a variational formulation in the context of hyperbolic heat equation was considered in \citep{DhGa2024}, also relying on the thermal displacement. In that context, the evolution equation governs the gradient of the thermal displacement, which, unlike the evolution equation for the heat flux $\boldsymbol{q}$ considered in the extended irreversible thermodynamics framework, does not give rise to entropy production.
Cattaneo-type heat conduction, although of a different nature than the formulations discussed here, also emerges from a Godunov-type structure, as shown in \citep{PePaRoGr2018}.

\begin{remark}[Causality]{\rm It will be important to analyze under which conditions the extended heat-conducting viscous fluid model developed in this paper - namely equations \eqref{NSF_spat_ext_standard} and \eqref{expression_stresses}, together with the phenomenological expressions derived in \S\ref{subsec_phenom} - is hyperbolic, as this would ensure the causal character of the system.
Unlike in the simpler setting of pure heat conduction, where a damped Fourier law directly yields a hyperbolic temperature equation, the general case of heat-conducting fluids considered here involves a strong coupling between the temperature and heat flux evolution equations. This coupling originates from the structure of the entropy balance equation in extended irreversible thermodynamics (see, for instance, the second equation in \eqref{NSF_spat_ext_standard}), which - when rewritten in terms of temperature - retains significant dependence on both the heat flux and its time derivative. Although the resulting temperature equation still resembles a hyperbolic equation, this entanglement with the flux dynamics complicates a straightforward interpretation of causality. The causal or non-causal character of the system may also depend on the specific form of the nonequilibrium internal energy. The closest work in the literature appears to be \citep{An2022}, which, however, demonstrates the non-hyperbolic nature of heat-conducting non-viscous fluid models that couple the standard mass, momentum, and energy balance equations with the Cattaneo-Christov fluid model. As discussed above, that model differs from ours and lacks thermodynamic consistency. We postpone the detailed analysis of hyperbolicity and causality of our model for future work. Alternatively, a numerical investigation of this property could be undertaken by leveraging the variational formulation developed here to construct thermodynamically consistent numerical schemes, as was done in \citep{GaGB2024} for the classical irreversible thermodynamics of heat-conducting fluids.}
\end{remark}

\subsection{Key enhancements}\label{sec_relevant_ext}

\subsubsection{Entropy flux beyond local equilibrium}

The classical local equilibrium relation 
$\boldsymbol{j}=\boldsymbol{q}/T$ for the entropy flux 
$\boldsymbol{j}$ does not necessarily hold in extended irreversible thermodynamics, as the heat flux 
$\boldsymbol{q}$ evolves dynamically and is no longer an instantaneous function of the temperature gradient. Consequently, this relation may be insufficient to fully describe entropy transport, a fact that is supported by kinetic theory. We demonstrate that our variational framework can be naturally extended to account for this more general setting.

\medskip

\noindent\textbf{Variational principle and evolution equations.} Since only the flux of entropy is changed, it suffices to add in our variational principle an internal entropy source in the form of a flux, denoted as $\mathcal{J}= \operatorname{div}\boldsymbol{J}'$, where $\boldsymbol{J}'$ represents the part of the entropy flux not accounted for in the heat conduction process.

In the Lagrangian description, this modification is achieved by preserving the variational condition \eqref{VCond_continuum_ext} while including the new effect in the phenomenological and variational constraints \eqref{PC_continuum_ext} and \eqref{VC_continuum_ext}. The \textit{phenomenological constraint} \eqref{PC_continuum_ext} is modified to
\begin{equation}\label{PC_continuum_ext_Jsmod}
\frac{\partial \widehat{\mathfrak{L}}}{\partial S}\dot \Sigma+\frac{\partial \widehat{\mathfrak{L}}}{\partial \boldsymbol{S} }:\dot{\boldsymbol{S}} +\frac{\partial \widehat{\mathfrak{L}}}{\partial \boldsymbol{Q} }\cdot  \dot{\boldsymbol{Q}}= - \boldsymbol{S} : \boldsymbol{D} (\dot{\boldsymbol{\varphi}} )+ \boldsymbol{J}  \cdot  \nabla  \dot \Gamma  - \mathcal{J} \dot  \Gamma 
\end{equation}
which leads to the corresponding  \textit{variational constraint}
\begin{equation}\label{VC_continuum_ext_Jsmod}
\frac{\partial \widehat{\mathfrak{L}}}{\partial S} \delta  \Sigma +\frac{\partial \widehat{\mathfrak{L}}}{\partial \boldsymbol{S} }:\delta \boldsymbol{S} +\frac{\partial \widehat{\mathfrak{L}}}{\partial \boldsymbol{Q} }\cdot  \delta  \boldsymbol{Q}= - \boldsymbol{S} : \boldsymbol{D} (\delta{\boldsymbol{\varphi} } )+ \boldsymbol{J}  \cdot\nabla  \delta  \Gamma - \mathcal{J} \delta \Gamma  .
\end{equation}
For $\mathcal{J}=0$, the previous case is recovered. Instead of \eqref{NSF_material_ext}, the system now takes the form
\begin{equation}\label{NSF_material_ext_ext} 
\hspace{-0.3cm}\left\{\!\!
\begin{array}{l}
\vspace{0.2cm}\displaystyle\frac{d}{dt}\frac{ \partial \widehat{\mathfrak{L}} }{ \partial   \dot{\boldsymbol{\varphi}} }   - \frac{ \partial \widehat{\mathfrak{L}}}{\partial \boldsymbol{\varphi}  }= \operatorname{DIV}\Big( - \frac{\partial\widehat{\mathfrak{L}}}{\partial \nabla{\boldsymbol{\varphi} } } +  \boldsymbol{P}  \Big) \\
\displaystyle
-\frac{\partial \widehat{\mathfrak{L}}}{ \partial    S}(\dot S + \operatorname{DIV} (\boldsymbol{J} + \boldsymbol{J}') )- \frac{\partial \widehat{\mathfrak{L}}}{\partial \boldsymbol{S} }:\dot{\boldsymbol{S}} -\frac{\partial \widehat{\mathfrak{L}}}{\partial \boldsymbol{Q} } \cdot \dot{ \boldsymbol{Q}}  =  \boldsymbol{S} : \boldsymbol{D} (\dot{\boldsymbol{\varphi}} ) + \boldsymbol{J}  \cdot \nabla \frac{\partial  \widehat{\mathfrak{L}} }{\partial    S}- \frac{\partial \widehat{\mathfrak{L}}}{ \partial    S} \operatorname{DIV} \boldsymbol{J}',\hspace{-0.3cm}
\end{array}
\right.
\end{equation}
along with the conditions
\begin{equation}\label{Gamma_Sigma_ext_ext} 
\dot \Gamma = - \frac{\partial  \widehat{\mathfrak{L}} }{ \partial   S} \qquad\text{and}\qquad \dot \Sigma = \dot S + \operatorname{DIV} (\boldsymbol{J} + \boldsymbol{J}').
\end{equation} 
The key effect is the emergence of the modified entropy flux $\boldsymbol{J}_s= \boldsymbol{J} + \boldsymbol{J}'$ in the expression of the internal entropy production rate $\dot \Sigma$.

Consequently, in the Eulerian variational formulation, the variational condition \eqref{VCond_continuum_spat_ext} remains unchanged, but the \textit{phenomenological constraint} \eqref{PC_continuum_spat_ext} is adjusted to
\begin{equation}\label{PC_continuum_spat_ext_ext}
\frac{\partial \widehat{\ell}}{\partial s}\bar D_t \varsigma+\frac{\partial \widehat{\ell}}{\partial \boldsymbol{\sigma} }:\mathfrak{D}_t \boldsymbol{\sigma} +\frac{\partial \widehat{\ell}}{\partial \boldsymbol{q} } \cdot \mathfrak{D}_t \boldsymbol{q}   = - \boldsymbol{ \sigma } : \nabla \boldsymbol{u }  + \boldsymbol{j}  \cdot  \nabla  D_t \gamma - j D_t\gamma,
\end{equation}
with corresponding \textit{variational constraint} given as
\begin{equation}\label{VC_continuum_spat_ext_ext}
\frac{\partial \widehat{\ell}}{\partial s}\bar D_ \delta  \varsigma+\frac{\partial \widehat{\ell}}{\partial \boldsymbol{\sigma} }:\mathfrak{D}_\delta \boldsymbol{\sigma} +\frac{\partial \widehat{\ell}}{\partial \boldsymbol{q} } \cdot \mathfrak{D}_\delta \boldsymbol{q}  = - \boldsymbol{ \sigma } : \nabla \boldsymbol{\zeta }  + \boldsymbol{j}  \cdot  \nabla  D_\delta  \gamma - j D_\delta \gamma
\end{equation}
in place of \eqref{VC_continuum_spat_ext}.
Here, $j=\operatorname{div}\boldsymbol{j}'$ is the Eulerian counterpart of $\mathcal{J}= \operatorname{DIV}\boldsymbol{J}'$, where $\boldsymbol{j}'$ represents the additional entropy flux contribution. The resulting system is
\begin{equation}\label{NSF_spat_ext_ext} 
\hspace{-0.5cm}\left\{
\begin{array}{l}
\vspace{0.2cm}\displaystyle \partial _t \frac{\partial \widehat{\ell}}{\partial \boldsymbol{u} }  + \pounds _ {\boldsymbol{u}} \frac{\partial \widehat{\ell}}{\partial \boldsymbol{u} }= \rho  \nabla \frac{\partial \widehat{\ell}}{\partial \rho   }+ s \nabla \frac{\partial \widehat{\ell}}{\partial s} - \frac{\partial \widehat{\ell}}{\partial \boldsymbol{\sigma} } : \nabla \boldsymbol{\sigma} - \frac{\partial \widehat{\ell}}{\partial \boldsymbol{q} } \cdot  \nabla \boldsymbol{q} + \operatorname{div} (\boldsymbol{\sigma} +\boldsymbol{\tau}_{\boldsymbol{\sigma }}+ \boldsymbol{\tau}_{\boldsymbol{q}})\\
\vspace{0.2cm}\displaystyle
-\frac{\partial  \widehat{\ell} }{ \partial  s}(\bar D_t s + \operatorname{div}( \boldsymbol{j}+\boldsymbol{j}')  ) = \boldsymbol{ \sigma } : \nabla \boldsymbol{u}   + \boldsymbol{j}  \cdot \nabla \frac{\partial   \widehat{\ell} }{\partial   s}+\frac{\partial \widehat{\ell}}{\partial \boldsymbol{\sigma} }:\mathfrak{D}_t \boldsymbol{\sigma} +\frac{\partial \widehat{\ell}}{\partial \boldsymbol{q} } \cdot \mathfrak{D}_t \boldsymbol{q} -\frac{\partial  \widehat{\ell} }{ \partial  s} \operatorname{div}\boldsymbol{j}'   \\
\displaystyle\bar D_t \rho  =0,\hspace{-0.3cm}
\end{array}
\right.
\end{equation}
with internal entropy production rate
\begin{equation}\label{varsigma_ext_ext}
D_t \varsigma = \bar D_t s + \operatorname{div}( \boldsymbol{j}+ \boldsymbol{j}').
\end{equation}

This system and the expression for  $D_t \varsigma$ should be compared with \eqref{NSF_spat_ext} and the first relation in \eqref{interm2}, respectively. The key distinction from the earlier case lies in relation \eqref{varsigma_ext_ext}, which governs internal entropy production -- after all, the terms involving $\boldsymbol{j}'$ simplify in the second equation of \eqref{NSF_spat_ext_ext}. Ultimately, the difference between \eqref{NSF_spat_ext} and \eqref{NSF_spat_ext_ext} comes down to which quantity is interpreted as the internal entropy production $\bar D_t\varsigma\geq 0$, as this differs in both cases.

For the Lagrangian density of the standard form \eqref{ell_ext}, one gets from \eqref{NSF_spat_ext_ext} the system
\begin{equation}\label{NSF_ext_ext}
\left\{
\begin{array}{l}
\vspace{0.2cm}\displaystyle \rho  ( \partial _t \boldsymbol{u}+ \boldsymbol{u} \cdot \nabla \boldsymbol{u} )  = - \nabla p + \operatorname{div}( \boldsymbol{\tau }_{ \boldsymbol{q}}+\boldsymbol{\tau }_{ \boldsymbol{ \sigma }}+ \boldsymbol{\sigma} )   \\
\vspace{0.2cm}\displaystyle
T(\bar D_t s + \operatorname{div} (\boldsymbol{j} +\boldsymbol{j}' ) = \boldsymbol{ \sigma } : \nabla \boldsymbol{u}   - \boldsymbol{j}  \cdot \nabla T-\frac{\partial \widehat{\varepsilon }}{\partial \boldsymbol{\sigma} }:\mathfrak{D}_t \boldsymbol{\sigma} -\frac{\partial \widehat{\varepsilon }}{\partial \boldsymbol{q} } \cdot \mathfrak{D}_t \boldsymbol{q} + T \operatorname{div}\boldsymbol{j}' \\
\displaystyle\bar D_t \rho  =0.
\end{array}
\right.
\end{equation}

\medskip

\noindent\textbf{Balance laws and thermodynamic consistency.} We note that this modification doesn't affect the energy balance derived earlier, as $\boldsymbol{j}'$ simply does not appear in it.
The entropy balance \eqref{2nd_law_spat_EIT} is however changed into 
\begin{equation}\label{2nd_law_spat_EIT_ext} 
\bar D_t s + \operatorname{div} \boldsymbol{j} _s =  \frac{1}{T} \bigg( \boldsymbol{ \sigma } : \nabla \boldsymbol{u}   - \boldsymbol{j}  \cdot \nabla T + \frac{\partial \widehat{\ell}}{\partial \boldsymbol{\sigma} }:\mathfrak{D}_t \boldsymbol{\sigma} +\frac{\partial \widehat{\ell}}{\partial \boldsymbol{q}} \cdot \mathfrak{D}_t \boldsymbol{q}-\frac{\partial  \widehat{\ell} }{ \partial  s} \operatorname{div}\boldsymbol{j}'  \bigg) \geq 0,
\end{equation} 
giving the integrated form
\begin{equation}
\frac{d}{dt} \int_ \mathcal{D} \!s\, {\rm d} \mathbf{x} = \int_ \mathcal{D} \! \frac{1}{T} \bigg( \boldsymbol{ \sigma } : \nabla \boldsymbol{u}   - \boldsymbol{j}  \cdot \nabla T + \frac{\partial \widehat{\ell}}{\partial \boldsymbol{\sigma} }:\mathfrak{D}_t \boldsymbol{\sigma} +\frac{\partial \widehat{\ell}}{\partial \boldsymbol{q}} \cdot \mathfrak{D}_t \boldsymbol{q}-\frac{\partial  \widehat{\ell} }{ \partial  s} \operatorname{div}\boldsymbol{j}' \bigg)  {\rm d} \mathbf{x} - \int_{ \partial \mathcal{D} }\! \boldsymbol{j}_s \cdot \boldsymbol{n} \,{\rm d} A .
\end{equation}
where $\boldsymbol{j}_s= \boldsymbol{j}+\boldsymbol{j'}$.

For completeness, we briefly sketch how thermodynamic consistency is achieved in this more general setting. We focus on the standard Lagrangian given in equation \eqref{ell_ext} and assume conditions \eqref{assumptions_ne}, as before. The entropy flux can now incorporate higher-order dependencies on $\boldsymbol{q}$ and $\boldsymbol{\sigma}$, such as $\boldsymbol{j}_s= \boldsymbol{q}/T + \boldsymbol{\sigma}\cdot \mathbb{M}\cdot \boldsymbol{q}$, where $\mathbb{M}_{ijk}^\ell$ is a tensor bilinearly converting the contravariant tensor densities $\boldsymbol{\sigma}$ and $\boldsymbol{q}$ into a vector field density. In the isotropic case, we have $\boldsymbol{j}_s= \boldsymbol{q}/T + \gamma' p^v \boldsymbol{q} + \gamma'' \boldsymbol{\sigma}^{(0)}\cdot \boldsymbol{q}$, \citep{JoCVLe2010}. By explicitly writing equation \eqref{2nd_law_spat_EIT_ext} in that case, thermodynamic consistency now requires
\begin{equation}
\boldsymbol{ \sigma } : \big(\operatorname{Def} \boldsymbol{u}   -\beta \mathfrak{D}_t \boldsymbol{\sigma} + \mathbb{M}\cdot \nabla \boldsymbol{q}\big) +\boldsymbol{q}\cdot \big(-  \frac{1}{T}\nabla T - \alpha  \mathfrak{D}_t \boldsymbol{q} +\nabla\boldsymbol{\sigma}\cdot  \mathbb{M}\big)\geq 0.
\end{equation}
It is achieved by assuming linear flux-force relations
\begin{equation}
\boldsymbol{\sigma} = \mathbb{L} \big(\operatorname{Def} \boldsymbol{u}   -\beta \mathfrak{D}_t \boldsymbol{\sigma} + \mathbb{M}\cdot \nabla \boldsymbol{q}\big),\qquad \boldsymbol{q}= \mathbb{K} \big(-  \frac{1}{T}\nabla T - \alpha  \mathfrak{D}_t \boldsymbol{q} +\nabla\boldsymbol{\sigma}\cdot  \mathbb{M}\big)
\end{equation}
naturally extending the Maxwell type evolution equations in \eqref{flux_force}. 
Note that the tensorial contractions involved are $(\mathbb{M}\cdot \nabla \boldsymbol{q})_{ij}=\mathbb{M}_{ijk}^\ell \partial_\ell \boldsymbol{q}^k$ and $(\nabla\boldsymbol{\sigma}\cdot  \mathbb{M})^k=\partial_\ell\boldsymbol{\sigma}^{ij}\mathbb{M}_{ijk}^\ell$ with $\mathbb{M}$ assumed constant. The isotropic case is readily carried out and yields the desired relations in extended irreversible thermodynamics, while incorporating objective rates.

\subsubsection{Higher-order fluxes}

To account for the complexity of some fast phenomena, it is necessary to introduce
additional variables beyond the thermodynamic fluxes, such as higher-order fluxes (\citep{DeCVJo1996}). A representative example is the Maxwell-Cattaneo model for the heat flux $\boldsymbol{q}=\boldsymbol{q}^{(1)}$ enriched to includes the divergence of an extra flux $\boldsymbol{q}^{(2)}$, leading to the equation $\tau_1\partial_t\boldsymbol{q}^{(1)} +  \boldsymbol{q}^{(1)} = - \kappa\nabla T + \operatorname{div} \boldsymbol{q}^{(2)}$. 
To simplify the exposition, we do not include the viscosity process here. However, higher-order fluxes for viscosity can be incorporated in a similar manner.

\medskip

\noindent\textbf{Variational principle and evolution equations.}  In this case, the nonequilibrium Lagrangian density in the material description takes the form $\mathfrak{L}( \boldsymbol{\varphi}, \dot{\boldsymbol{\varphi}}, \nabla \boldsymbol{\varphi}, \varrho , S, \boldsymbol{Q}^{(1)},..., \boldsymbol{Q}^{(n)})$, where $\boldsymbol{Q}^{(k)}$ is a $k$-times contravariant tensor field density. The higher-order fluxes are naturally incorporated in the \textit{phenomenological constraint} \eqref{PC_continuum_ext_Jsmod} as
\begin{equation}\label{PC_continuum_ext_higher}
\frac{\partial \widehat{\mathfrak{L}}}{\partial S}\dot \Sigma +\sum_{k=1}^n\frac{\partial \widehat{\mathfrak{L}}}{\partial \boldsymbol{Q} ^{(k)}}:  \dot{\boldsymbol{Q}}^{(k)}=  \boldsymbol{J}  \cdot  \nabla  \dot \Gamma  - \mathcal{J} \dot  \Gamma 
\end{equation}
which leads to the corresponding  \textit{variational constraint}
\begin{equation}\label{VC_continuum_ext_higher}
\frac{\partial \widehat{\mathfrak{L}}}{\partial S} \delta  \Sigma  +\sum_{k=1}^n\frac{\partial \widehat{\mathfrak{L}}}{\partial \boldsymbol{Q} ^{(k)}}:  \delta  \boldsymbol{Q}^{(k)}=  \boldsymbol{J}  \cdot\nabla  \delta  \Gamma - \mathcal{J} \delta \Gamma  .
\end{equation}
From this, we obtain the system \eqref{NSF_material_ext_ext} with the expected higher-order modifications, as the reader can readily verify. Recall that we do not explicitly include the treatment of 
$\boldsymbol{S}$ here to avoid unnecessarily complicating the notation. We refer to Table 1 for the energy and entropy balance equations.

Similarly, in the Eulerian description, the variational condition \eqref{VCond_continuum_spat_ext} remains unchanged, except that it now involves a nonequilibrium Lagrangian density $\widehat{\ell}(\boldsymbol{u}, \rho, s,  \boldsymbol{q}^{(1)},..., \boldsymbol{q}^{(n)})$, which depends on the Eulerian higher-order fluxes $\boldsymbol{q}^{(k)}=\boldsymbol{\varphi}_*\boldsymbol{Q}^{(k)} J_{\boldsymbol{\varphi}^{-1}}$. Naturally, this requires the Lagrangian density $\mathfrak{L}$ to be material covariant, see \eqref{mat_cov_ext}.  The \textit{phenomenological constraint} \eqref{PC_continuum_spat_ext_ext} is modified to
\begin{equation}\label{PC_continuum_spat_ext_higher}
\frac{\partial \widehat{\ell}}{\partial s}\bar D_t \varsigma +\sum_{k=1}^n\frac{\partial \widehat{\ell}}{\partial \boldsymbol{q}^{(k)} } : \mathfrak{D}_t \boldsymbol{q}   ^{(k)}=  \boldsymbol{j}  \cdot  \nabla  D_t \gamma - j D_t\gamma,
\end{equation}
with the corresponding \textit{variational constraint}
\begin{equation}\label{VC_continuum_spat_ext_higher}
\frac{\partial \widehat{\ell}}{\partial s}\bar D_ \delta  \varsigma+\sum_{k=1}^n\frac{\partial \widehat{\ell}}{\partial \boldsymbol{q}^{(k)} } : \mathfrak{D}_\delta \boldsymbol{q}   ^{(k)} = \boldsymbol{j}  \cdot  \nabla  D_\delta  \gamma - j D_\delta \gamma
\end{equation}
which replaces \eqref{VC_continuum_spat_ext_ext}. They incorporate the rates $\mathfrak{D}_t$ and variations $\mathfrak{D}_\delta$ associated with the type of tensorial density for $k=1,...,n$, and follow from \eqref{PC_continuum_ext_higher} and \eqref{VC_continuum_ext_higher} via the Lagrange-to-Euler transformation, as before. The resulting system is
\begin{equation}\label{NSF_spat_ext_higher} 
\hspace{-0.5cm}\left\{
\begin{array}{l}
\vspace{0.2cm}\displaystyle \partial _t \frac{\partial \widehat{\ell}}{\partial \boldsymbol{u} }  + \pounds _ {\boldsymbol{u}} \frac{\partial \widehat{\ell}}{\partial \boldsymbol{u} }= \rho  \nabla \frac{\partial \widehat{\ell}}{\partial \rho   }+ s \nabla \frac{\partial \widehat{\ell}}{\partial s} -\sum_{k=1}^n \frac{\partial \widehat{\ell}}{\partial \boldsymbol{q}^{(k)} } :  \nabla \boldsymbol{q}^{(k)} + \operatorname{div} \Big(\sum_{k=1}^n\boldsymbol{\tau}_{\boldsymbol{q}^{(k)}}\Big)\\
\vspace{0.2cm}\displaystyle
-\frac{\partial  \widehat{\ell} }{ \partial  s}(\bar D_t s + \operatorname{div}( \boldsymbol{j}+\boldsymbol{j}')  ) =   \boldsymbol{j}  \cdot \nabla \frac{\partial   \widehat{\ell} }{\partial   s}+\sum_{k=1}^n\frac{\partial \widehat{\ell}}{\partial \boldsymbol{q}^{(k)} } : \mathfrak{D}_t \boldsymbol{q}^{(k)} -\frac{\partial  \widehat{\ell} }{ \partial  s} \operatorname{div}\boldsymbol{j}'   \\
\displaystyle\bar D_t \rho  =0,\hspace{-0.3cm}
\end{array}
\right.
\end{equation}
where the higher-order thermal nonequilibrium stresses are given by
\begin{equation}\label{higher_NES}
\boldsymbol{\tau}_{\boldsymbol{q}^{(k)}}=\Big(\frac{\partial\widehat{\ell}}{\partial \boldsymbol{q}^{(k)}}:\boldsymbol{q}^{(k)}\Big) \delta+\frac{\partial\widehat{\ell}}{\partial \boldsymbol{q}^{(k)}}\therefore\widehat{\boldsymbol{q}}^{(k)},\quad k=1,...,n.
\end{equation}
Here $\widehat{\boldsymbol{q}}$ is the tensor density defined as $\widehat{\boldsymbol{q}}^{i_1...i_k}{}^i_j:=\sum_r\boldsymbol{q}^{i_1...i_{r-1} i i_{r+1}...i_n}\delta^{i_r}_j$ and  $\therefore$ denotes the contraction $\big(\frac{\partial\widehat{\ell}}{\partial \boldsymbol{q}}\therefore\widehat{\boldsymbol{q}}\big)^i_j= \frac{\partial\widehat{\ell}}{\partial \boldsymbol{q}^{i_1...i_k}}\widehat{\boldsymbol{q}}^{i_1...i_k}{}^i_j$. We leave the details to the reader, see \citep{GB2024} for similar formulas. For the standard Lagrangian \eqref{ell_ext} with nonequillibrium internal energy $\widehat{\varepsilon}(\rho, s, \boldsymbol{q}^{(1)},...,\boldsymbol{q}^{(n)})$, one obtains a natural extension of \eqref{NSF_ext_ext} which includes the sum over higher-order fluxes.

\medskip

\noindent\textbf{Balance laws and thermodynamic consistency.} The energy balance is the generalization of \eqref{energy_balance_EIT_spat} to higher-order fluxes, while the entropy balance extends \eqref{2nd_law_spat_EIT_ext} as follows:
\begin{equation}\label{2nd_law_spat_EIT_higher}
\bar D_t s + \operatorname{div} \boldsymbol{j} _s =  \frac{1}{T} \bigg(  - \boldsymbol{j}  \cdot \nabla T  +\sum_{k=1}^n\frac{\partial \widehat{\ell}}{\partial \boldsymbol{q}^{(k)}} : \mathfrak{D}_t \boldsymbol{q}^{(k)}-\frac{\partial  \widehat{\ell} }{ \partial  s} \operatorname{div}\boldsymbol{j}'  \bigg) \geq 0,
\end{equation}
where $\boldsymbol{j}_s= \boldsymbol{j}+ \boldsymbol{j}'$, and we assume $\boldsymbol{\sigma}$ absent.

As before, we briefly outline how thermodynamic consistency is achieved in the presence of higher-order fluxes. We focus on the standard Lagrangian given in \eqref{ell_ext} and assume, analogously to \eqref{assumptions_ne}, that
\begin{equation}
\frac{\partial \widehat{\varepsilon}}{\partial \boldsymbol{q}^{(k)}}= \alpha_k \boldsymbol{q}^{(k)},\qquad k=1,...,n.
\end{equation}
The entropy flux can now incorporate higher-order fluxes as follows:
\begin{equation}\label{j_s_higher}
\boldsymbol{j}_s=  \frac{1}{T}\boldsymbol{q}^{(1)} + \sum_{k=1}^{n-1} \gamma_k\boldsymbol{q}^{(k+1)} \cdot \boldsymbol{q}^{(k)},
\end{equation}
as in \citep{DeCVJo1996}. The contraction is taken over all indices except one of $\boldsymbol{q}^{(k+1)}$, say, the first one. For a fully geometric and intrinsic treatment, a Riemannian metric would be used. Using these relations, thermodynamic consistency \eqref{2nd_law_spat_EIT_higher} requires that
\begin{equation}
\sum_{k=1}^n \boldsymbol{q}^{(k)}\cdot \Big( - \frac{\nabla T}{T}\delta ^k_1 - \alpha_k \mathfrak{D}_t \boldsymbol{q}^{(k)} + \gamma_k T \operatorname{div}(\boldsymbol{q}^{(k+1)}) + T\gamma_{k-1}\nabla \boldsymbol{q}^{(k-1)}\Big) \geq 0,
\end{equation}
where we adopt the convention that $\boldsymbol{q}^{(i)}$ for $i=0$ and $i=n+1$ are absent. This condition is achieved by assuming the linear flux-force relations $\boldsymbol{q}^{(k)}= L^k \big( - \frac{\nabla T}{T}\delta ^k_1 - \alpha_k \mathfrak{D}_t \boldsymbol{q}^{(k)} + \gamma_k T \operatorname{div}(\boldsymbol{q}^{(k+1)}) + T\gamma_{k-1}\nabla \boldsymbol{q}^{(k-1)}\big) $, where $L_k\geq 0$.
Choosing $L^k= \frac{\kappa_k...\kappa_1}{\tau_{k-1}...\tau_1}T$, $\alpha_k = L_k/\tau_k$, and $\gamma_k=-\alpha_k/T$, one gets the Christov-Cattaneo model for higher-order heat fluxes
\begin{equation}
\tau_k\big( \mathfrak{D}_t \boldsymbol{q}^{(k)} + \operatorname{div}( \boldsymbol{q}^{(k+1)})\big) + \boldsymbol{q}^{(k)}= - \kappa_k \nabla\boldsymbol{q}^{(k-1)},\quad k=1,...,n,
\end{equation}
where we adopt the convention that $\boldsymbol{q}^{(0)}$ represents $T$ and $\boldsymbol{q}^{(n+1)}$ is absent.

\section{Conclusion}

In this paper, we have presented a variational framework for extended irreversible thermodynamics of continua, with a focus on heat-conducting viscous fluids. The key feature of this framework is that it consistently extends:

(i) The Hamilton principle of reversible continuum mechanics;

(ii) The variational formulation for classical irreversible thermodynamics.\\
This extension follows a systematic construction achieved through the following two steps:

(i) Introducing a nonequilibrium Lagrangian that incorporates thermodynamic fluxes as independent variables.

(ii) Extending the thermodynamic constraints to account for these new dependencies.

The transition from Hamilton's principle to the proposed variational framework is illustrated in Figure \ref{figure_1}.
As shown in this figure, these principles take a particularly simple form in the material (Lagrangian) description of continuum mechanics. Under material covariance (or relabeling symmetries), they can be equivalently expressed in the spatial (Eulerian) frame. Although the Eulerian formulation is more intricate, it follows systematically from its material counterpart.

This process, based on the ideas of reduction by symmetries, naturally determines the appropriate choice of the objective rate in the Eulerian description. In particular, it leads to the Cattaneo-Christov model for heat flux and the upper-convected Maxwell model (with the Truesdell rate) for viscous stress. The consistency of these models necessitates the inclusion of additional nonequilibrium viscous and thermal stresses, which are directly obtained from the variational principle. Regarding viscosity, this result stems from our initial choice of the second Piola-Kirchhoff stress tensor as the privileged stress description in the material formulation, which in turn implies the occurrence of the Truesdell rate in the Eulerian description. Choosing a different material stress tensor would have led to a different objective rate. In classical irreversible thermodynamics, these choices are all equivalent since these stresses are not time-differentiated.

The variational setting introduced in this paper has potential applications in the development of structure-preserving and thermodynamically consistent numerical methods, as demonstrated in \citep{GaGB2024}. It is also relevant for thermodynamically consistent modeling in areas where entropy production must be carefully accounted for (see \citep{GB2019,ElGB2021,GBPu2022}). Additionally, recasting this formulation within a Riemannian manifold framework could elucidate the role of the metric choice. We plan to explore these directions in future work.




\begin{thebibliography}{xx}



\bibitem[Angeles(2022)]{An2022}
Angeles, F.  [2022], Non-hyperbolicity of the inviscid Cattaneo-Christov system for compressible fluid flow in several space dimensions, \textit{The Quarterly Journal of Mechanics and Applied Mathematics} \textbf{75}(2), 147--170.

\bibitem[Angeles, M\'alaga, and Plaza(2020)]{AnMaPl2020}
Angeles, F., M\'alaga, C. and Plaza G. [2020], Strict dissipativity of Cattaneo-Christov systems for compressible fluid flow, \textit{Journal of Physics A: Mathematical and Theoretical}, \textbf{53} 065701.

\bibitem[Arnold(1966)]{Ar1966}
Arnold, V.~I. [1966], Sur la g\'eom\'etrie diff\'erentielle des groupes de Lie de dimension infinie et ses applications \`a l'hydrodynamique des fluides parfaits, \textit{Annales de l'Institut Fourier} \textbf{16}(1), 319--361.

\bibitem[Beris and Edwards(1994)]{BeEd1994}
Beris, A. N., and Edwards, B. J. [1994], \textit{Thermodynamics of Flowing Systems}. Oxford University Press, New York.


\bibitem[Bollada and Phillips(2012)]{BoPh2012}
Bollada, P. C. and T. N. Phillips, On the mathematical modelling of a compressible viscoelastic fluid, \textit{Arch. Rat. Mech.
Anal.} \textbf{205} (1), 1--26.


 
\bibitem[Cattaneo(1948)]{Ca1948}
Cattaneo, C.~R. [1948], Sulla conduzione del calore, \textit{Atti Seminario Mat. Fis. Univ. Modena}, \textbf{3}, 83--101.


\bibitem[Cattaneo(1958)]{Ca1958}
Cattaneo, C.~R. [1958], Sur une forme de l'\'equation de la chaleur \'eliminant le paradoxe d'une propagation instantan\'ee, \textit{Comptes Rendus} \textbf{247}(4): 431.


\bibitem[Christov(2009)]{Ch2009}
Christov, C.~I. [2009], On frame indifferent formulation of the Maxwell-Cattaneo model of finite-speed heat conduction, \textit{Mechanics Research Communications}, \textbf{36} , 481--486.



\bibitem[Cimmelli, Jou, Ruggeri, and V\'an(2014)]{CiJoRuVa2014}
Cimmelli, V.~A., D. Jou, T. Ruggeri, P. V\'an [2014], Entropy principle and recent results in nonequilibrium
theories, \textit{Entropy} \textbf{16}, 1756.

\bibitem[Crin-Barat, Kawashima, and Xu(2024)]{CBKaXu2024}
Crin-Barat, T., Kawashima, S., and Xu, J. [2024], The Cattaneo-Christov approximation of Fourier heat-conductive compressible fluids, \url{https://arxiv.org/abs/2404.07809}

\bibitem[Dedeurwaerdere, Casas-Vazquez, and Jou(1984)]{DeCVJo1996}
Dedeurwaerdere, T., Casas-Vázquez, J., Jou, D., and Lebon, G. [1996], Foundations and applications of a mesoscopic thermodynamic theory of fast phenomena, \textit{Physical Review E.}, \textbf{53}(1), 498--506.

\bibitem[Dhaouadi F, Gavrilyuk(2024)]{DhGa2024}
Dhaouadi F. and Gavrilyuk S. [2024], An Eulerian hyperbolic model for heat transfer derived via Hamilton's principle: analytical and numerical study, \textit{Proceedings of the Royal Society A}, \textbf{480}(2283), 20230440.

\bibitem[de Groot and Mazur(1969)]{dGMa1969}
de Groot, S.~R. and Mazur, P., \textit{Nonequilibrium Thermodynamics}, North-Holland.

\bibitem[Edwards and Beris(1991a)]{EdBe1991a}
Edwards, B.~J. and A.~N. Beris (1991a), Noncanonical Poisson bracket for nonlinear elasticity with extensions to viscoelasticity, \textit{Phys. A: Math. Gen.}, \textbf{24}, 2461--2480.

\bibitem[Edwards and Beris(1991b)]{EdBe1991b}
Edwards, B.~J. and A.~N. Beris (1991b), Unified view of transport phenomena based on the generalized bracket formulation, \textit{Ind. Eng. Chem. Res.}, \textbf{30}, 873--881.

\bibitem[Eldred and Gay-Balmaz(2020)]{ElGB2020}
Eldred, C. and Gay-Balmaz F. [2020], Single and double generator bracket formulations of multicomponent fluids with irreversible processes, \textit{J. Phys. A: Mathematical and Theoretical} \textbf{53}(39), 395701.

\bibitem[Eldred and Gay-Balmaz(2021)]{ElGB2021}
Eldred, C. and Gay-Balmaz F. [2021], Thermodynamically consistent semi-compressible fluids: a variational perspective, \textit{Journal of Physics A: Mathematical and Theoretical}, \textbf{54}, 345701.

\bibitem[Eldred, Gay-Balmaz, and Wu(2024)]{ElGBWu2024}
Eldred, C., Gay-Balmaz F., and Wu, M [2024], Geometric, variational, and bracket descriptions of fluid motion with open boundaries, \textit{Geometric Mechanics}, \textbf{1}(4), 325--381.

\bibitem[Gawlik and Gay-Balmaz(2024)]{GaGB2024}
Gawlik, E.~S. and Gay-Balmaz, F. [2024], Variational and thermodynamically consistent finite element discretization for heat conducting viscous fluid, \textit{Mathematical Models and Methods in Applied Sciences}, \textbf{34}(2), 243--284.

\bibitem[Gay-Balmaz(2019)]{GB2019}
Gay-Balmaz, F., [2019], A variational derivation of the nonequilibrium thermodynamics of a moist atmosphere with rain
process and its pseudoincompressible approximation, \textit{Geophysical \& Astrophysical Fluid Dynamics}, \textbf{113}:5-6, 428--465.


\bibitem[Gay-Balmaz(2024)]{GB2024}
Gay-Balmaz, F., [2024], General relativistic Lagrangian continuum theories -- Part I: reduced variational principles and junction conditions for hydrodynamics and elasticity,  \textit{Journal of nonlinear science} \textbf{34}(46). 


\bibitem[Gay-Balmaz, Marsden, and Ratiu(2012)]{GBMaRa2012}
Gay-Balmaz, F., Marsden, J.~E., and Ratiu, T.~S. [2012],
Reduced variational formulations in free boundary continuum mechanics, \textit{Journal of nonlinear science} \textbf{22}(4), 463--497.

\bibitem[Gay-Balmaz and Putkaradze(2022)]{GBPu2022}
Gay-Balmaz F. and V. Putkaradze [2022], Variational geometric approach to the thermodynamics of porous media,  \textit{Zeitschrift f\"ur Angewandte Mathematik und Mechanik}, \textbf{102}(11).

\bibitem[Gay-Balmaz and Putkaradze(2024)]{GBPu2024}
Gay-Balmaz F. and V. Putkaradze [2024], Thermodynamically consistent variational theory of porous media with a breaking component, \textit{Continuum Mech. Thermodyn.}, \textbf{36}, 75--105.

\bibitem[Gay-Balmaz and Yoshimura(2017a)]
{GBYo2017a}
Gay-Balmaz, F. and H.~Yoshimura [2017a], A {L}agrangian variational formulation for nonequilibrium thermodynamics. {P}art {I}: discrete systems, \emph{J. Geom. Phys.}, \textbf{111}, 169--193.

\bibitem[Gay-Balmaz and Yoshimura(2017b)]
{GBYo2017b}
Gay-Balmaz, F. and H.~Yoshimura [2017b], A {L}agrangian variational formulation for nonequilibrium thermodynamics. {P}art {II}: continuum systems, \emph{J. Geom. Phys.}, \textbf{111}, 194--212.


\bibitem[Gay-Balmaz and Yoshimura(2018a)]
{GBYo2018a} Gay-Balmaz, F. and H.~Yoshimura [2018a], A variational formulation of nonequilibrium thermodynamics for discrete open systems with mass and heat transfer, \emph{Entropy}, \textbf{20}  (3), 163.




\bibitem[Gay-Balmaz and Yoshimura(2019a)]{GBYo2019a}
Gay-Balmaz, F. and H.~Yoshimura [2019a], From {L}agrangian mechanics to nonequilibrium thermodynamics: a variational perspective,  \emph{Entropy}, \textbf{21}(1).

\bibitem[Gay-Balmaz and Yoshimura(2019b)]{GBYo2019b}
Gay-Balmaz, F. and H.~Yoshimura [2019b], A free energy Lagrangian variational formulation of the Navier-Stokes-Fourier system, \textit{Int. J. Geom. Methods Mod. Phys.}, \textbf{16}, 




\bibitem[Gay-Balmaz and Yoshimura(2023)]{GBYo2023}
Gay-Balmaz, F. and H.~Yoshimura [2023], Systems, variational principles and
interconnections in non-equilibrium thermodynamics, \textit{Phil. Trans. R. Soc. A} 381:20220280.


\bibitem[Glansdorff and Prigogine(1971)]
{GlPr1971}
Glansdorff, P. and I. Prigogine [1971], {\it Thermodynamic Theory of Structure, Stability, and Fluctuations}, Wiley-Interscience.

\bibitem[Green and Naghdi(1991)]
{GrNa1991}
Green, A. and P. Naghdi [1991], A re-examination of the basic postulates of thermomechanics,
\textit{Proc. Roy. Soc., London A} \textbf{432}(1885), 171-194.



\bibitem[Grmela and \"Ottinger(1997)]{GrOt1997}
Grmela, M. and H.-C. \"Ottinger [1997], Dynamics and thermodynamics of complex fluids. I. Development of a general formalism, \textit{Phys. Rev. E}, \textbf{56}, 6620--6632.


\bibitem[Gyarmati(1970)]
{Gyarmati1970}
Gyarmati, I. [1970], \textit{Nonequilibrium Thermodynamics: Field Theory and
Variational Principles}, Springer-Verlag, New York.

\bibitem[Herivel(1955)]{He1955}
Herivel, J. W. [1955], The derivation of the equations of motion of an ideal fluid by Hamilton's principle, \textit{Mathematical Proceedings of the Cambridge Philosophical Society}, \textbf{51}(2). Cambridge University Press.

\bibitem[Holm, Marsden, Ratiu(1998)]{HoMaRa1998}
Holm, D.~D., J.~E. Marsden, T.~S. Ratiu [1998], The Euler-Poincar\'e equations and semidirect products with applications to continuum theories, \textit{Adv. Math.} \textbf{137}, 1--81.


\bibitem[Ichiyanagi(1994)]
{Ichiyanagi1994}
Ichiyanagi M. [1994], Variational principles in irreversible processes, \textit{Phys. Rep.} \textbf{243}, 125--182.

\bibitem[Israel(1976)]{Is1976}
Israel, W. [1976], Nonstationary irreversible thermodynamics: a causal relativistic theory, \textit{Ann. Phys.} \textbf{100}, 310.

\bibitem[Jou(2020)]{Jo2020}
Jou, D. [2020], Relationships between rational extended thermodynamics and extended irreversible
thermodynamics, \textit{Phil. Trans. R. Soc. A} \textbf{378}: 20190172.


\bibitem[Jou, Casas-V\'azquez, and Lebon(2010)]{JoCVLe2010}
Jou, D., J. Casas-V\'azquez, and G. Lebon [2010], \textit{Extended Irreversible Thermodynamics}, Springer Netherlands.




\bibitem[Kaufman(1984)]{Ka1984}
Kaufman.A. [1984], Dissipative Hamiltonian systems: A unifying principle, \textit{Phys. Lett. A}, \textbf{100}, 419--422.

\bibitem[Kondepudi and Prigogine(1998)]{KoPr1998}
Kondepudi, D. and Prigogine, I. [1998], \textit{Modern Thermodynamics}, John Wiley \& Sons.

\bibitem[Landau and Lifshitz(1969)]{LaLi1969}
Landau, L.~D. and Lifshitz, E.~M. [1969], \textit{Mechanics}; Volume 1 of A Course of Theoretical Physics; Pergamon Press:
Oxford, UK, 1969.




\bibitem[Lavenda(1978)]
{Lavenda1978}
Lavenda, B.~H. [1978], \textit{Thermodynamics of Irreversible Processes}, Macmillan, London.


\bibitem[Lebon, Jou, and Casas-V\'azquez (1980)]{LeJoCV1980}
Lebon, G., D. Jou, and J. Casas-V\'azquez [1980], An extension of the local equilibrium hypothesis, \textit{J. Phys. A: Math. Gen.}, \textbf{13}, 275--290.



\bibitem[Lebon, Jou, and Grmela(2016)]{LeJoGr2016}
Lebon, G., D. Jou, Grmela, M. [2007], Extended Reversible and Irreversible Thermodynamics: A Hamiltonian Approach with Application to Heat Waves, \textit{Journal of Non-Equilibrium Thermodynamics}, \textbf{42}(2), 153--168.



\bibitem[Marsden and Hughes(1983)]{MaHu1983}
Marsden, J.~E. and T.~J.~R. Hughes [1983], \textit{Mathematical Foundations of Elasticity}, Prentice Hall, New York (reprinted by Dover, New York, 1994).

\bibitem[M\"uller(1966)]{Mu1966}
M\"uller, I. [1966], \textit{Zur Ausbreitungsgeschwindigkeit von St\"orungen in kontinuierlichen Medien}. Dissertation
TH, Aachen (1966).


\bibitem[M\"uller(1967)]{Mu1967}
M\"uller, I. [1967], Zum Paradoxon der W\"armeleitungstheorie, \textit{Zeitschrift f\"ur Physik} \textbf{198}.

\bibitem[M\"uller and Ruggeri(1998)]{MuRu1998}
Müller, I. and Ruggeri, T. [1998], \textit{Rational Extended Thermodynamics}, 2nd edn., Springer, New York.

\bibitem[M\"uller and Weiss(2012)]{MuWe2012}
M\"uller, I. and W. Weiss [2012], Thermodynamics of irreversible processes -- past and present, \textit{Eur. Phys. J. H.} \textbf{37}, 139 (2012).





\bibitem[Morrison(1986)]{Mo1986}
Morrison, P. [1986], A paradigm for joined Hamiltonian and dissipative systems, \textit{Physica D}, \textbf{18}, 410--419.


\bibitem[Onsager(1931)]
{Onsager1931}
Onsager, L. [1931], Reciprocal relations in irreversible processes I, \textit{Phys. Rev.} \textbf{37}, 405--426; Reciprocal relations in irreversible processes II, \textit{Phys. Rev.} \textbf{38}, 2265--2279.  

\bibitem[Onsager and Machlup(1953)]
{OnMa1953}
Onsager, L. and S. Machlup [1953], Fluctuations and irreversible processes, \textit{Phys. Rev.} \textbf{91}, 1505--1512.

\bibitem[Machlup  and Onsager(1953)]
{MaOn1953}
Onsager, L. and S. Machlup [1953], Fluctuations and irreversible processes II. Systems with kinetic energy. \textit{Phys. Rev.} \textbf{91}, 1512--1515.


\bibitem[\"Ottinger and Grmela(1997)]{OtGr1997}
\"Ottinger, H.-C. and M. Grmela [1997], Dynamics and thermodynamics of complex fluids. II. Illustrations of a general formalism, \textit{Phys. Rev. E}, \textbf{56}, 6633--6655.


\bibitem[Peshkov, Pavelka, Romenski, and Grmela(2018)]{PePaRoGr2018}
Peshkov I., M. Pavelka, E. Romenski, and M. Grmela [2018], Continuum mechanics and thermodynamics in the Hamilton and the Godunov-type formulations, \textit{Continuum Mechanics and Thermodynamics}, \textbf{30}(6), 1343--1378.


\bibitem[Prigogine(1947)]
{Prigogine1947}
Prigogine [1947], \textit{Etude Thermodynamique des Ph\'enom\`enes Irr\'eversibles}, Thesis, Paris: Dunod and Li\`ege: Desoer.

\bibitem[Ruggeri and Sugiyama(2015)]{RuSu2015}
Ruggeri,T. and  Sugiyama, M. [2015], \textit{Rational Extended Thermodynamics beyond the Monatomic Gas}, Springer.

\bibitem[Straughan(2010)]{St2010}
Straughan, B. [2010], Acoustic waves in a Cattaneo-Christov gas, \textit{Physics Letters A} \textbf{374}, 2667-2669.



\bibitem[Stueckelberg and Scheurer(1974)]{StSc1974}
Stueckelberg, E.~C.~G. and Scheurer, P.~B. \textit{Thermocin\'etique Ph\'enom\'enologique Galil\'eenne}; Birkh\"auser: Basel, Switzerland, 1974.

\bibitem[Taub(1949)]{Ta1949}
Taub, A. H. [1949], On Hamilton's principle for perfect compressible fluids, \textit{Proceedings of symposia in applied mathematics}, volume 1.


\bibitem[Tokatly and Pankratov(1999)]{ToPa1999}
Tokatly, I. and O. Pankratov, Hydrodynamic theory of an electron gas, \textit{Physical Review B}, \textbf{60}(23)

\bibitem[Van and Nyiri(1999)]{VaNy1989}
Van, P. and B. Nyiri [1999], Hamilton formalism and variational principle construction, \textit{Ann. Phys. (Leipzig)} \textbf{8} 4, 331--354.


\bibitem[von Helmholtz(1884)]{vH1884}
von Helmholtz, H. [1884], Studien zur Statik monocyklischer Systeme, \textit{Sitzungsberichte der K\"oniglich Preussischen Akademie der Wissenschaften zu Berlin}, 159--177.

 
\bibitem[von Laue(1921)]{vL1921}
von Laue, M. [1921], \textit{Relativit\"atstheorie}, Braunschweig, Vieweg, vol. 1, p. 248.
 
\bibitem[Vernotte(1958)]{Ve1958}
Vernotte, P. [1958], La v\'eritable \'equation de la chaleur, \textit{Compt. Rend. Acad. Sci. Paris}, \textbf{247}, 
2103--2107.


\bibitem[Woods(1975)]{Wo1975}
Woods, L.~C. [1075], \textit{The Thermodynamics of Fluid Systems}, Clarendon Press Oxford.

\bibitem[Ziegler(1968)]
{Ziegler1968}
Ziegler, H. [1968],  A possible generalization of Onsager's theory, in H. Barkus and L.I. Sedov (eds.), \textit{Irreversible Aspects of Continuum Mechanics}, Springer, New York.


\end{thebibliography}

\end{document}